\documentclass[twocolumn,tighten,times]{aastex62}

\usepackage{bm}
\usepackage[utf8]{inputenc}
\usepackage{amsmath}
\usepackage{csquotes}
\usepackage{longtable}
\usepackage{multirow}
\usepackage{soul} 
\usepackage{fancyvrb}
\VerbatimFootnotes

\usepackage[ruled]{algorithm2e}
\usepackage[hang,flushmargin]{footmisc}

\usepackage{listings}

\definecolor{codegreen}{rgb}{0,0.6,0}
\definecolor{codegray}{rgb}{0.5,0.5,0.5}
\definecolor{codemauve}{rgb}{0.58,0,0.82}

\usepackage{graphicx}
\usepackage{booktabs}
\usepackage{appendix}

\newcommand{\morpheus}{\textit{Morpheus}}
\newcommand{\JWST}{\textit{JWST}}
\newcommand{\LSST}{\textit{LSST}}

\SetArgSty{textrm}
\DontPrintSemicolon

\makeatletter
\newcommand{\algorithmfootnote}[2][\footnotesize]{%
  \let\old@algocf@finish\@algocf@finish
  \def\@algocf@finish{\old@algocf@finish
    \leavevmode\rlap{\begin{minipage}{\linewidth}
    #1#2
    \end{minipage}}%
  }%
}
\makeatother

\newcommand{\bc}{c}
\newcommand{\bndmax}{nd_{max}}
\newcommand{\bq}{q}
\newcommand{\bp}{p}
\newcommand{\br}{r}
\newcommand{\bx}{x}

\newcommand{\by}{y}
\newcommand{\bb}{b}
\newcommand{\bmask}{m}
\newcommand{\bs}{s}
\newcommand{\bsm}{sm}
\newcommand{\bh}{h}
\newcommand{\bdb}{db}
\newcommand{\bw}{w}
\newcommand{\IOU}{\mathcal{I}_\mathcal{U}}
\newcommand{\ACC}{\mathcal{A}}

\submitjournal{AAS Journals}
\accepted{April 6, 2020}

\begin{document}

\title{\textit{Morpheus}: A Deep Learning Framework For Pixel-Level Analysis
       of Astronomical Image Data}

\correspondingauthor{Ryan Hausen}
\email{rhausen@ucsc.edu, brant@ucsc.edu}

\author[0000-0002-8543-761X]{Ryan Hausen}
\affiliation{Department of Computer Science and Engineering, University of
             California, Santa Cruz, 1156 High Street, Santa Cruz, CA 95064 USA}

\author[0000-0002-4271-0364]{Brant E. Robertson}
\affiliation{Department of Astronomy and Astrophysics, University of California,
             Santa Cruz, 1156 High Street, Santa Cruz, CA 95064 USA}
\affiliation{Institute for Advanced Study, 1 Einstein Drive, Princeton, NJ
             08540 USA}

\begin{abstract}
We present \morpheus{}, a new model for generating pixel-level
morphological classifications of astronomical sources. \morpheus{} leverages
advances in deep learning to perform source detection, source segmentation, and
morphological classification pixel-by-pixel via a semantic segmentation
algorithm adopted from the field of computer vision. By utilizing morphological
information about the flux of real astronomical sources during object detection,
\morpheus{} shows resiliency to false-positive identifications of
sources. We evaluate \morpheus{} by performing source detection, source
segmentation, morphological classification on the {\it Hubble Space Telescope}
data in the five CANDELS fields with a focus on the GOODS South field,
and demonstrate a high completeness in recovering known GOODS South
3D-HST sources with $H<26$ AB. We release the code publicly, provide
online demonstrations, and present an interactive visualization of the
\morpheus{} results in GOODS South.
\end{abstract}


\section{Introduction} \label{section:intro}

Morphology represents the structural end state of the galaxy formation process.
Since at least \citet{hubble1926a}, astronomers have connected the morphological
character of galaxies to the physics governing their formation. Morphology can
reflect the initial conditions of galaxy formation, dissipation, cosmic
environment and large-scale tidal fields, merger and accretion history, internal
dynamics, star formation, the influence of supermassive black holes, and a range
of other physics
\citep[e.g.,][]{binney1978a,dressler1980a,binney1987a,djorgovski1987a,dressler1987a,bender1992a,tremaine2002a}.
The development of morphological measures for galaxies, therefore,
comprises an important task in observational astronomy. To help realize the
potential of current and future surveys for understanding galaxy formation
through morphology, this work presents \morpheus{}, a deep learning-based model
for the simultaneous detection and morphological classification of objects
through the pixel-level semantic segmentation of large astronomical image
datasets.

The established connections between morphology and the physics of galaxy
formation run deep, and the way these connections manifest themselves
observationally depends on the measures of morphology used. Galaxy size
and surface brightness profile shape have served as common
proxies for morphology, as quantitatively measured from the light distribution
of objects \citep{devaucouleurs1959a,sersic1968a,peng2010a}. Size, radial
profile, and isophotal shape or ellipticity vary with stellar mass and
luminosity
\citep[e.g.,][]{kormendy1977a,roberts1994a,shen2003a,sheth2010a,bruce2012a,vanderwel2012a,vanderwel2014a,morishta2014a,huertas-company2015a,allen2017a,jiang2018a,miller2019a,zhang2019a}.
When controlled for other variables, these measures of galaxy morphology may
show variations with cosmic environment
\citep{dressler1997a,smail1997a,cooper2012a,huertas-company2016a,kawinwanichakij2017a},
redshift
\citep{abraham2001a,trujillo2004a,conselice2005a,elmegreen2005a,trujillo2006a,lotz2008a,vandokkum2010a,patel2013a,shibuya2015a},
color \citep{franx2008a,yano2016a}, star formation rate or quiescence
\citep{toft2007a,zirm2007a,wuyts2011a,bell2012a,lee2013a,whitaker2015a},
internal dynamics \citep{bezanson2013a}, the presence of active galactic nuclei
\citep{kocevski2012a,bruce2016a,powell2017a}, and stellar age
\citep{williams2017a}. The presence and size of bulge, disk, and bar components
also vary with mass and redshift
\citep{sheth2008a,simmons2014a,margalef-bentabol2016a,dimauro2018a}, and provide
information about the merger rate \citep[e.g.,][]{lofthouse2017a,weigel2017a}.
Galaxy morphology encodes a rich spectrum of physical processes and
can augment what we learn from other galaxy properties.

While complex galaxy morphologies may be easily summarized with qualitative
descriptions (e.g., ``disky'', ``spheroidal'', ``irregular''), providing
quantitative descriptions of this complexity represents a long-standing goal for
the field of galaxy formation and has motivated ingenuitive analysis methods
including measures of galaxy asymmetry, concentration, flux distribution
\citep[e.g.,][]{abraham1994a,abraham1996a,conselice2000a,conselice2003a,lotz2004a},
shapelet decompositions \citep{kelly2004a,kelly2005a},
morphological principal component analyses \citep{peth2016a}, and
unsupervised morphological hierarchical classifications \citep{hocking2017a}.
These measures provide well-defined characterizations of the surface brightness
distribution of galaxies and can be connected to their underlying physical state
by, e.g., calibration through numerical simulation \citep{huertas-company2018a}.
The complementarity between these quantitative measures and qualitative
morphological descriptions of galaxies means that developing both classes of
characterizations further can continue to improve our knowledge of galaxy
formation physics.

Characterizing large numbers of galaxies with descriptive classifications
simultaneously requires domain knowledge of galaxy morphology (``expertise''),
the capability to evaluate quickly each galaxy (``efficiency''), a capacity to
work on significant galaxy populations (``scalability''), some analysis of the
data to identify galaxy candidates for classification (``pre-processing''), a
presentation of galaxy images in a format that enables the characteristic
structures to be recognized (``data model''), and an output production of
reliable classifications (``accuracy''). Methods for the descriptive
classification of galaxy morphology have addressed these challenges in
complementary ways.

Perhaps the most important and influential framework for galaxy morphological
classification to date has been the Galaxy Zoo project
\citep{lintott2008a,lintott2013a,lintott2017a}, which enrolls the public in the
analysis of astronomical data including morphological classification. This
project has addressed the expertise challenge by training users in the
classification of galaxies and statistically accounting for the distribution of
users' accuracies. The efficiency of users varies, but by leveraging the power
of the public interest and enthusiasm, and now machine learning
\citep{beck2018a,walmsley2019a}, the project can use scalability to offset
variability in the performance of individual users. The pre-processing and
delivery of suitable images to the users has required significant investment and
programming, but has led to a robust data model for both the astronomical data
and the data provided by user input. Science applications of Galaxy Zoo include
quantitative morphological descriptions of $\sim$50,000 galaxies
\citep{simmons2017a} in the CANDELS survey \citep{grogin2011a,koekemoer2011a},
probes of the connection between star formation rate and morphology in spiral
galaxies \citep{willett2015a}, and measuring galaxy merger rates
\citep{weigel2017a}.

Other efforts have emphasized different dimensions of the morphological
classification task. \citet{kartaltepe2015a} organized the visual classification
of $\sim$10,000 galaxies in CANDELS by a team of dozens of professional
astronomers. This important effort performed object detection and source
extraction on the CANDELS science data, assessed their completeness, and
provided detailed segmentation maps of the regions corresponding to classified
objects. The use of high expertise human classifiers leads to high accuracy, but
poses a challenge for scalability to larger samples. The work of
\citet{kartaltepe2015a} also leveraged a significant investment in the
preprocessing and presentation of the data to their users with a custom
interface with a high-quality data model for the results.

Leveraging human classifiers, be they highly expert teams or well-calibrated
legions, to provide descriptive morphologies for forthcoming datasets will prove
challenging. These challenges motivate a consideration of other
approaches, and we present two salient examples in {\it James Webb Space
Telescope} \citep[\JWST;][]{gardner2006a} and the Large Synoptic Survey
Telescope \citep[\LSST;][]{ivezic2019a,lsst2009a}.

\JWST{} enables both sensitive infrared imaging with {\it NIRCam} and
multiobject spectroscopy with {\it NIRSpec} free of atmospheric attenuation. The
galaxy population discovered by \JWST{} will show a rich range of morphologies,
star formation histories, stellar masses, and angular sizes
\citep{williams2018a}, which makes identifying {\it NIRCam}-selected samples for
spectroscopic follow-up with {\it NIRSpec} challenging. The efficiency gain of
parallel observations with {\it NIRCam} and {\it NIRSpec} will lead to programs
where the timescale for constructing {\it NIRCam}-selected samples will be very
short ($\sim$2 months) to enable well-designed parallel survey geometries. For
this application, the ability to generate quick morphological classifications
for thousands of candidate sources will enhance the spectroscopic target
selection in valuable space-based observations.

\LSST{} presents a challenge of scale, with an estimated 30 billion astronomical
sources, including billions of galaxies over $\sim$17,000 $\deg^{2}$
\citep{lsst2009a}. The morphological classification of these galaxies will
require the development of significant analysis methods that can both scale to
the enormity of the \LSST{} dataset and perform well enough to allow imaging
data to be reprocessed in pace with the \LSST{} data releases. Indeed,
morphological classification methods have been identified as keystone
preparatory science tasks by in the \LSST{} Galaxies Science Roadmap
\citep[][see also \citealt{robertson2019a}.]{robertson2017a}.

Recently, advances in the field of machine learning called deep learning have
enjoyed success in morphological classification. \citet{dielman2015a} (D15) and
\citet{dai2018b} use deep learning to classify the Galaxy Zoo Survey.
\citet{huertas-company2015a} used a deep learning model derived from D15 and the
classifications from K15 to classify the CANDELS survey. \citet{gonzales2018a}
used deep learning to perform galaxy detection and morphological classification,
an approach that has also been used to characterize Dark Energy Survey galaxy
morphologies \citep{tarsitano2018a}. Deep learning models have been further
applied to infer the surface brightness profiles of galaxies
\citep{tuccillo2018a} and measure their fluxes  \citep{boucard2019a}, and now to
simulate entire surveys \citep{smith2019a}.

Here, we extend previous efforts by applying a semantic segmentation algorithm
to both classify pixels and identify objects in astronomical images using our
deep learning framework called \morpheus. The software architecture of the
\morpheus{} framework is described in Section \ref{section:framework}, with the
essential convolutional neural network and deep learning components reviewed in
Appendix \ref{appendix:dl}. The \morpheus{} framework has been engineered by
using TensorFlow \citep{tensorflow2015} implementations of these components to
perform convolutions and tensorial operations, and is not a port of existing
deep learning frameworks or generated via ``transfer
learning''\citep[e.g.,][]{pratt1993a} of existing frameworks pre-trained on
non-astronomical data such as ImageNet \citep{deng2009a}.

We train \morpheus{} using multiband Flexible Image Transport System
\citep[FITS;][]{wells1981a} images of CANDELS galaxies visually classified by
\citet{kartaltepe2015a} and their segmentation maps derived from standard {\it
sextractor} analyses \citep{bertin1996a}. The training procedure is described in
Section \ref{section:training}, including the ``loss function'' used to optimize
the \morpheus{} framework. Since \morpheus{} provides local estimates of whether
image pixels contain source flux, the \morpheus{} output can be used to perform
source segmentation and deblending. We present fiducial segmentation and
deblending algorithms for \morpheus{} in Section \ref{section:segmentation}.

We then apply \morpheus{} to the Hubble Legacy Fields \citep{illingworth2016a}
reduction of the CANDELS and GOODS data in the GOODS South region, the
v1.0 data release \citep{grogin2011a,koekemoer2011a} for the other four CANDELS
regions, and generate FITS data files of the same pixel format as the input FITS
images, each containing the pixel-by-pixel model classifications of the image
data into {\it spheroid}, {\it disk}, {\it irregular},
{\it point source/compact}, and {\it background} classes, as described in
Section \ref{section:products}. We release publicly these \morpheus{}
pixel-level classification data products and detailed them in Appendix
\ref{appendix:data_release}. We evaluate the performance of \morpheus{} in
Section \ref{section:performance}, including tests that use the catalog of
3D-HST photometric sources \citep{skelton2014a,momcheva2016a} to measure the
completeness of \morpheus{} in recovering sources as a function of source
magnitude. We find that \morpheus{} is highly complete (${>}90\%$) for sources
up to one magnitude fainter than objects used to train the model. Using the
\morpheus{} results, we provide estimates of the morphological classification of
3D-HST sources as a public value-added catalog, described in Section
\ref{section:vac}. In Section \ref{section:discussion}, we discuss applications
of \morpheus{} and semantic segmentation, which extend well beyond morphological
classification, and connect the capabilities of \morpheus{} to other research
areas in astronomical data analysis. We publicly release the \morpheus{} code,
provide on-line tutorials for using the framework via Jupyter notebooks, and
present an interactive website to visualize the \morpheus{} classifications and
segmentation maps in the context of the HLF images and 3D-HST catalog. These
software and data releases are described in Appendices
\ref{appendix:code_release}, \ref{appendix:tutorial}, and
\ref{appendix:data_release}. A summary of our work is presented with some
conclusions in Section \ref{section:summary}. Throughout the paper, we have used
the AB magnitude system \citep{oke1983a} and assumed a flat $\Lambda$CDM
universe ($\Omega_m=0.3$, $\Omega_\Lambda=0.7$) with a Hubble parameter $H_0=70$
km/s/Mpc when necessary.

\section{\morpheus{} Deep Learning Framework}
\label{section:framework}


\morpheus{} provides a deep learning framework for analyzing astronomical images
at the pixel level. Using a semantic segmentation algorithm, \morpheus{}
identifies which pixels in an image are likely to contain source flux and
separates them from ``background'' or sky pixels. \morpheus{},
therefore, allows for the definition of corresponding segmentation
regions or ``segmentation maps'' by finding contiguous regions of source pixels
distinct from the sky. Within the same framework, \morpheus{} enables for
further classification of the source pixels into additional ``classes''. In this
paper, we have trained \morpheus{} to classify the source pixels into
morphological categories ({\it spheroid}, {\it disk}, {\it irregular}, {\it
point source/compact}, and {\it background}) approximating the visual
classifications performed by the CANDELS collaboration in K15. These source
pixel classes identified by \morpheus{} could, in principle, be
trained to reproduce other properties of the galaxies, such as, e.g.,
photometric redshift, provided a sufficient training dataset is available. In
the sections below, we describe the architecture of the \morpheus{} deep
learning framework. Readers unfamiliar with the primary computational elements
of deep learning architectures may refer to Appendix \ref{appendix:dl} where
more details are provided.

\subsection{Input Data}
\label{section:input_data}

We engineered the \morpheus{} deep learning framework to accept astronomical
image data as direct input for pixel-level analysis. \morpheus{} operates on
science-quality FITS images, with sufficient pipeline processing (e.g., flat
fielding, background subtraction, etc.) to enable photometric analysis.
\morpheus{} accepts multiband imaging data, with a FITS file for each of the
$n_b$ bands used to train the model (see Section \ref{section:training}). The
pixel format of the input FITS images (or image region) matches the format of
FITS images used to perform training, reflecting the size of the convolutional
layers of the neural network determined before training. \morpheus{} allows for
arbitrarily large images to be analyzed by subdividing them into regions that
the model processes in parallel, as described in Section
\ref{section:large_images} below.

For the example application of morphological classification presented in this
paper, we use the $F606W$($V$), $F850LP$($z$), $F125W$($J$), and $F160W$($H$)
band images from
Hubble Space Telescope for training, testing, and our final analysis. Our
training and testing images were FITS thumbnails and segmentation maps provided
by \citet{kartaltepe2015a}. Once trained, \morpheus{} can be applied to
arbitrarily large images via a parallelization scheme described below in Section
\ref{section:large_images}. We have used the CANDELS public release data
\citep{grogin2011a,koekemoer2011a} in additional performance tests and the
Hubble Legacy Fields v2.0 data \citep{illingworth2016a} for our \morpheus{} data
release.

We note that the approach taken by \morpheus{} differs from deep learning models
that use traditional image formats, e.g., three-color Portable
Network Graphics (PNG) or Joint Photographic Experts Group (JPEG) images as
input. Using PNG or JPEG files as input is convenient because deep learning
models trained on existing PNG or JPEG datasets, such as ImageNet
\citep{deng2009a, russakovsky2015a}, can be retrained via transfer learning  to
classify galaxies. However, the use of these inputs requires additional
pre-processing beyond the science pipeline, including arbitrary decisions
about how to weight the FITS images to represent the channels of the multicolor
PNG or JPEG. With the goal of including \morpheus{} framework analyses as part
of astronomical pipelines, we have instead used FITS images directly as input to
the neural network.

\subsection{Neural Network}
\label{section:nn}

\begin{figure}
\centering
\includegraphics[width=\linewidth]{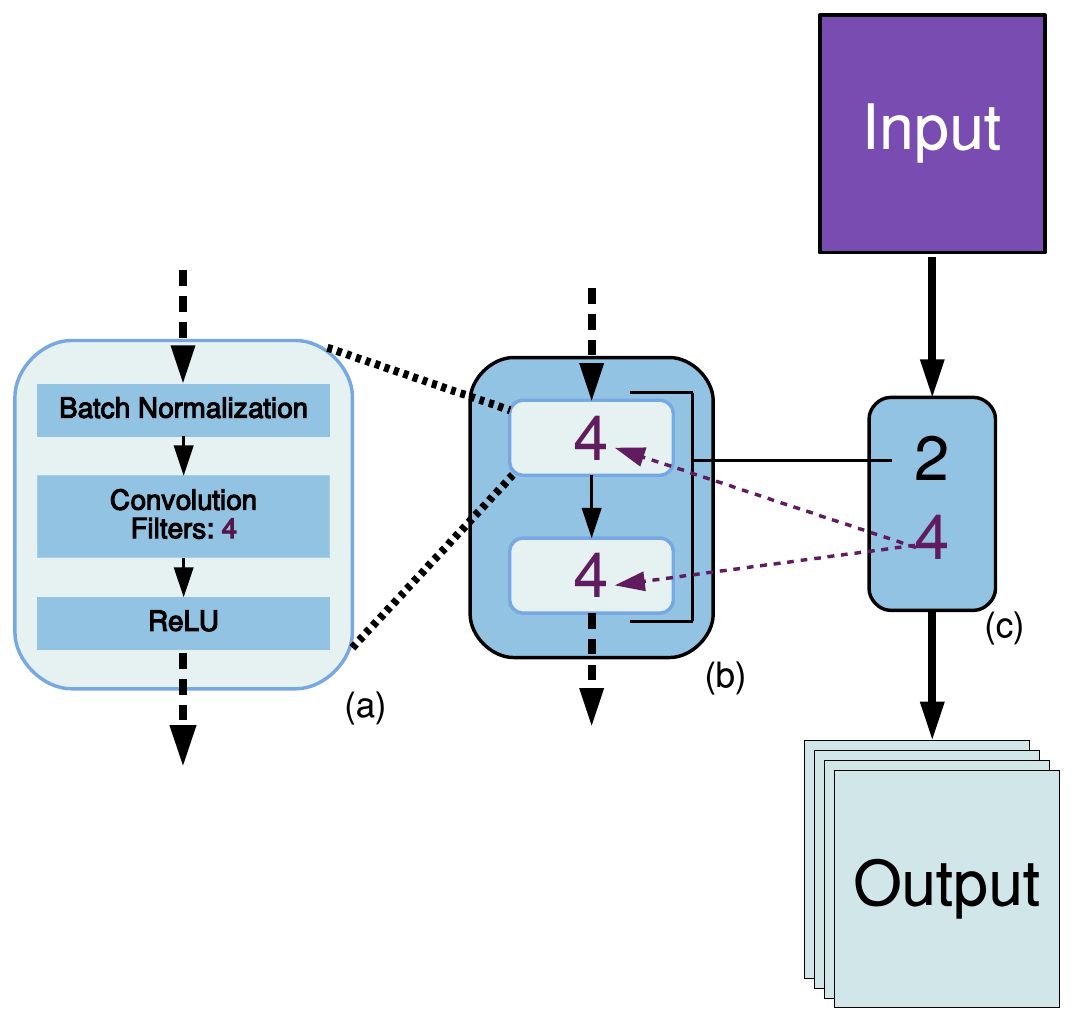}
\caption{Diagram of a single block in the \morpheus{}
         neural network architecture (Figure \ref{fig:architecture}). Panel (c)
         shows a single block from the architecture, parameterized by the number
         $P$ (black) of block operations and the number $Q$ (purple) of
         convolutional artificial neurons (CANs; Section
         \ref{section:convolution}) in all of the convolutional layers within
         the block. Panel (b) shows an example zoom-in where there are $P=2$
         groups of $Q=4$ block operations. Panel (a) shows a zoom-in on a block
         operation, which consists of batch normalization, $Q=4$ CANs, and a
         rectified linear unit (ReLU). In the notation of Equation
         \ref{eq:block_op}, this block operation would be written as
         $\text{OP}_4(\mathbf{X})$.}
\label{fig:block}
\end{figure}

\morpheus{} uses a neural network inspired by the U-Net architecture
\citep[][See Section \ref{section:unet}]{ronneberger2015a}
and is implemented using Python 3 \citep{rossum1995} and the TensorFlow library
\citep{tensorflow2015}. We construct \morpheus{} from a series of ``blocks''
that combine multiple operations used repeatedly by the model.
Each block performs a sequence of ``block operations''. Figure \ref{fig:block}
provides an illustration of a \morpheus{} block and its block operations. Block
operations are parameterized by the number $Q$ of convolved output images, or
``feature maps'', they produce, one for each convolutional artificial neuron in
the layer. We describe this process in more detail below.

Consider input data $\mathbf{X}$, consisting of $K$ layers of images with
$N\times M$ pixels. We define a block operation on $\mathbf{X}$ as

\begin{equation} \label{eq:block_op}
    \text{OP}_Q(\mathbf{X}) = \text{ReLU}(\text{CONV}_Q(\text{BN}(\mathbf{X})),
\end{equation}
\noindent
where ReLU is the Rectified Linear Unit activation function \citep[ReLU;][See
also Appendix \ref{section:an}]{hahnloser2000a,lecun2015a}, $\text{CONV}_Q$ is a
convolutional layer (see Appendix \ref{section:convolution}) with a number $Q$
convolutional artificial neurons (see Appendix \ref{section:convolution}), and
BN is the batch normalization procedure \citep[][and Appendix
\ref{section:batch_normalization}]{ioffe2015a}. Note that the values of $Q$
appearing in $\text{OP}_Q$ and $\text{CONV}_Q$ are equal. For example,
$\text{OP}_4$ would indicate that the convolutional layer within the
$\text{OP}_4$ function has 4 convolutional artificial neurons. Unless stated
otherwise, all inputs into a convolutional layer are zero-padded to
preserve the width and height of the input, and all convolutional
artificial neurons have kernel dimensions $3 \times 3$. Given Equation
\ref{eq:block_op}, for an input $\mathbf{X}$ with dimensions $N \times M \times
K$ the output of the function $\text{OP}_4(\mathbf{X})$ would have dimensions $N
\times M \times 4$.

Equation \ref{eq:block_op} allows for a recursive definition of a
function describing a series of block operations, where the input data to one
block operation consist of the output from a previous block operation. This
recursion can be written as
\begin{equation} \label{eq:block_op_recursive}
    \text{OP}^P_Q(\mathbf{X}) =
    \begin{cases}
        \mathbf{X}, & \text{ if } P = 0 \\
        \text{ReLU}(\text{CONV}_Q(\text{BN}(\text{OP}_Q^{P-1}(\mathbf{X}))) & \text{ if } P > 0
    \end{cases}.
\end{equation}
\noindent
Equation \ref{eq:block_op_recursive} introduces a new parameter $P$, shown with
a superscript in $\text{OP}_Q^P$. The parameter $P$ establishes the conditions
of a base case for the recursion. Note that in Equation
\ref{eq:block_op_recursive} the input $\mathbf{X}$ is processed directly when
$P=1$, and when $P>1$ the input to the $\text{OP}_Q^P$ function is the output
from $\text{OP}_Q^{P-1}$. It can be seen from the formulation of Equations
\ref{eq:block_op} and \ref{eq:block_op_recursive} that $OP_Q(\mathbf{X}) =
OP_Q^1(\mathbf{X})$.

Since a block performs a number $P$ block operations, we
can define a block mathematically as
\begin{equation} \label{eq:block}
    \text{BLOCK}(Q, P, \mathbf{X}) = \text{OP}^P_Q(\mathbf{X}).
\end{equation}
\noindent
An example block and its block operations can be seen diagrammatically in Figure
\ref{fig:block}. With these definitions, we can present the neural network
architecture used in \morpheus{}.

\begin{figure*}
\centering
\includegraphics[width=\textwidth]{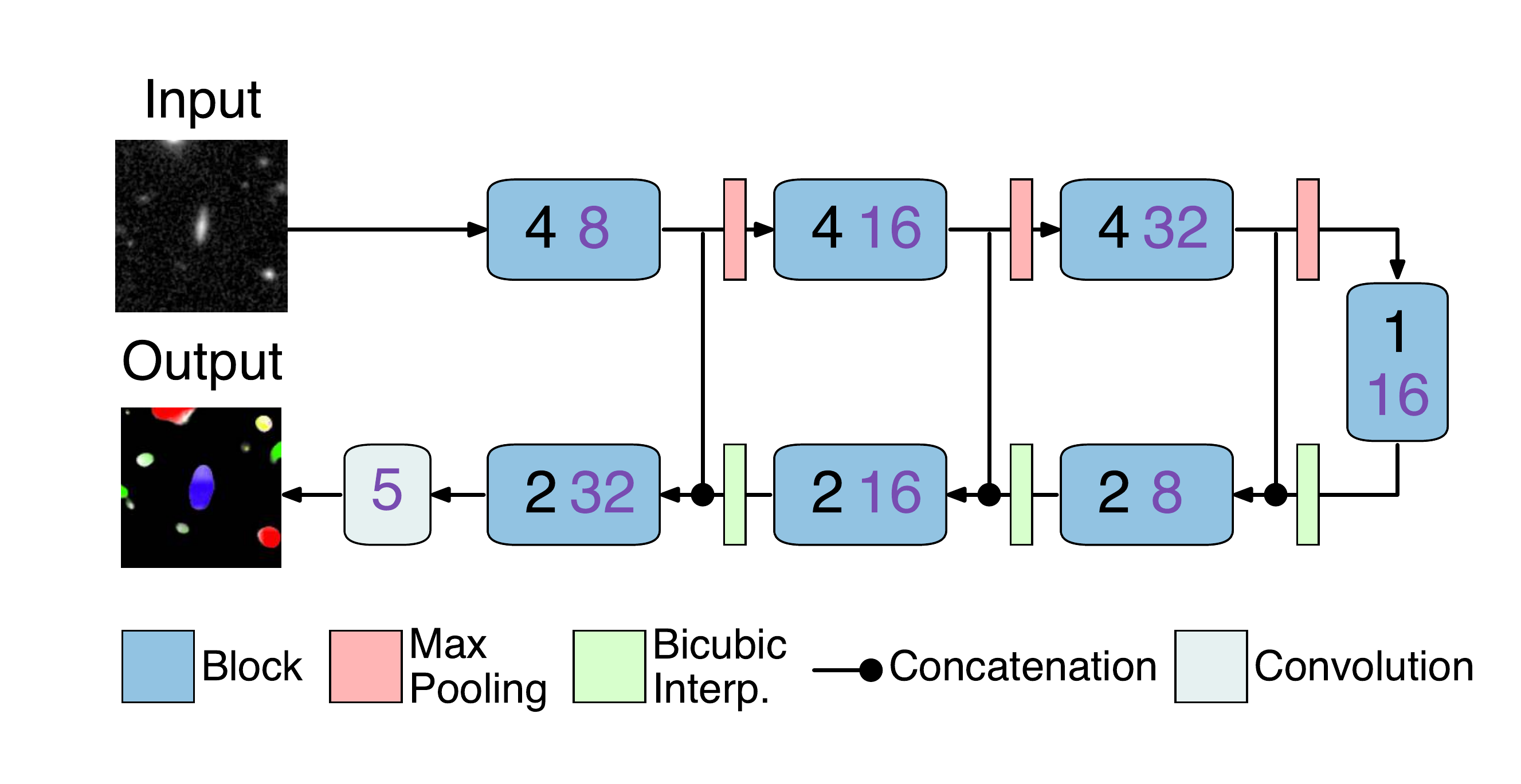}
\caption{Neural network architecture of the \morpheus{} deep learning framework,
         following a U-Net \citep{ronneberger2015a} configuration. The input to
         the model \morpheus{} consists of astronomical FITS images in $n_b$
         bands (upper left). These images are processed through a series of
         computational blocks (sky blue rectangles), each of which
         applies $P$ (black numbers) block operations
         consisting of a batch normalization and multiple convolutional layers
         producing $Q$ (purple numbers) feature maps. The blocks are described
         in more detail in Figure \ref{fig:block}. During the contraction phase
         of the model, max-pooling layers (salmon rectangles) are applied
         to the data to reduce the pixel size of the images by taking local
         maxima of $2\times2$ regions. The contraction phase is followed by an
         expansion phase where the output
         feature maps from each block are expanded
         by a $2\times2$ factor via bicubic interpolation (green rectangles) and
         concatenated with the  output from the corresponding block in the
         contraction phase. The output from the last block is processed through
         a set of convolutional layers (light blue box with $Q=5$) that result
         in a feature map for each classification
         in the model. These ``classification images'' are normalized to
         sum to unity pixel-by-pixel. In this paper, the classification images
         are {\it spheroid}, {\it disk}, {\it irregular}, {\it point
         source/compact}, and {\it background}.}
\label{fig:architecture}
\end{figure*}

Like the U-Net architecture, the \morpheus{} architecture consists of a
contraction phase and an expansion phase. The contraction phase consists of
three blocks with parameters $(P=4, Q=8)$, $(P=4, Q=16)$, and $(P=4, Q=32)$.
Each block is followed by a max-pooling operation with size=($2 \times
2$) (see Section \ref{section:pooling}), halving the width and height of its
input. After the contraction phase there is a single intermediary block
preceding the expansion phase with the parameters $(P=1, Q=16)$. The expansion
phase consists of three blocks with the parameters $(P=2, Q=8)$, $(P=2, Q=16)$,
$(P=2, Q=32)$. Each block is preceded by a bicubic interpolation operation that
doubles the width and the height of its input. Importantly, the output from each
block in the contraction phase is concatenated (see Section
\ref{sec:concatenation}) with the output from the bicubic interpolation
operation in the expansion phase whose output matches its width and height (see
Figure \ref{fig:architecture}). The output from the final block in the expansion
phase is passed through a single convolutional layer with 5 convolutional
artificial neurons. A softmax operation (see Equation \ref{eqn:softmax}) is
performed on the values in each pixel, ensuring the values sum to unity. The
final output is a matrix with the same width and height as the input into the
network, but where the last dimension, 5, now represents a classification
distribution describing the confidence the corresponding pixel from the input
belongs to one of the 5 specified morphological classes.

The blocks in \morpheus{} are organized into the U-Net structure, shown in
Figure \ref{fig:architecture}. The model proceeds clockwise, starting from
``Input'' on the upper left through to ``Output'' on the lower left. The very
first step involves the insertion of the input FITS images into the model. Each
FITS image is normalized to have a mean of 0 and unit variance before processing
by \morpheus{}. We will refer to the number of input bands as $n_b$, and in the
application presented here, we take $n_b=4$ (i.e., $VzJH$). The input
images each have pixel dimensions $N\times M$, and we can,
therefore, consider the astronomical input data to have dimensions
$N\times M \times n_b$. Only the first block operation takes the FITS images as
input, and every subsequent block operation in the model takes {the}
output from previous blocks as input.

The first convolution in the first block operation convolves the normalized
$N\times M \times n_b$ astronomical data with three-dimensional kernels of size
$n_k^2 \times n_b$, and each element of the kernel is a variable parameter of
the model to be optimized. The convolutions operate only in the two pixel
dimensions, such that $n_b$ convolutions are performed, one for each $N\times M$
pixel image, using a different $n_k \times n_k$ kernel for each convolution. The
$n_b$ convolved images are then summed pixel by pixel to create an output
feature map of size $N\times M$. The convolutional layer repeats this process
$Q$ times with different kernels, generating $Q$ output feature maps and an
output dataset of size $N \times M \times Q$. For the first block in \morpheus{}
we use $Q=8$ (see Figure \ref{fig:architecture}). After the first convolution on
the astronomical data, every subsequent convolution in the first block has both
input and output data of size $N \times M \times Q$.

Each block performs a number $P$ block operations, resulting in output data with
dimensions of $N\times M \times Q$ emerging from the block. The number of
feature maps $Q$ changes with each block. For a block producing $Q$ filters, if
the data incoming into the block has size $N\times M \times Q'$ with $Q'\ne Q$,
then the first convolutional layer in the first block operation will have $Q$
kernels of size $n_k^2 \times Q'$. All subsequent convolutional layers in the
block will then ingest and produce data of size $N\times M \times Q$ by using
kernels of size $n_k^2 \times Q$.

We can apply further operations on the data in between the blocks, and the
character of these operations can affect the dimensions of the data. The first
half of the model is a contraction phase, where each block is followed by a
max-pooling operation \citep[][and Appendix
\ref{section:pooling}]{ciresan2012a}. The max-pooling is applied to each
feature map output by the block, taking the local maximum over small areas
within each feature map (in the version of \morpheus{} presented here, a
$2\times2$ pixel region) and reducing the size of the data input to the next
block by the same factor. For this paper, the contraction phase in the
\morpheus{} framework uses three pairs of blocks plus max-pooling layers.

After the contraction phase, the model uses a series of blocks,
bicubic interpolation layers, and data concatenations in an expansion phase to
grow the data back to the original format. Following each block in the expansion
phase, a bicubic interpolation layer expands the feature maps by the same areal
factor as the max-pooling layers applied in the contraction phase
($2\times2$ in the version of \morpheus{} presented here). The output feature
maps from the interpolation layers are concatenated with the output feature maps
from the contraction phase blocks where the data have the same format. Finally,
the output from the last block in the expansion phase is input into a
convolutional layer that produces the final output images that we call
``\morpheus{} classification images'', one image for each class. The pixel
values in these images contain the model estimates for their classification,
normalized such that the element-wise sum of the classification images equals
unity.  For this paper, where we are performing galaxy morphological
classification, there are five classification images ({\it spheroid}, {\it
disk}, {\it irregular}, {\it point source /compact}, and {\it background}).

As the data progresses through the model, the number of feature maps and their
shapes change owing to the max-pooling and interpolation layers. For
reference, in Table \ref{table:architecture}, we list the dimensions of
the data at each stage in the model, assuming input images in $n_b$
bands, each with $N\times M$ pixels, and a total of $n_c$ classification images
produced by the model.

\begin{table}[]
	\centering
	\begin{tabular}{c c c}
	\toprule
    Layer & Input & Output Dimensions \\
	\midrule
	Input Images 	& $n_b$ Bands, $N\times M$ Pixels& [$N$, $M$, $n_b$] \\
	Block 1a 		& Input Images 				& [$N$, $M$, 8] \\
    Block 1b 		& Block 1a 	  				& [$N$, $M$, 8] \\
    Block 1c 		& Block 1b     				& [$N$, $M$, 8] \\
    Block 1d 		& Block 1c 	  				& [$N$, $M$, 8] \\
    Max Pooling 1 	& Block 1d	  				& [$N/2$, $M/2$, 8] \\
    Block 2a 		& Max Pooling 1				& [$N/2$, $M/2$, 16] \\
    Block 2b 		& Block 2a 	  				& [$N/2$, $M/2$, 16] \\
    Block 2c 		& Block 2b     				& [$N/2$, $M/2$, 16] \\
    Block 2d 		& Block 2c 	  				& [$N/2$, $M/2$, 16] \\
    Max Pooling 2	& Block 2d	  				& [$N/4$, $M/4$, 16] \\
    Block 3a 		& Max Pooling 2 			& [$N/4$, $M/4$, 32] \\
    Block 3b 		& Block 3a 	  				& [$N/4$, $M/4$, 32] \\
    Block 3c 		& Block 3b     				& [$N/4$, $M/4$, 32] \\
    Block 3d 		& Block 3c 	  				& [$N/4$, $M/4$, 32] \\
    Max Pooling 3	& Block 3d	  				& [$N/8$, $M/8$,  32] \\
	Block 4a 		& Max Pooling 3& [$N/8$, $M/8$,  16] \\
    Interpolation 1	& Block 4a& [$N/4$, $M/4$, 16] \\
    Block 5a		& Interp. 1 + Block 3d& [$N/4$, $M/4$, 8] \\
    Block 5b		& Block 5a& [$N/4$, $M/4$, 8] \\
    Interpolation 2 & Block 5b					& [$N/2$, $M/2$, 8] \\
    Block 6a		& Interp. 2 + Block 2d& [$N/2$, $M/2$, 16] \\
    Block 6b		& Block 6a					& [$N/2$, $M/2$, 16] \\
    Interpolation 3 & Block 6b					& [$N$, $M$, 16] \\
    Block 7a		& Interp. 3 + Block 1d& [$N$, $M$, 32] \\
    Block 7b		& Block 7a					& [$N$, $M$, 32] \\
    Convolution		& Block 7b					& [$N$, $M$, $n_c$] \\
	\bottomrule
	\end{tabular}
    \caption{Computational steps in the \morpheus{} deep learning framework. For
             each Layer (left column), we list its Input (center column), and
             the Output Shape of its data (right column). The model takes as its
             starting input a set of images in $n_b$ bands, each with $N\times
             M$ pixels. The final output of the model is a set of $n_c$
             classification images, each with $N\times M$ pixels. The
             \morpheus{} block structures are illustrated in
             Figure \ref{fig:block}. The ``+'' symbol denotes a concatenation
             between two layer outputs, as shown in Figure
             \ref{fig:architecture}.}
    \label{table:architecture}
\end{table}

\subsection{Parallelization for Large Images}
\label{section:large_images}

While the \morpheus{} neural network performs semantic segmentation on pixels in
FITS images with a size determined by the training images, the model can process
and classify pixels in arbitrarily large images. To process large
images, \morpheus{} uses a sliding window strategy by breaking the
input FITS files into thumbnails of size $N\times M$ (the size of the training
images) and classifying them individually. \morpheus{} proceeds through the
large format image, first column by column, and then row by row, shifting
the active $N\times M$ window by a unit pixel stride and then recomputing the
classification for each pixel.

As the classification process continues with unit pixel shifts, each
 pixel is deliberately classified many times. We noticed
heuristically that the output \morpheus{} classification of pixels depended on
their location within the image, and that the pixel classifications were more
accurate relative to our training data when they resided in the inner $n_p =
(N-B)\times(M-B)$ region of the classification area, where the lesser accuracy
region consisted of a border about $B\sim5$ pixels wide on each side. Outside of
the very outer $B$ pixels in the large format image, \morpheus{} classifies each
pixel $n_p$ times. For the large FITS data images used in this paper, this
repetition corresponds to $n_p = 900$ separate classifications per pixel per
output class, where each classification occurs when the pixel lies at a
different location within the active window. This substantial additional
information can be leveraged to improve the model, but storing the full
``distribution'' of classifications produced by this method would increase our
data volume by roughly three orders of magnitude.

While \morpheus{} would enable  full use of these distributions,
for practical considerations, we instead record some statistical
information as the computation proceeds and do not store the entire set of $n_p$
samples. To avoid storing the full distribution, we track running estimates of
the mean and variance of the distribution\footnote{See, e.g.,
http://people.ds.cam.ac.uk/fanf2/hermes/doc/antiforgery/stats.pdf for an example
of running mean and variance estimators.}. Once the mean for each class for each
pixel is computed, we normalize the means across classes to sum to unity. We
further record a statistic we call \textit{rank voting}, which is a tally of the
number of times each output class was computed by the model to be the top class
for each pixel. The sum of rank votes across classes for a single pixel equals
the number of times \morpheus{} processed the pixels (i.e., $n_p$ for most
pixels). After the computation, the rank votes are normalized to sum to unity
across the classes for each pixel.

The strips of classified regions produce fifteen output images, containing the
mean and variance estimators for the classification distribution and normalized
rank votes for each class. This striped processing of the image can be performed
in parallel across multiple \morpheus{} instances and then stitched back
together. The weak scaling of this processing is, in principle,
trivial and is limited only by the number of available GPUs and the
total memory of the computer used to perform the calculation.

\section{Model Training}
\label{section:training}

The training of deep learning frameworks involves important decisions
about the training data, the metrics used to optimize the network, numerical
parameters of the model, and the length of training. We provide some rationale
for these choices below.

\subsection{Training Data}
\label{subsection:training_data}

To train a model to perform semantic segmentation, we require a dataset that
provides both information on the segmentation of regions of interest and
classifications associated with those regions. For galaxy morphological
classification, we use 7,629 galaxies sampled from the K15 dataset. Their
2-epoch CANDELS data provide an excellent combination of multiband FITS
thumbnails, segmentation maps in FITS format, and visually-classified
morphologies in tabulated form. The K15 classifications consisted of votes by
expert astronomers, between $3-60$ per object, who inspected images of galaxies
and then selected from several morphological categories to
assign to the object. The number of votes for each category for each object are
provided, allowing \morpheus{} to use the distribution of votes across
classifications for each object when training. We downloaded and used the
publicly available K15 thumbnail FITS files for the $F606W$, $F850LP$, $F125W$,
and $F160W$ bands as input into the model for training and testing. In
training Morpheus to reproduce the K15 classifications, multiband data approximates
the information provided to the astronomers who performed
the K15 classifications. Morpheus is trained using the same $V$, $z$, $J$, and
$H$-band image thumbnails used in the K15 classification process. Other bands
or different numbers of bands could be used for training as necessary, and
\morpheus{} allows for reconfiguration and retraining depending on the available
training images. Of the K15 data set, we used $80\%$ of the objects to form our
training sample and $20\%$ to form our test sample. Various statistical
properties of the test and training samples are described throughout the rest of
the paper.

The primary K15 classifications \textit{spheroid, disk, irregular}, and
\textit{point source/compact} were used in the example \morpheus{}
application presented here. We added one additional class, \textit{background},
to represent sky pixels absent significant source flux. We classify pixels as
belonging to the background category if those pixels fell outside the K15
segmentation maps. Pixels inside the segmentation maps were assigned the
distribution of classifications provided by the K15 experts.

The K15 classification scheme also included an \textit{unknown} class for
objects. Since \morpheus{} works at the pixel level and could provide individual
pixel classifications that were locally accurate within a source but that
collectively could sum to an object whose morphology expert astronomers might
classify as \textit{unknown}, we were posed with the challenge of how to treat
the K15 \textit{unknown} class. Given our addition of the \textit{background}
class constructed from large image regions dominated by sky, one might expect
overlap in the features of regions that are mostly noise and amorphous regions
classified as \textit{unknown}. Since one might also expect overlap between
\textit{unknown} and \textit{irregular} classifications, we wanted to preserve
some distinction in the object classes. We, therefore, removed the
\textit{unknown} class by removing any sources that had \textit{unknown} as
their primary classification from the training sample (213 sources). For any
sources where the non-dominant K15 classifications included \textit{unknown}, we
redistributed the \textit{unknown} votes proportionally to the other classes.


\subsection{Data Augmentation}
\label{section:augmentation}

To increase the effective size of the training dataset, \morpheus{} uses a data
augmentation method. Augmentation supplements the input training data set by
performing transformations on the training images to alter them with the intent
of adding similar but not identical images with known classifications.
Augmentation has been used successfully in the context of galaxy morphological
classification \citep[e.g.,][]{dielman2015a}, and \morpheus{} adopts a
comparable approach to previous implementations.

\begin{figure*}
\centering
\includegraphics[width=\textwidth]{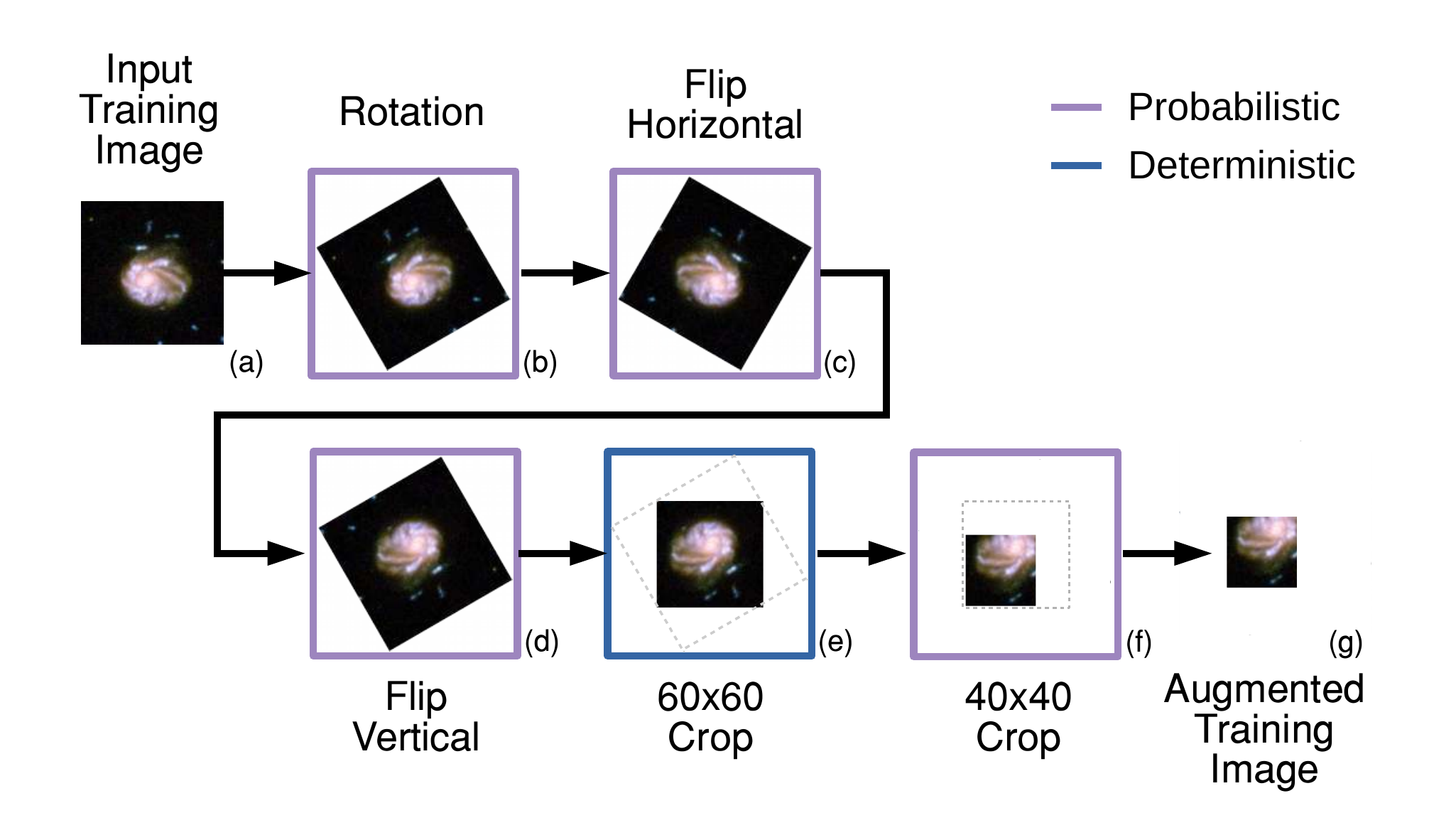}
\caption{Data augmentation pipeline used during neural network training. Each
         training image is processed by the data augmentation pipeline before
         being presented to the neural network during training. The pipeline can
         be described in 7 stages (annotated `(a)-(g)' above). First, an
         image from the training set is selected (Panel a). A number of
         augmentation operations are then applied to the image. The image is
         rotated by a random angle $\theta{\in}[0,2\pi]$ (Panel b), flipped
         horizontally with 50\% probability (Panel c), and flipped vertically
         with a 50\% probability (Panel d). The centermost $60 \times 60$ subset
         of the resulting image is cropped (Panel e), and then a random $40
         \times 40$ subset is selected from the cropped image (Panel f). The
         output $40 \times 40$ rotated, flipped, and cropped image is then used
         for training. This procedure increases the available images for
         training by a factor of ${\sim}574,400$. Using this process helps
         reduce overfitting, particularly in cases of datasets with limited
         training sample sizes.}
\label{fig:train}
\end{figure*}

During training, \morpheus{} produces a series of $40\times40$ pixel augmented
versions of the training images. The augmentation approach is illustrated in
Figure \ref{fig:train}. For each band in the original training image, the image
is collectively rotated by a random angle $\phi\in[0,2\pi]$, flipped
horizontally with a random $50\%$ probability, and then flipped vertically with
a random $50\%$ probability. A crop of the inner $60\times 60$ pixels of the
resulting image is produced, and then a random $40\times 40$ pixel subset of the
image is selected and passed to the model for training. This method allows us to
increase the effective number of images available for training by a factor of
${\sim}574,400$ and helps ameliorate over-training on the original
training image set.

\subsection{Loss Function}
\label{section:loss_function}

A standard method for training deep learning frameworks is to define a loss
function that provides a statistic based on the output classifications to
optimize via stochastic gradient descent with gradients computed using
back-propagation \citep{rumelhart1986a}. Here, we describe how the
\morpheus{} loss function is constructed.

The first task is to assign a distribution of input classifications on a
per-pixel basis, choosing between the $n_c$ classes available to the
\morpheus{} model. For this work, we choose $n_c=5$ ({\it background}, {\it
disk}, {\it spheroid}, {\it irregular}, and {\it point source/compact}), but
\morpheus{} can adopt an arbitrary number of classes. We use the index $k$ to
indicate a given class, with $k\in[1,n_c]$. Consider an $N \times M$ image of an
astronomical object that has been visually classified by a collection of
experts, and a segmentation map defining the extent of the object in the image.
Outside the segmentation map of the object, the pixels are assumed to belong to
the sky and are assigned the {\it background} class. Inside the segmentation
map, pixels are assigned the distribution of {\it disk}, {\it spheroid}, {\it
irregular}, and {\it point source/compact} classifications determined by the
experts for the entire object. For each pixel $ij$, with $i\in[1,N]$ rows and
$j\in[1,M]$ columns, we then have the vector $\bq_{ij}$ whose elements $q_{ijk}$
contain the input distribution of classifications. Here, the index $k$ runs over
the number of classes $n_c$ and $\sum_k q_{ijk} = 1$ for each pixel with indices
$ij$. The goal of the model is to reproduce this normalized distribution
$\bq_{ij}$ of discrete classes for each pixel of the training images. We wish to
define a {\it total loss function} $L_{tot}$ that provides a single per-image
statistic for the model to optimize when attempting to reproduce $\bq_{ij}$.
\morpheus{} combines a weighted {\it cross entropy} loss function with a Dice
loss \citep{novikov2015a,milletari2016a} for its optimization statistic,
which we describe below.

At the end of the \morpheus{} data flow, as outlined in Figure
\ref{fig:architecture}, the raw output of the model consists of $N\times M$ vectors
$\bx_{ij}$ with $n_c$ elements per-pixel estimates that represent unnormalized
approximations to the input per-pixel distributions $\bq_{ij}$. The model
outputs $\bx_{ij}$ for each pixel are then normalized to form a probability
distribution $\bp_{ij}$ using the {\it softmax} function
\begin{equation}
\label{eqn:softmax}
\bp_{ij}=\frac{\exp(\bx_{ij})}{\sum_{k=1}^{n_c}\exp(x_{ijk})}\textnormal{, for } k\in[1,n_c].
\end{equation}
\noindent
The distribution $\bp_{ij}$ then represents the pixel-by-pixel classifications
computed by \morpheus{} for each of the $k\in[1,n_c]$ classes. For a pixel with
indices $ij$, we can define the per-pixel cross entropy loss function as
\begin{equation}
\label{eqn:cross_entropy}
L_{ij}(\bp_{ij},\bq_{ij})=-\sum_{k=1}^{n_c} p_{ijk} \log(q_{ijk})
\end{equation}
\noindent
where $\bp_{ij}$ and $\bq_{ij}$ are again the two per-pixel probability
distributions, with $\bq_{ij}$ representing the true distribution of the input
classifications for the pixel $ij$ and $\bp_{ij}$ representing the model output.

Equation \ref{eqn:cross_entropy} provides the per-pixel contribution to the
entropy loss function. However, for many images, the majority of pixels
lie outside the segmentation maps of sources identified in the training data and
are therefore labeled as {\it background}. To overcome this imbalance and
disincentivize the model from erroneously learning to classify pixels containing
source flux as {\it background}, we apply a weighting to the per-pixel loss. We
define an index $k_{ij}^{max, \bq}$ that indicates which class is the maximum of
the input classification distribution for each pixel, written as
\begin{equation}
\label{eqn:argmax}
k_{ij}^{max, \bq} = \mathrm{argmax}~\bq_{ij}
\end{equation}
with $1\leq k_{ij}^{max, \bq}\leq n_c$. For each class $k$, we then define a
weight $w_k$ that is inversely proportional to the number of pixels with
$k_{ij}^{max, \bq} = k$. We can write
\begin{equation}
\label{eqn:w_class}
w_k =\left[\sum_{i=1}^N \sum_{j=1}^M \max(\bq_{ij}) \delta_{k,k_{ij}^{max, \bq}}\right]^{-1}.
\end{equation}
Here, $\delta_{i,j}$ is the Kronecker delta function. The vector $\bw$ has size
$n_c$ and each of its elements $w_k$ contain the inverse of the sum of
$\max(\bq_{ij})$ for pixels with $k_{ij}^{max, \bq}=k$. In a given image, we
ignore any classes that do not appear in the input classification distribution
(i.e., any class $k$ for which $\sum_{i} \sum_{j} q_{ijk} = 0$).

Using $\bw$, we define a weighted cross entropy loss for each pixel as
\begin{equation}
\label{eqn:weighted_loss}
L^{w}_{ij} = w_{k_{ij}^{max}} L_{ij}(\bp_{ij},\bq_{ij}).
\end{equation}
\noindent
A mean weighted loss function is then computed by averaging Equation
\ref{eqn:weighted_loss} over all pixels as
\begin{equation}
\label{eqn:mean_weighted_loss}
\bar{L}^{w} = \frac{1}{N\times M} \sum_{i=1}^{N} \sum_{j=1}^{M} L^{w}_{ij}.
\end{equation}
\noindent
This mean weighted loss function serves as a summary statistic of the cross
entropy between the output of \morpheus{} and the input classification
distribution.

When segmenting images primarily comprised of {\it background} pixels, the
classification distributions of the output pixels should be highly
unbalanced, with the majority having {\it background}$\approx1$. In this
case, the mean loss function statistic defined by Equation
\ref{eqn:mean_weighted_loss} will be strongly influenced by a single class. A
common approach to handle unbalanced segmentations is to employ a Dice loss
function to supplement the entropy loss function
\citep[e.g.,][]{milletari2016a,sudre2017a}. The Dice loss function used by
\morpheus{} is written as
\begin{equation}
L^{D}(\bm{b}, \bm{m})=1 - 2\frac{\sum_i\sum_j(S(\bm{b})
                      \circ \bm{m})_{ij}}{\sum_i\sum_j(S(\bm{b}) + \bm{m})_{ij}}.
\end{equation}
\noindent
Here, $S(\bm{b})=(1+\exp(-\bm{b}))^{-1}$ is the sigmoid function (see Equation
\ref{eq:sigmoid}) applied pixel-wise to the {\it background} classification
image output by the model. The image $\bm{m}$ is the input mask with values
$m{=}1$ denoting {\it background} pixels and $m{=}0$ indicating source pixels,
defined, e.g., by a segmentation map generated using {\it sextractor}. The
$\circ$ symbol indicates a Hadamard product of the matrices $S(\bm{b})$ and
$\bm{m}$. Note that the output {\it background} matrix $\bm{b}$ has not yet been
normalized using a softmax function, and so $b_{ij}{\in}[-\infty,\infty]$ and
$S(b_{ij}){\in}[0,1]$. The Dice loss then ranges from $L^D=0$ if
$S(\bm{b}){\approx}\bm{m}$ and $L^D\sim1$ when $S(\bm{b})$ and $\bm{m}$ differ
substantially. The addition of this loss function helps to maximize the spatial
coincidence of the output {\it background} pixels assigned $b_{ij}{\approx}1$
with the non-zero elements of the input segmentation mask $\bm{m}$.

To define the total loss function optimized during the training of \morpheus{},
the cross entropy and Dice losses are combined as a sum weighted by two
parameters $\lambda_{w}$ and $\lambda_{D}$. The total loss function is written
as
\begin{equation}
\label{eqn:total_loss}
L_{tot} = \lambda_{w}L^{w} + \lambda_{D}L^{D}
\end{equation}
\noindent
For the implementation of \morpheus{} used in this paper, the entropy and Dice
loss functions are weighted equally by setting $\lambda_w {=}1$ and
$\lambda_D{=}1$.

\subsection{Optimization Method}

To optimize the model parameters, the {\it Adam} stochastic gradient descent
method \citep{Kingma2014a} was used. The {\it Adam} algorithm uses the first and
second moments of first-order gradients computed via backpropagation to find
the minimum of a stochastic function (in this case, our loss function,
see Section \ref{section:loss_function}, which depends on the many parameters of
the neural network). The {\it Adam} optimizer, in turn, depends on
hyper-parameters that determine how the algorithm iteratively finds a minimum.
Since the loss function is stochastic, the gradients change each
iteration, and {\it Adam} uses an exponential moving average of the
gradients ($\hat{m}$) and squared gradients ($\hat{v}$) when searching for a
minimum. Two dimensionless hyper-parameters ($\beta_1$ and $\beta_2$) set the
decay rates of these exponential averages \citep[see Algorithm 1 of
][]{Kingma2014a}. As the parameters $\theta$ of the function being optimized are
iterated between steps $t-1$ and $t$, they are updated according to
\begin{equation}
\theta_t \leftarrow \theta_{t-1} - \alpha \cdot \hat{m}_t / (\sqrt{\hat{v}_t} + \epsilon).
\end{equation}
Here, $\epsilon$ is a small, dimensionless safety hyper-parameter that prevents
division by zero, and $\alpha$ is a small, dimensionless hyper-parameter that
determines the magnitude of the iteration step. Table \ref{tbl:adam_params}
lists the numerical values of the {\it Adam} optimizer hyper-parameters used by
\morpheus{}. We use the default suggested values for $\beta_1$, $\beta_2$, and
$\epsilon$. After some experimentation, we adopted a more conservative step size
for $\alpha$ than used by \citet{Kingma2014a}.

\begin{table}
  \centering
  \begin{tabular}{c c}
  \multicolumn{2}{c}{{\it Adam} Optimizer Hyper-parameters} \\
  \toprule
  Hyper-parameter & Value \\
  \midrule
  $\beta_1$  &  $0.9$ \\
  $\beta_2$  &  $0.999$ \\
  $\epsilon$ &  $10^{-8}$ \\
  $\alpha$   &  $9.929\times10^{-5}$ \\
  \bottomrule
   \end{tabular}
   \label{tbl:adam_params}
   \caption{{\it Adam} optimizer \citep{Kingma2014a} hyper-parameter values used
            during the training of the neural network used in
            \morpheus{}. See the text for definitions of the hyper-parameters.}
\end{table}

\subsection{Model Evaluation}
\label{section:evaluation}

\begin{table}
    \centering
    \begin{tabular}{c c c}
    \multicolumn{3}{c}{\morpheus{} Training and Test Results} \\
    \toprule
    Metric & Training & Test \\
    \midrule
    \textbf{Accuracy} $\ACC$ & & \\
    Background & 91.5\% & 91.4\% \\
    Disk & 74.9\% & 75.1\% \\
    Irregular & 80.6\% & 68.6\% \\
    Point source/compact & 91.0\% & 83.8\% \\
    Spheroid & 72.3\% & 71.4\% \\
    All Classes & 86.8\% & 85.7\% \\
    \textbf{Intersection-Over-Union} $\IOU$ & & \\
    $B{>}0.5$ & 0.899 & 0.888 \\
    $B{>}0.6$ & 0.900 & 0.891\\
    $B{>}0.7$ & 0.902 & 0.893\\
    $B{>}0.8$ & 0.902 & 0.895\\
    $B{>}0.9$ & 0.900 & 0.896\\
    \bottomrule
    \end{tabular}
    \caption{\morpheus{} training and test results for accuracy $\ACC$, and
             intersection-over-union $\IOU$ as a function of {\it background}
             threshold $B$.}
    \label{table:train_results}
\end{table}

As training proceeds, the performance of the model can be quantified using
various metrics and monitored to determine when training has effectively
completed. The actual performance of \morpheus{} will vary depending on the
classification scheme used, and here we report the performance of the model
relative to the CANDELS images morphologically classified in K15. Performance
metrics reported in this Section refer to pixel-level quantities, and we discuss
object-level comparisons of morphological classifications relative to K15 in
Section \ref{section:morpheus_output}.

While the model training proceeds by optimizing the loss function defined in
Section \ref{section:loss_function}, we want to quantify the accuracy of the
model in recovering the per-pixel classification and the overlap of contiguous
regions with the same classification. First, we will need to define the index
$k_{ij}^{max}$ with maximum probability to reflect either the input
classification $\bq_{ij}$ or the output classification $\bp_{ij}$. We define an
equivalent of Equation \ref{eqn:argmax} for $\bp_{ij}$ as
\begin{equation}
k_{ij}^{max, \bp} = \mathrm{argmax}~\bp_{ij}.
\end{equation}
\noindent
We can then define a percentage accuracy
\begin{equation}
\ACC = \frac{100}{N \times M} \sum_{i=1}^{N} \sum_{j=1}^{M} \delta_{k_{ij}^{max, \bp}, k_{ij}^{max, \bq}}.
\end{equation}
\noindent
The accuracy $\ACC$ then provides the percentage of pixels for which the maximum
probability classes of the input and output distributions match.

In addition to accuracy, the intersection-over-union $\IOU$ of pixels with {\it
background} probabilities above some threshold is computed between the input
$\bq_{ij}$ and output $\bp_{ij}$ distributions. If we define the index $b$ to
represent the {\it background} class, we can express the input {\it background}
probabilities as $\bq_{b}{=}q_{ijb}$ for $i{\in}[1,N]$ and $j{\in}[1,M]$, and
the equivalent for the output {\it background} probabilities $\bp_{b}$. We can
refer to $\bq_b$ and $\bp_b$ as the input and output {\it background} images,
and the regions of these images with values above some threshold  $B$ as
$\bq_b({>}B)$ and $\bp_b({>}B)$, respectively. Note that the input $\bq_b$ only
contains values of zero or one, whereas the output $\bp_b$ has continuous values
between zero and one. We can then define the $\IOU$ metric for threshold $B$ as
\begin{equation}
\label{eqn:iou}
\IOU(B) = \frac{\bp_b({>}B)\cap\bq_b({>}B)}{\bp_b({>}B)\cup\bq_b({>}B)}.
\end{equation}
\noindent
Intuitively, this $\IOU$ metric describes how well the pixels assigned by
\morpheus{} as belonging to a source match up with the input source segmentation
maps. A value of $\IOU=1$ indicates a perfect match between source pixels
identified by \morpheus{} and the input segmentation maps, while a value of
$\IOU=0$ would indicate no pixels in common between the two sets.

As training proceeds, the accuracy $\ACC$ and intersection-over-union $\IOU$ are
monitored until they plateau with small variations. For the K15 training data,
the model plateaued after about 400 epochs. The training then continues for
another 100 epochs to find a local maximum in $\ACC$ and $\IOU$, and the model
parameters at this local maximum adopted for testing. Table
\ref{table:train_results} summarizes the per-pixel performance of \morpheus{} in
terms of $\ACC$ for each class separately, $\ACC$ for all classes, and $\IOU(B)$
for $B{=}[0.5,0.6,0.7,0.8,0.9]$. We also report the performance
of the training and testing samples separately. The
pixel-level classifications are $70-90\%$ accurate depending on the class, and
the intersection-over-union is $\IOU{\sim}0.9$ for all thresholds $B{\ge}0.5$.
The model shows some evidence for overfitting as accuracy declines slightly from
the training to test sets for most classes.

\section{Segmentation and Deblending}
\label{section:segmentation}

\begin{figure*}
\centering
\includegraphics[width=\linewidth]{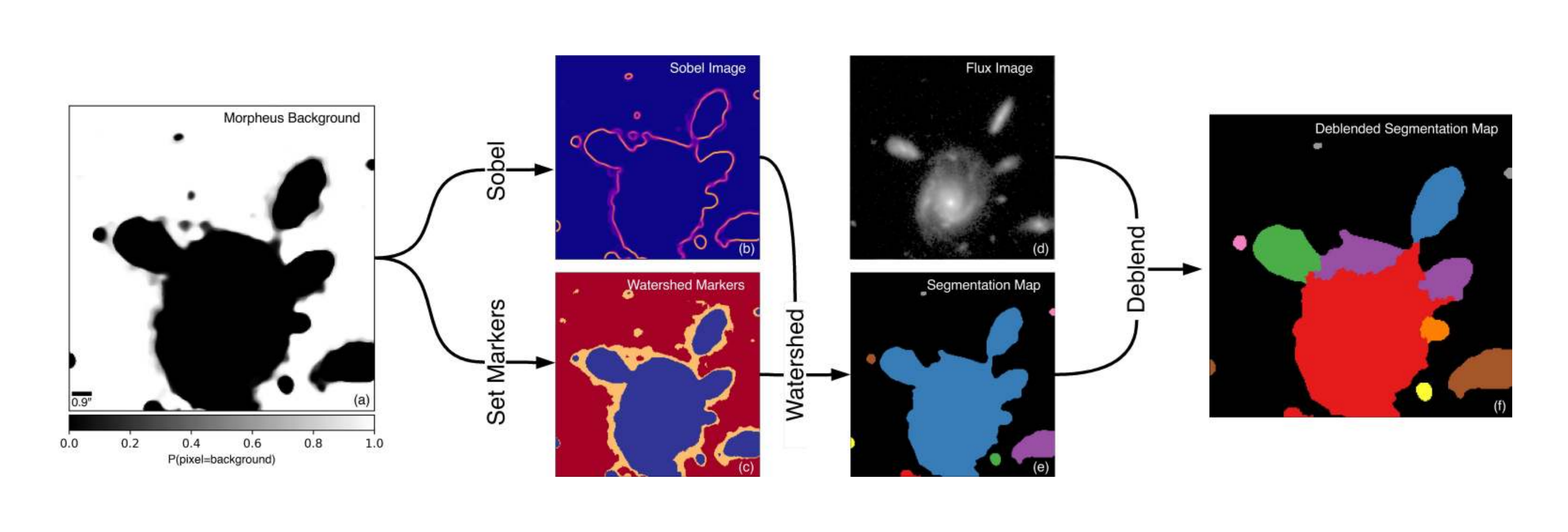}
\caption{Segmentation and deblending process used by \morpheus{}, illustrating
         Algorithms \ref{alg:segmentation} and \ref{alg:deblending}. The {\it
         background} image (Panel a) output from the \morpheus{} neural network
         is used as input to a Sobel-filtered image (Panel b) and a discretized
         map marking regions of high and low {\it background} (Panel c). These
         two images are input to a watershed algorithm to identify and label
         distinct, connected regions of low {\it background} that serve as the
         highest-level \morpheus{} segmentation map (Panel e) This segmentation
         map represents the output of Algorithm \ref{alg:segmentation}. A flux
         image and a list of object locations (Panel d) are combined with the
         high-level segmentation map to deblend multicomponent objects using an
         additional watershed algorithm by using the source locations in the
         flux image as generating points. The end result is a deblended
         segmentation map (Panel f), corresponding to the output of Algorithm
         \ref{alg:deblending}.}
\label{fig:segmap_process}
\end{figure*}

To evaluate the completeness of \morpheus{} in object detection and to compute
an object-level classification, segmentation maps must be constructed and then
deblended from the \morpheus{} pixel-level output. \morpheus{} uses the {\it
background} class from the output of the neural network described in Section
\ref{section:nn} to create a segmentation map. The segmentation algorithm uses a
watershed transform to separate {\it background} pixels from source
pixels and then assigns contiguous source pixels a unique label.
The deblending algorithm uses the flux from the input science images and the
output of the segmentation algorithm to deblend under-segmented regions
containing multiple sources. We summarize these procedures as Algorithms
\ref{alg:segmentation} and \ref{alg:deblending}. Figure \ref{fig:segmap_process}
illustrates the process for generating and deblending segmentation maps.

\subsection{Segmentation}
\label{subsection:segmentation}

The segmentation algorithm operates on the output {\it background}
classification image and identifies contiguous regions of low
background as sources. The algorithm begins with the {\it background} image
$\bb{\equiv}\bp_b$ defined in Section \ref{section:evaluation} and an initially
empty mask $\bmask{=}\mathbf{0}$ of the same size. For every pixel in the image,
if $b_{ij}{=}1$ we set $m_{ij}=1$ and if $b_{ij}=0$ we set $m_{ij}=2$. The {\it
background} mask $\bmask$ then indicates extreme regions of $\bb$. The
\cite{sobel1968a} algorithm is applied to the {\it background} image $\bb$ to
produce a Sobel edge image $\bs$. \morpheus{} then applies the watershed
algorithm of \citet{couprie1997a}, using the Sobel image $\bs$ as the ``input
image'' and the {\it background} mask $\bmask$ as the ``marker set''. We refer
the reader to \citet{couprie1997a} for more details on the watershed algorithm,
but in summary, the watershed algorithm collects together regions with
the same marker set value within basins in the input image. The Sobel image
$\bs$ provide these basins by identifying edges in the
background, and the {\it background} mask $\bmask$ provides the marker
locations for generating the individual sheds. The output of the watershed
algorithm is then an image $\bsm$ containing distinct regions generated from
areas of low {\it background} that are bounded by edges where the {\it
background} is changing quickly. The algorithm then visits each of the distinct
regions in $\bsm$ and assigns them a unique $id$, creating the segmentation map
$\bsm$ before deblending.

\begin{algorithm}[tb]
    \caption{Segmentation}\label{alg:segmentation}
    \algorithmfootnote{Where \textsc{Sobel} is the Sobel algorithm \citep{sobel1968a}
                       and \textsc{Watershed} is the watershed algorithm
                       \citep{couprie1997a}. Optional parameter $p$ allows for
                       pixel locations to be specified, such as the locations of
                       known sources, and used as generating points for the
                       watersheding operation.}

    \KwIn{Background probability map $\bb$, Specified marker set $\bp$ (optional,
            same size as $\bb$)}

    \KwOut{Labelled segmentation map $\bsm$}
    $\bmask \gets$ zero matrix same size as $\bb$

    \For{\upshape $m_{ij}$ in $\bmask$}{
        \If{$b_{ij}=1$}{
            $m_{ij} \gets 1$
        }
        \ElseIf{$b_{ij}=0$ or $p_{ij}>0$}{
            $m_{ij} \gets 2$
        }
    }

    $\bs \gets$ \textsc{Sobel}($\bb$)\\
    $\bsm \gets$ \textsc{Watershed}($\bs$, $\bmask$)\\
    $id \gets 1$\\
    \For{each contiguous set of pixels $\by>0$ in $\bsm$}{
        \For{pixel $y_{ij}$ in $\by$}{
            $y_{ij} \gets id$
        }
        $id \gets id + 1$
    }

    \textbf{return} $\bsm$
\end{algorithm}

\subsection{Deblending}
\label{subsection:deblending}

The algorithm described in Section \ref{subsection:segmentation} provides a
collection of segmented regions of contiguous areas, each with a unique index.
Since this algorithm identifies contiguous regions of low {\it background},
neighboring sources with overlapping flux in the science images will be blended
by the segmentation algorithm. The deblending algorithm used in \morpheus{} is
{\it ad hoc} and is primarily designed to separate the segmented
regions into distinct subregions containing a single pre-defined object. The
locations of these objects may be externally specified, such as catalog entries
from a source catalog (e.g., 3D-HST sources), or they may be internally derived
from the science images themselves (e.g., local flux maxima).

The deblending algorithm we use applies another round of the watershed operation
on each of the distinct regions identified by the segmentation algorithm, using
the local flux distributions from the negative of a science image (e.g.,
$F160W$) as the basins to fill and object locations as the marker set. We assign
the resulting subdivided segmentations a distinct $subid$ in addition to their
shared $id$, allowing us to keep track of adjacent deblended regions that share
the same parent segmentation region. The $subid$ of deblended sources is
indicated by decimal values and the parent $id$ is indicated by the whole number
of the $id$. For example, if a source with $id=8$ was actually two sources,
after deblending the two deblended sources would have id values $8.1$ and $8.2$.

In testing \morpheus{, we find that the deblending algorithm may shred
extended sources like large disks or point source diffraction spikes. However,
the Morpheus algorithm successfully deblends some small or faint sources
proximate to bright sources that are missing from the 3D-HST catalog.}

\begin{algorithm}
    \caption{Deblending} \label{alg:deblending}
    \algorithmfootnote{Where \textsc{Watershed} is the watershed algorithm
                       \citep{couprie1997a}. \textsc{PeakLocalMaxima}($x$, $y$,
                       $z$) returns a list of tuples marking the pixel locations
                       of at most $z$ local maxima in $x$ that lie at least $2y$
                       pixels apart, as implemented by \citet{scikit-image}.
                       \textsc{Count} returns the number of elements in a
                       collection. \textsc{Max} returns the maximum element from
                       a matrix. Optional parameter $p$ allows for pixel
                       locations to be specified, such as the locations of known
                       sources, and used as generating points for the
                       watersheding.}

    \KwIn{Segmentation map $\bsm$, flux image $\bh$, minimum radius between
    flux peaks $\br$, maximum number of deblended subregions $\bndmax$,
    Specified marker set $\bp$ (optional, same size as $\bsm$)}

    \KwOut{Deblended segmentation map $\bdb$}

    \If{$\bp$ is not specified}{
        $idc \gets 10^{\lceil \log_{10} \bndmax \rceil}\,\,\,$ ($\lceil\rceil$ indicate ceiling operation)\\
        $\bsm \gets idc \times  \bsm$ \\
    }
    \For{each contiguous set of source pixels $\bs>0$ in $\bsm$}{
        $h_{local} \gets $ subset of $\bh$ corresponding to $\bs$\\

        \If{$\bp$ is specified}{
            $\bp_{local} \gets $ subset of $\bp$ corresponding to $\bs$\\

            \If{$\bp_{local}$ contains more than one id}{
                $\bs \gets $ \textsc{Watershed}($-h_{local}$, $\bp_{local}$)
            }
            \Else{
                $\bs \gets $ \textsc{Max}($\bp_{local}$)
            }
        }
        \Else{
            $idx \gets $ \textsc{PeakLocalMaxima}($h_{local}$, $\br$, $\bc$)\\
            \If{\textsc{Count}($idx$) $>1$}{
                $subid \gets 1$\\
                $m \gets $ a zero matrix same size as $\bs$\\
                \For{indices $i$, $j$ in $idx$}{
                    $m_{ij} \gets subid$\\
                    $subid \gets subid + 1$\\
                }
                $\bs \gets $ \textsc{Watershed}($-h_{local}$, $m$)
            }
        }
    }
    \If{$\bp$ is not specified}{
        $\bdb \gets idc^{-1} \times \bsm$
    }
    \Else{
        $\bdb \gets  \bsm$
    }
    \textbf{return} $\bdb$
\end{algorithm}

\section{Object-Level Classification}
\label{section:morpheus_output}

While \morpheus{} uses a semantic segmentation model to enable pixel-level
classification of astronomical images using a deep learning framework, some
applications, like the morphological classification of galaxies, additionally
require object-level classification. \morpheus{} aggregates pixel-level
classifications into an object-level classification by using a flux-weighted
average.

\begin{figure*}
\centering
\includegraphics[width=\linewidth]{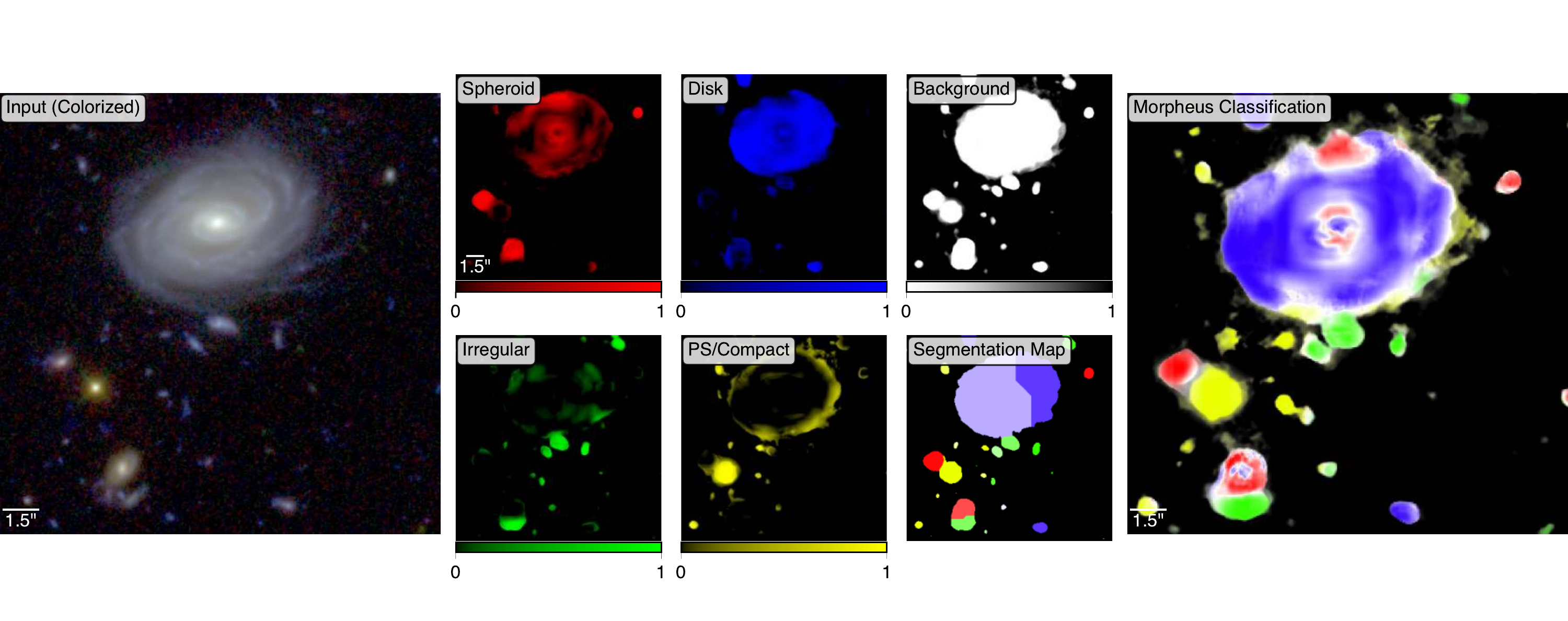}
\caption{\morpheus{} morphological classification results for a region of the
         GOODS South field. The far left panel shows a three-color composite
         $VzH$ image. The scale bar indicates 1.5''. The $V$, $z$, $J$, and $H$
         FITS images are supplied to the \morpheus{} framework, which then
         returns  images for the {\it spheroid} (red-black panel), {\it disk}
         (blue-black panel), {\it irregular} (green-black panel), {\it point
         source/compact} (yellow-black panel), and {\it background}
         (white-black panel) classifications. The pixel values of these images
         indicate the local dominant \morpheus{} classification, normalized to
         sum to one across all five classifications. The panel labeled
         ``Segmentation Map'' is also generated by \morpheus{}, using the 3D-HST
         survey sources as generating locations for the segmentation Algorithm
         \ref{alg:segmentation}. The regions in the segmentation map are
         color-coded by their flux-weighted dominant class computed from the
         \morpheus{} classification values. The far right panel shows the
         \morpheus{} ``classification color'' image, where the pixel hues
         indicate the dominant morphological classification, and the
         intensity indicates $1{-}${\it background}. The saturation of the
         \morpheus{} color image indicates the difference between the dominant
         classification value and the second most dominant classification, such
         that white regions indicate pixels where \morpheus{} returns a
         comparable result for multiple classes. See Section
         \ref{section:morphological_color_image} for more details.}
\label{fig:zoom}
\end{figure*}

Figure \ref{fig:zoom} shows the results of the \morpheus{} pixel-level
classification for an example area of the CANDELS region of GOODS South. The
leftmost panel shows a three-color $VzH$
composite of the example area for reference, though \morpheus{} operates
directly on the science-quality $VzJH$ FITS images. The central panels show the
output pixel classifications (i.e., $\bq$ from Section
\ref{section:loss_function}) for the {\it background}, {\it spheroid}, {\it
disk}, {\it irregular}, and {\it point source/compact} classes, with the
intensity of each pixel indicating the normalized probability
$q_{ijk}{\in}[0,1]$. The segmentation map resulting from the algorithms
described in Section \ref{section:segmentation} is also shown in as a central
panel. The rightmost panel shows a color composite
of the \morpheus{} pixel-level classification, with the color of each pixel
indicating its the dominant class and the saturation of the pixel being
proportional to the difference $\Delta \bq$ between the dominant and second most
dominant class. White pixels then indicate regions where the model did not
strongly distinguish between two classes, such as in transition regions in the
image between two objects with different morphological classes. The pixel
intensities in the pixel-level classification image are set to 1-{\it
background} and are not flux-weighted. The dominant classification
for each object, as determined by \morpheus{}, is often clear
visually. The brightest objects are well-classified and agree with the intuitive
morphological classifications an astronomer might assign based on the $VzH$
color composite image. Faint objects in the image have less morphological
information available and are typically classified as
{\it point source/compact}, in rough agreement with their classifications in the
K15 training set. However, these visual comparisons are qualitative, and we now
turn to quantifying the object-level classification from the pixel values.

Consider a deblended object $\by$ containing a total of $n_o$ contiguous pixels
of arbitrary shape within a flux image, and a single index $i{=}[1,n_o]$
scanning through the pixels in $\by$. Each class $k{\in}[1,n_c]$ in the
distribution of classification probabilities $\bm{Q}$ for the object is computed
as
\begin{equation}
\label{eqn:label_aggregation}
Q_{k} = \frac{\sum_{i=1}^{n_o} y_{i} q_{ik}}{\sum_{i=1}^{n_o} y_{i}}.
\end{equation}
\noindent
Here, $\by$ represents the pixel region in a science image assigned to the
object, and $y_i$ is the flux in the $i$th pixel of the object. The quantity
$q_{ik}$ is the $k$th classification probability of the $i$th pixel in $\by$.
Equation \ref{eqn:label_aggregation} represents object-level classification
computed as the flux-weighted average of the pixel-level classifications in the
object.

\section{\morpheus{} Data Products}
\label{section:products}

Before turning the quantifications of the object-level performance, we provide a
brief overview of the derived data products produced by \morpheus{}. A more
detailed description of the data products is presented in Appendix
\ref{appendix:data_release}, where we describe a release of pixel-level
morphologies for the 5 CANDELS fields and
3D-HST value-added catalog, including object-level
morphologies. The Hubble Legacy Fields \citep{illingworth2016a} GOODS South
v2.0 release and 3D-HST survey \citep{momcheva2016a} are the primary focus of the
analysis of the \morpheus{}' performance owing to their depth and completeness.

As described in Section \ref{section:morpheus_output}, \morpheus{} produces a
set of $n_c$ ``classification images'' that correspond to the pixel-by-pixel
model estimates $\bq_{ij}$ for each class, normalized across classes such that
$\sum_k q_{ijk} = 1$. The value of each pixel is, therefore,
bounded ($q_{ijk}{\in}[0,1]$). The classification images are stored in FITS
format, and inherit the same ($N\times M$) pixel dimensions as the input
FITS science images provided to \morpheus{}. When presenting classification
images used in this paper, we represent {\it background} images in negative
grayscale, {\it spheroid} images in black-red, {\it
disk} images in black-blue, {\it irregular} images in black-green, and {\it
point source/compact} images in black-yellow color scales. Figure
\ref{fig:zoom} shows {\it spheroid}, {\it disk}, {\it irregular}, {\it point
source/compact}, and {\it background} images (central panels) for a region of
CANDELS GOODS South.

Given the separate classification images, we can construct what we deem a
``\morpheus{} morphological color image" that indicates the local dominant class
for each pixel. To produce a Red-Blue-Green false color image to represent the
morphological classes visually, we use the Hue-Saturation-Value (HSV) color
space and convert from HSV to RGB via standard conversions. In the HSV color
space, the Hue image indicates a hue on the color wheel, Saturation provides the
richness of the color (from white or black to a deep color), and Value sets the
brightness of a pixel (from dark to bright). On a color wheel of hues,
$\mathcal{H}{\in}[0,360]$ ranges from red ($\mathcal{H}{=}0$) to red
($\mathcal{H}{=}360$) through yellow ($\mathcal{H}{=}120$), green
($\mathcal{H}{=}180$), and blue ($\mathcal{H}{=}240$), we can assign Hue pixel
values corresponding to the dominant morphological class ({\it spheroid} as red,
{\it disk} as blue, {\it irregular} as green, and {\it point source/compact}
as yellow). We set the Saturation of the image to be the $\Delta q_{ijk}$
between the dominant class and the second most dominant class, such that cleanly
classified pixels ($q_{ijk_{ij}^{max}}{\approx}1$, $\Delta q_{ijk}{\approx}1$)
appear as deep red, blue, green, or yellow, and pixels where \morpheus{}
produces an indeterminate classification ($\Delta q_{ijk}{\approx}0$) appear as
white or desaturated. The Value channel is set equal to $1 - ${\it background},
such that regions of low background containing sources are bright, and
regions with high background are dark. Figure \ref{fig:zoom} also shows the
\morpheus{} morphological color image (far right panel) for a region of CANDELS
GOODS South.

\subsection{Morphological Images for GOODS South}
\label{section:hfl_morphology}

As part of our data products, we have produced \morpheus{} morphological images
of the Hubble Legacy Fields \citep[HLF v2.0;][]{illingworth2016a} reduction of
GOODS South. These data products are used in Section \ref{section:performance}
to quantify the performance of \morpheus{} relative to standard astronomical
analyses, and we, therefore, introduce them here. The \morpheus{}
morphological classification images for the HLF were computed as described in
Section \ref{section:large_images}, feeding \morpheus{} subregions of the HLF
$VzJH$ images for processing and then tracking the distribution of output pixel
classifications to select the best classification for each. The $\sim10^8$
pixels in each classification image are then stitched back together to produce
contiguous {\it background}, {\it spheroid}, {\it disk}, {\it irregular}, and
{\it point source/compact} images for the entire HLF GOODS South.


\subsubsection{Background Image}
\label{section:background_image}

\begin{figure*}
\centering
\includegraphics[]{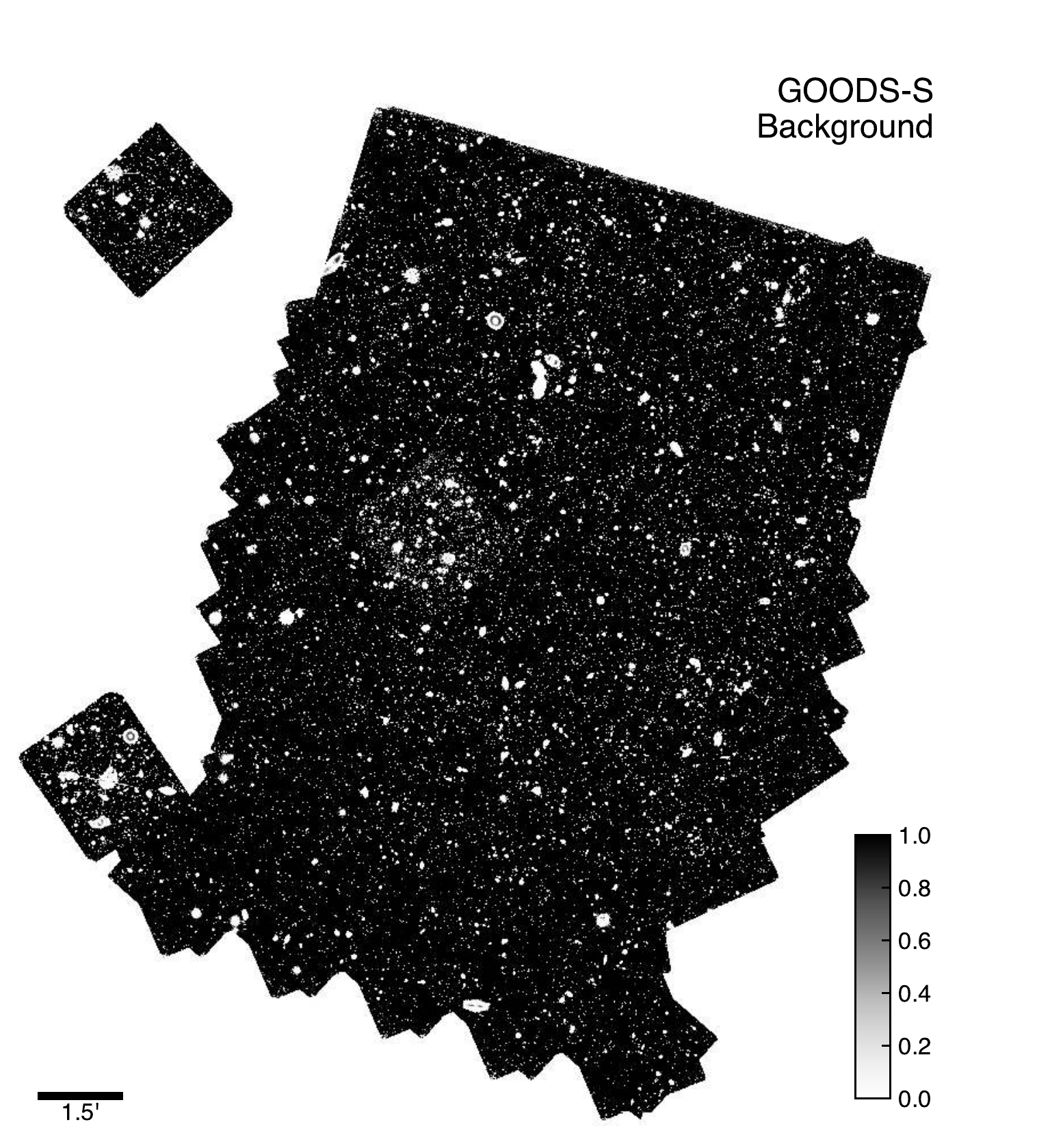}
\caption{\morpheus{} {\it background} classification image for the Hubble Legacy
         Fields \citep{illingworth2016a} reduction of the CANDELS survey data
         \citep{grogin2011a,koekemoer2011a} in GOODS South. Shown are the
         normalized model estimates that each of the $\sim10^8$ pixels belongs
         to the {\it background} class. The scale bar indicates 1.5 arcmin. The
         color bar indicates the {\it background}${\in}[0,1]$, increasing from
         white to black. Correspondingly, the bright areas indicate regions of
         low background where sources were detected by \morpheus{}.}
\label{fig:background}
\end{figure*}

Figure \ref{fig:background} shows the {\it background} image for the \morpheus{}
analysis of the HLF reduction of GOODS South. The {\it background}
classification for each pixel is shown in negative gray scale, with black
corresponding to {\it background}${=}1$ and white regions corresponding to {\it
background}${=}0$. The {\it background} image is used throughout Section
\ref{section:performance} to quantify the performance of \morpheus{} in object
detection.

\subsubsection{Spheroid Image}
\label{section:spheroid_image}

\begin{figure*}
\centering
\includegraphics[]{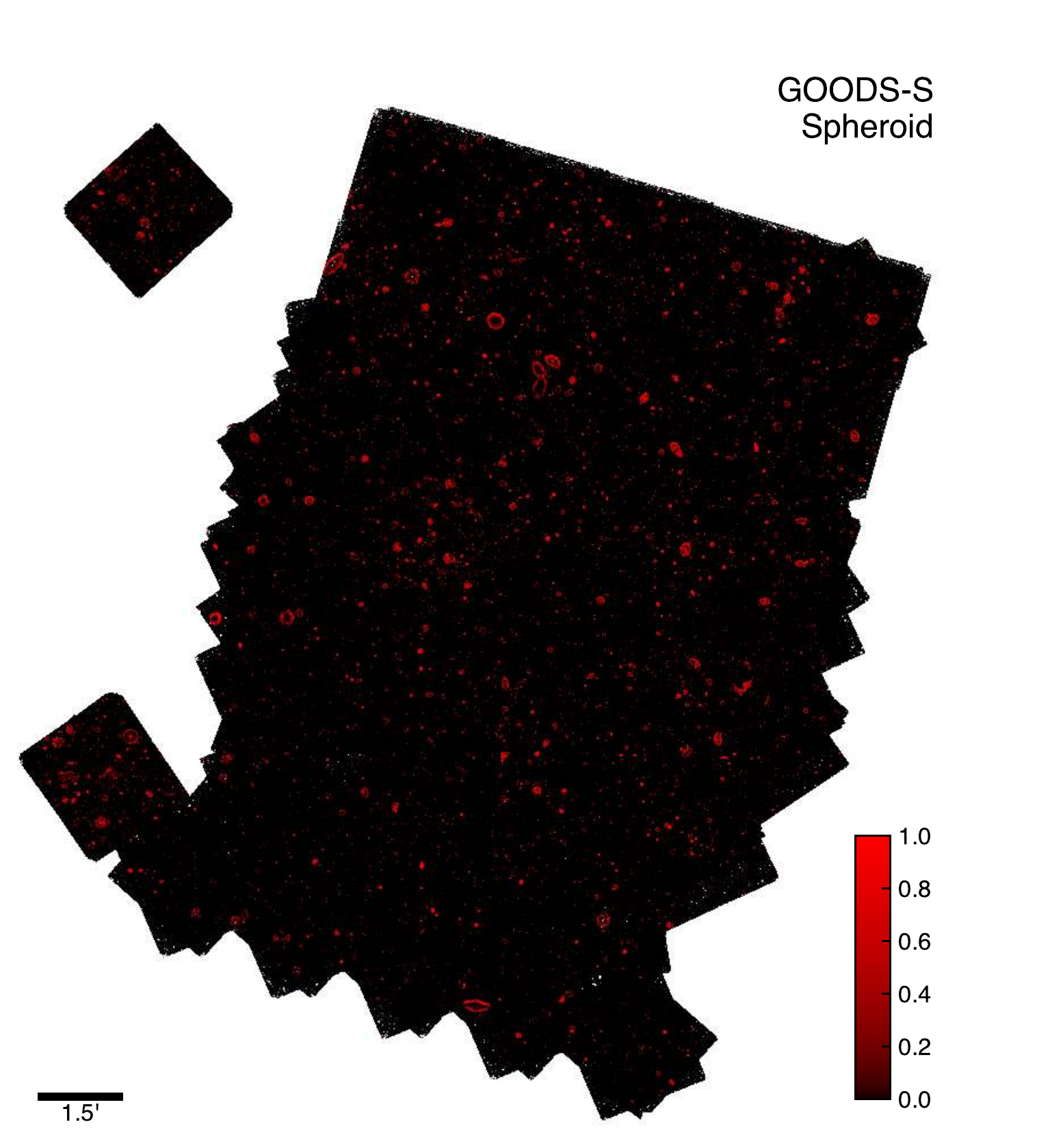}
\caption{\morpheus{} {\it spheroid} classification image for the Hubble Legacy
         Fields \citep{illingworth2016a} reduction of the CANDELS survey data
         \citep{grogin2011a,koekemoer2011a} in GOODS South. Shown are the
         normalized model estimates that each of the $\sim10^8$ pixels belongs
         to the {\it spheroid} class. The scale bar indicates 1.5 arcmin. The
         color bar indicates the {\it spheroid}${\in}[0,1]$, increasing from
         black to red. Correspondingly, the bright red areas indicate pixels
         where \morpheus{} identified {\it spheroid} objects.}
\label{fig:spheroid}
\end{figure*}

Figure \ref{fig:spheroid} shows the {\it spheroid} image for the \morpheus{}
analysis of the HLF reduction of GOODS South. The {\it spheroid} classification
for each pixel is shown on a black-to-red colormap, with black corresponding
to {\it spheroid}${=}0$ and red regions corresponding to {\it spheroid}${=}1$.

\subsubsection{Disk Image}
\label{section:disk_image}

\begin{figure*}
\centering
\includegraphics[]{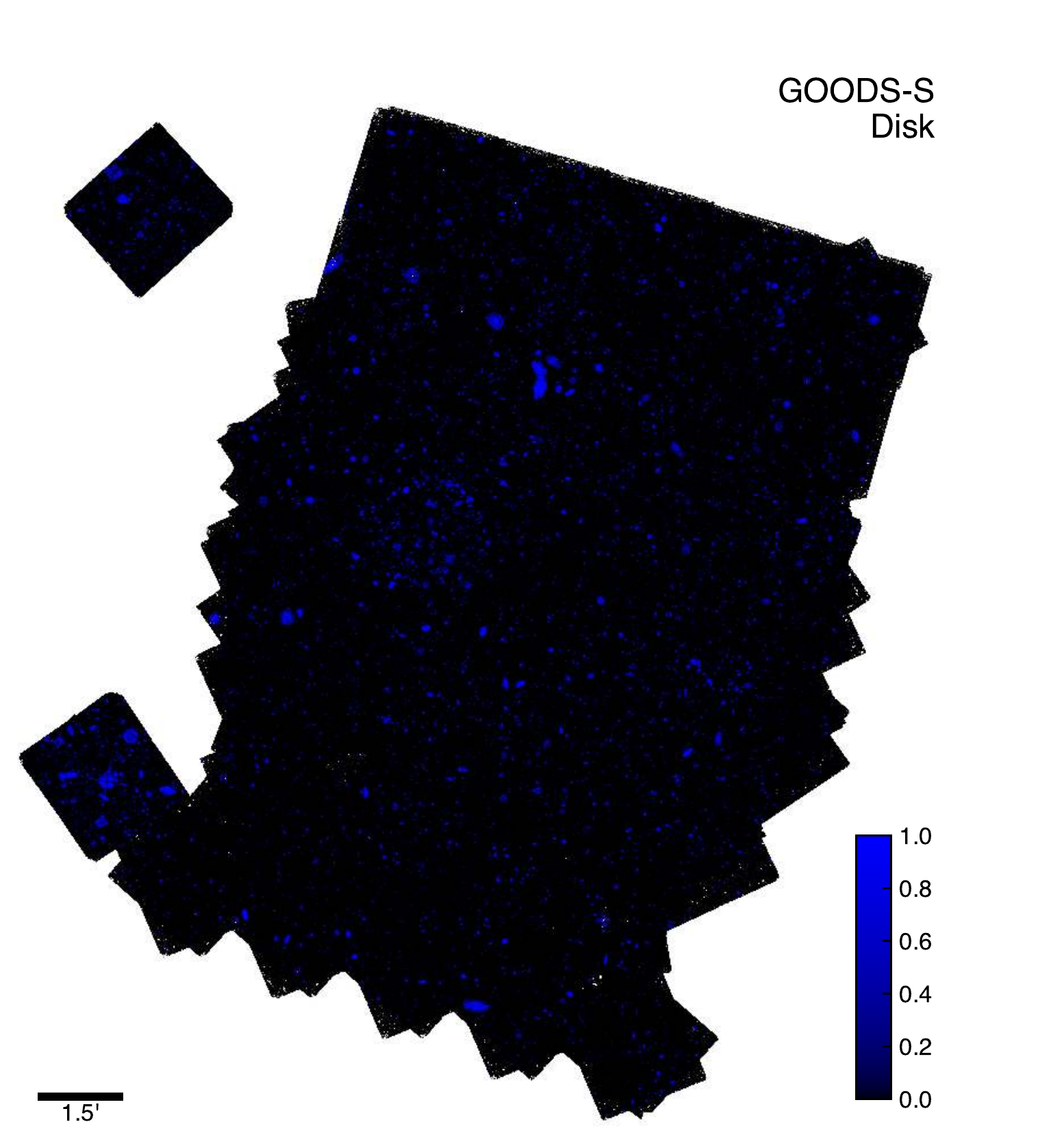}
\caption{\morpheus{} {\it disk} classification image for the Hubble Legacy
         Fields \citep{illingworth2016a} reduction of the CANDELS survey data
         \citep{grogin2011a,koekemoer2011a} in GOODS South. Shown are the
         normalized model estimates that each of the $\sim10^8$ pixels belongs
         to the {\it disk} class. The scale bar indicates 1.5 arcmin. The color
         bar indicates the {\it disk}${\in}[0,1]$, increasing from black to
         blue. Correspondingly, the bright blue areas indicate pixels where
         \morpheus{} identified {\it disk} objects.}
\label{fig:disk}
\end{figure*}

Figure \ref{fig:disk} shows the {\it disk} image for the \morpheus{} analysis of
the HLF reduction of GOODS South. The {\it disk} classification for each pixel
is shown on a black-to-blue colormap, with black corresponding to {\it
disk}${=}0$ and blue regions corresponding to {\it disk}${=}1$.

\subsubsection{Irregular Image}
\label{section:irregular_image}

\begin{figure*}
\centering
\includegraphics[]{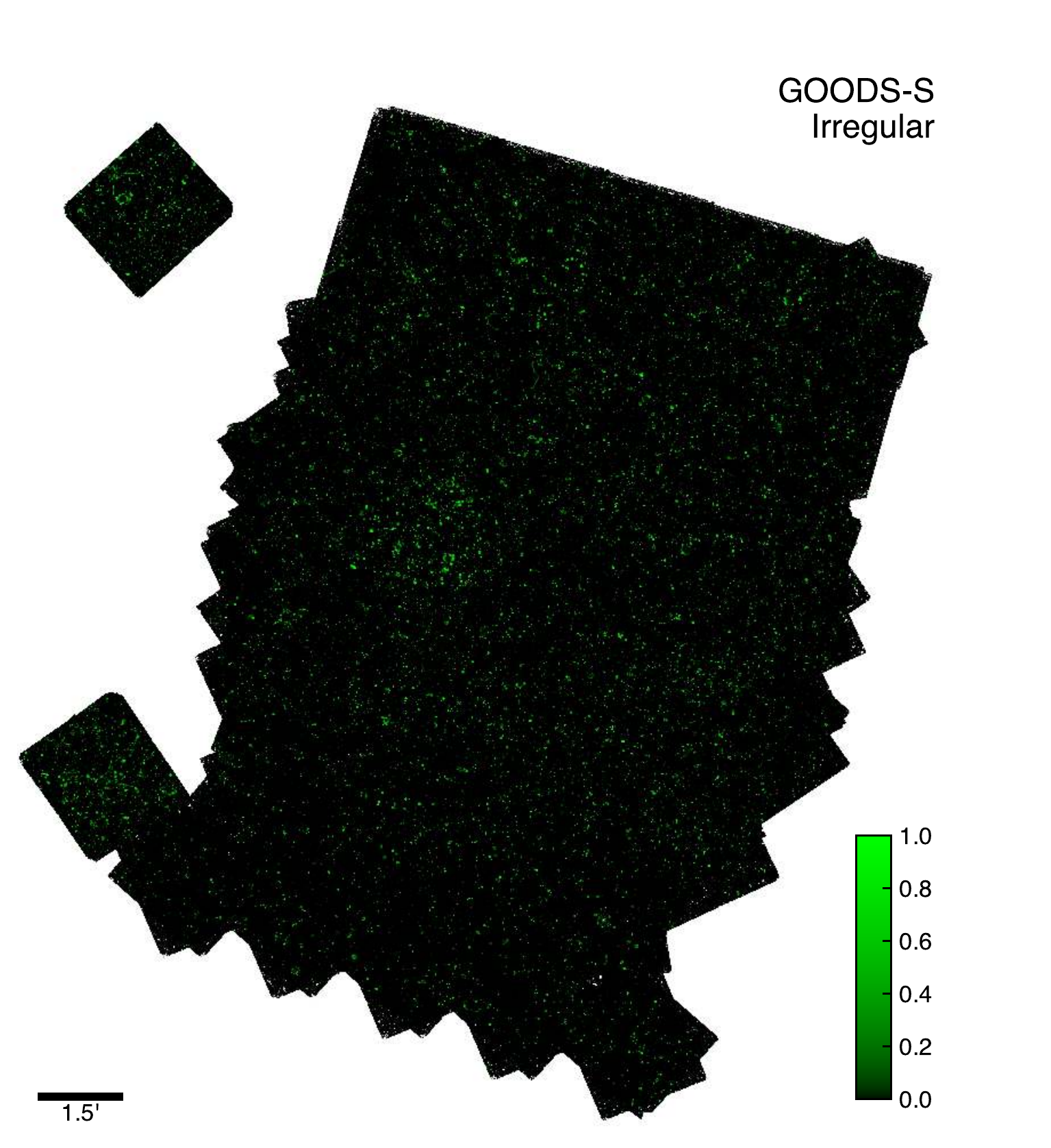}
\caption{\morpheus{} {\it irregular} classification image for the Hubble Legacy
         Fields \citep{illingworth2016a} reduction of the CANDELS survey data
         \citep{grogin2011a,koekemoer2011a} in GOODS South. Shown are the
         normalized model estimates that each of the $\sim10^8$ pixels belongs
         to the {\it irregular} class. The scale bar indicates 1.5 arcmin. The
         color bar indicates the {\it irregular}${\in}[0,1]$, increasing from
         black to green. Correspondingly, the bright green areas indicate pixels
         where \morpheus{} identified {\it irregular} objects.}
\label{fig:irregular}
\end{figure*}

Figure \ref{fig:irregular} shows the {\it disk} image for the \morpheus{}
analysis of the HLF reduction of GOODS South. The {\it irregular} classification
for each pixel is shown on a black-to-green colormap, with black corresponding
to {\it irregular}${=}0$ and green regions corresponding to {\it
irregular}${=}1$.

\subsubsection{Point source/compact Image}
\label{section:ps_compact_image}

\begin{figure*}
\centering
\includegraphics[]{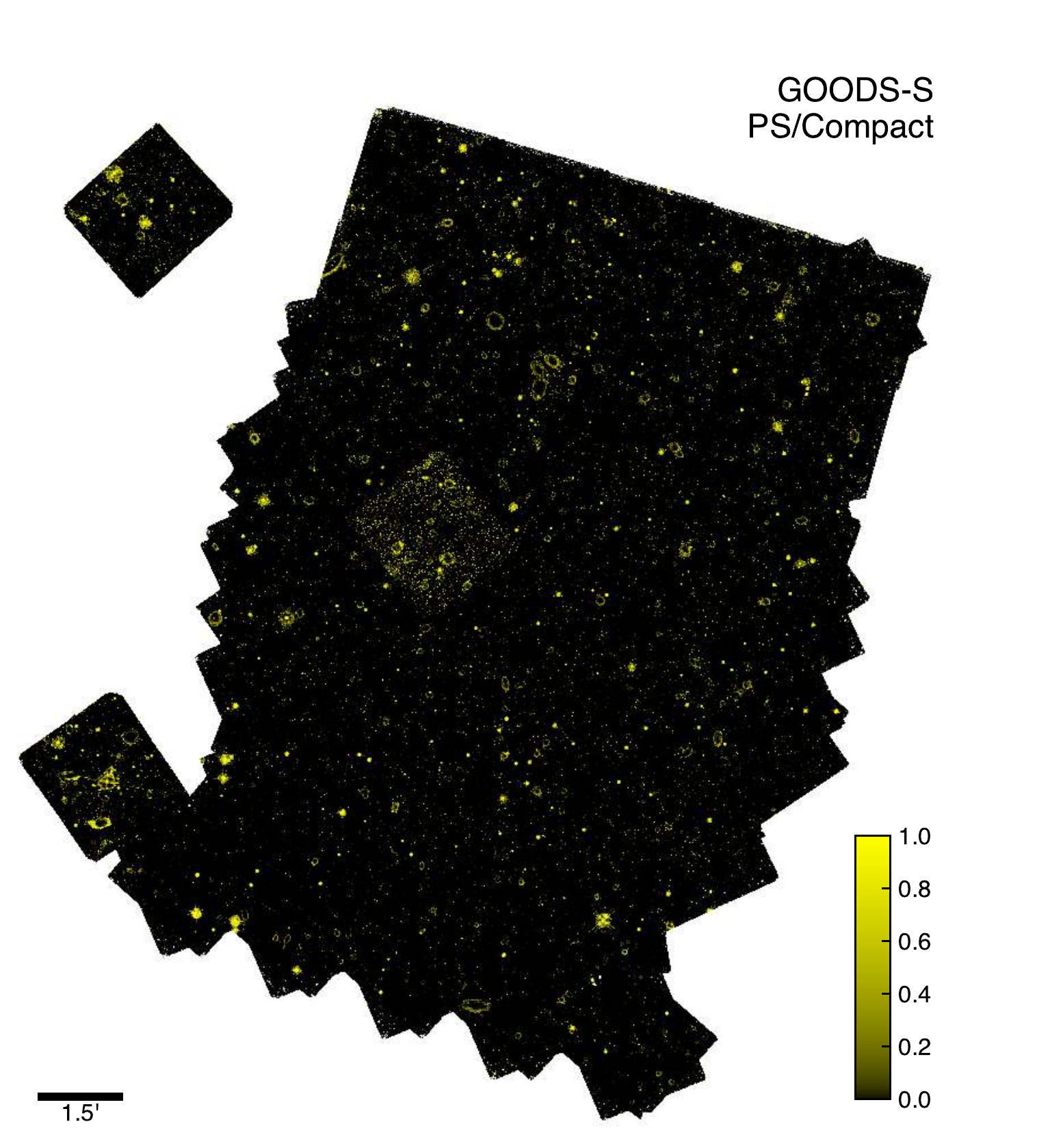}
\caption{\morpheus{} {\it point source/compact} classification image for the
         Hubble Legacy Fields \citep{illingworth2016a} reduction of the CANDELS
         survey data \citep{grogin2011a,koekemoer2011a} in GOODS South. Shown
         are the normalized model estimates that each of the $\sim10^8$ pixels
         belongs to the {\it point source/compact} class. The scale bar
         indicates 1.5 arcmin. The color bar indicates the {\it  point
         source/compact}${\in}[0,1]$, increasing from black to yellow.
         Correspondingly, the bright yellow areas indicate pixels where
         \morpheus{} identified {\it point source/compact} objects.}
\label{fig:point_source}
\end{figure*}

Figure \ref{fig:point_source} shows the {\it point source/compact} image for
the \morpheus{} analysis of the HLF reduction of GOODS South. The {\it point
source/compact} classification for each pixel is shown on a black-to-yellow
colormap, with black corresponding to {\it point source/compact}${=}0$ and
yellow regions corresponding to {\it point source/compact}${=}1$.

\subsubsection{Morphological Color Image}
\label{section:morphological_color_image}

Figure \ref{fig:point_source} shows the {\it morphological color} image for the
\morpheus{} analysis of the HLF reduction of GOODS South. The false color image
is constructed following Section \ref{section:products}, with the pixel
intensities scaling with $1-${\it background}, the pixel hues set according to
the dominant class, and the saturation indicating the indeterminacy of the pixel
classification. Pixels with a single dominant class appear as bright red, blue,
green, or yellow for {\it spheroid}, {\it disk}, {\it irregular}, or {\it point
source/compact} classifications, respectively. Bright white pixels indicate
regions of the image where the model results were indeterminate in selecting a
dominant class. Dark regions represent pixels the model classified as {\it
background}. We note that the pixel intensities are not scaled with the flux in
the image, and the per-object classifications require a local flux weighting
following Equation \ref{eqn:label_aggregation} and the process described in
Section \ref{section:morpheus_output}. This flux weighting usually results in a
distinctive class for each object, since the bright regions of objects often
have a dominant shared pixel classification. The outer regions of objects with
low flux show more substantial variation in the per-pixel classifications, but
these regions often do not contribute strongly to the flux-weighted per-object
classifications computed from this morphological color image.

\begin{figure*}
\centering
\includegraphics[]{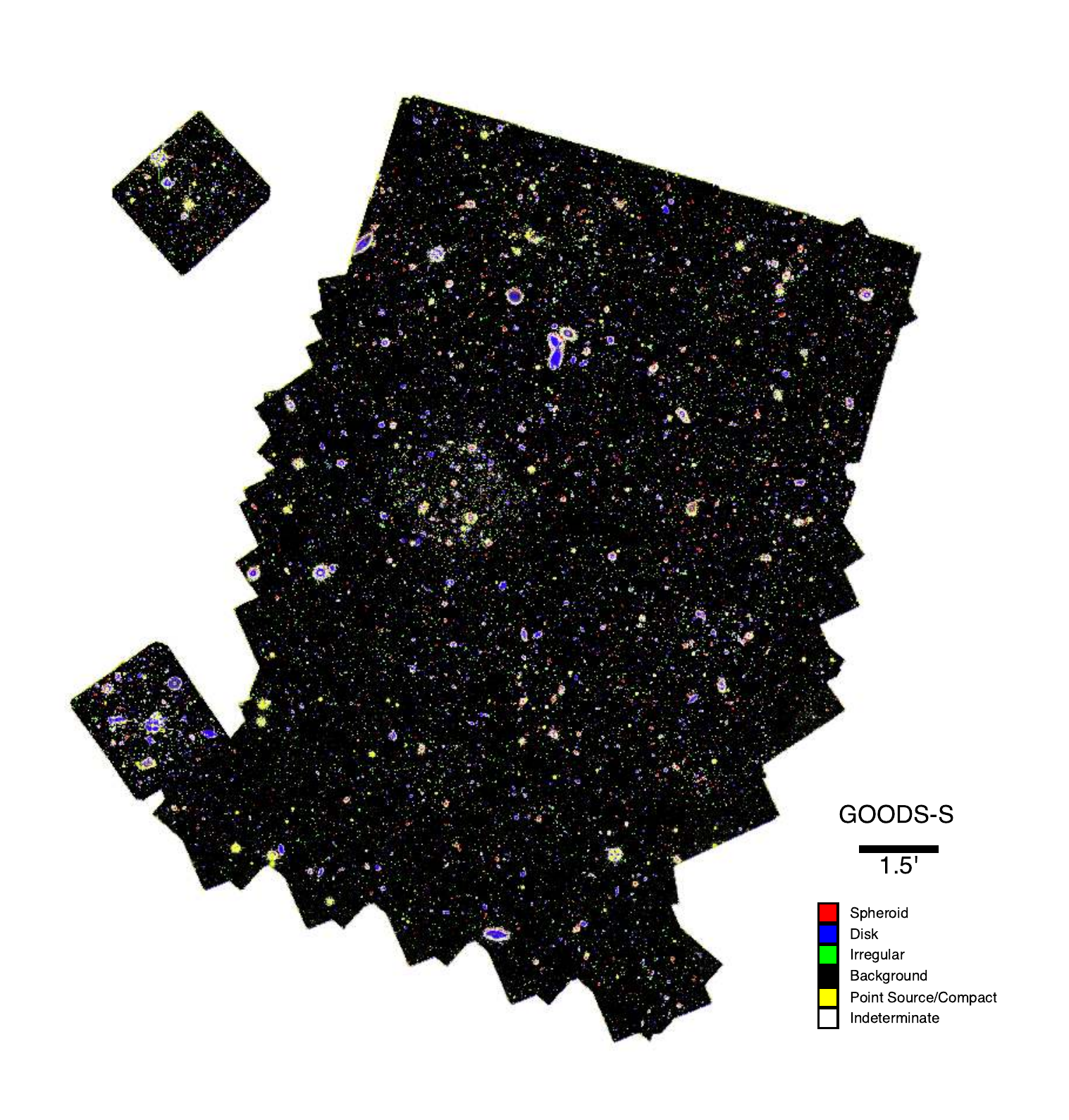}
\caption{\morpheus{} morphological color image for the Hubble Legacy Fields
         \citep{illingworth2016a} reduction of the CANDELS survey data
         \citep{grogin2011a,koekemoer2011a} in GOODS South. The image intensity
         is set proportional to $1{-}${\it background} for each pixel, such that
         regions of high background are black and regions with low background
         containing source pixels identified by \morpheus{} appear bright. The
         hue of each source pixel indicates its dominant classification, with
         {\it spheroid} shown as red, {\it disk} as blue, {\it irregular} as
         green, and {\it point source/compact} as yellow. The color saturation
         of each pixel is set to the difference between the first and second
         most dominant class values, such that regions with indeterminate
         morphologies as determined as \morpheus{} appear as white and regions
         with strongly determined classifications appear as deep colors. Note
         that the morphological color image is not flux-weighted, and the
         per-object classifications assigned by \morpheus{} include a
         flux-weighted average of the per-pixel classifications shown in this
         image.}
\label{fig:morphological_color}
\end{figure*}

\section{\morpheus{} Performance}
\label{section:performance}

Given the data products generated by \morpheus{}, we can perform a variety of
tests to quantify the performance of the model. There are basic performance
metrics relevant to how the model is optimized, reflecting the relative
agreement between the output of the model and the training data classifications.
However, given the semantic segmentation approach of \morpheus{} and the
pixel-level classification it provides, there are additional performance metrics
that can be constructed to mirror widely-used performance metrics in more
standard astronomical analyses including the completeness of sources detected by
\morpheus{} as regions of low background. In what follows, we attempt to address
both kinds of metrics and provide some ancillary quantifications to enable
translations between the performance of \morpheus{} as a deep learning framework
and as an astronomical analysis tool. In particular, we focus our analysis
on the 3D-HST catalog and HLF
reduction of the GOODS South region in the CANDELS Survey.

\subsection{Object-Level Morphological Classifications}
\label{section:test_morphology}

The semantic segmentation approach of \morpheus{} provides classifications for
each pixel in an astronomical image. These pixel-level classifications can then
be combined into object-level classifications $\bm{p}$ using the flux-weighted
average described by Equation \ref{eqn:label_aggregation}. The \morpheus{}
object-level classifications can then be compared directly with a test set of
visually-classified object morphologies provided by \citet{kartaltepe2015a}.

To understand the performance of \morpheus{} relative to the K15 visual
classifications, we present some summary statistics of the training and test
sets pulled from the K15 samples. During training, the loss function used by
\morpheus{} is computed relative to the distribution of input K15
classifications for each object and not only their dominant classification. The
goal is to retain a measure of the uncertainty in visual classifications for
cases where the morphology of an object is not distinct.

\subsubsection{Distribution of Training Sample Classifications}

\begin{figure}
\centering
\includegraphics[width=3.5in]{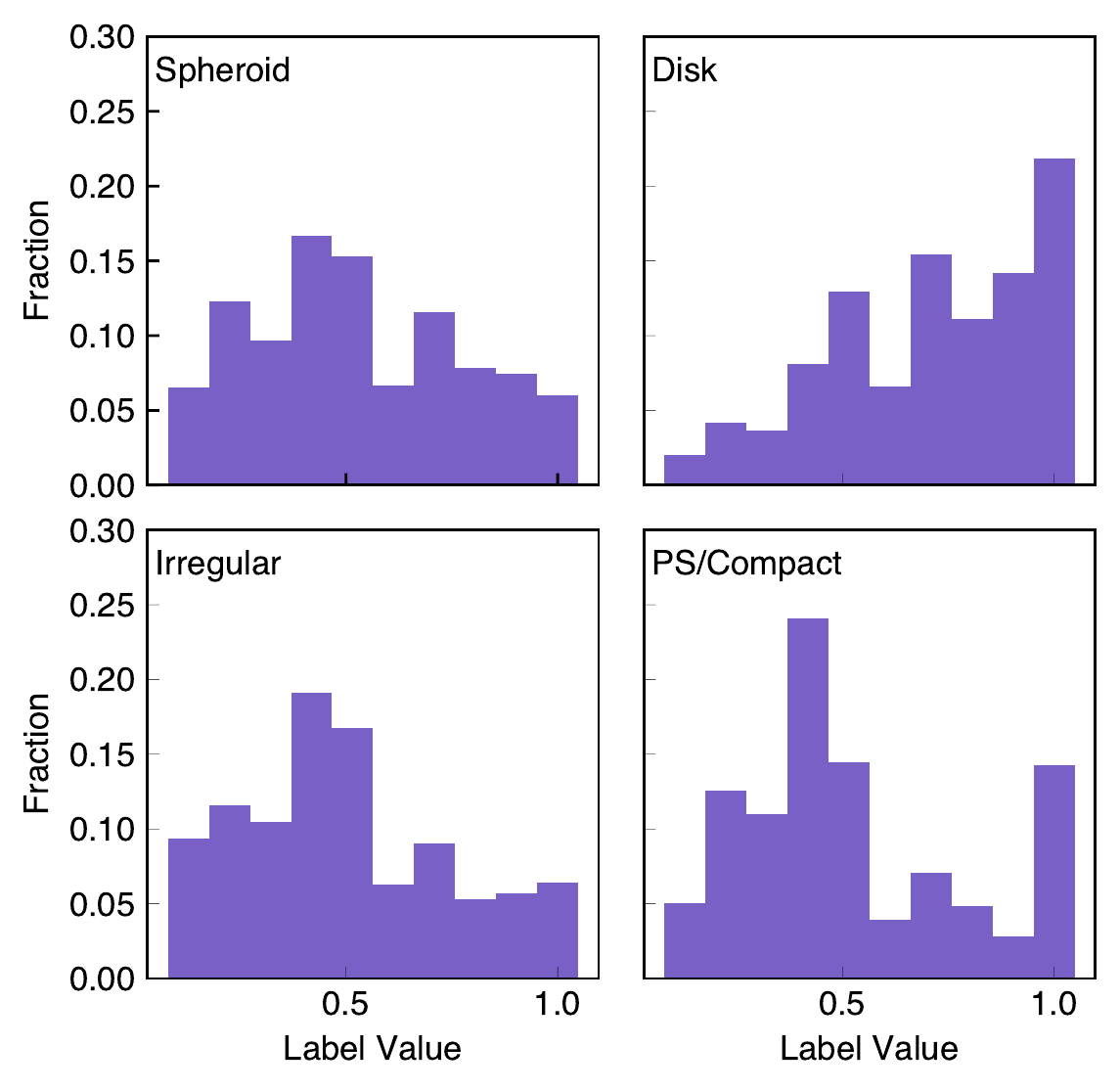}
\caption{Distribution of morphological classifications in the
         \citet{kartaltepe2015a} sample, which serve as a training sample for
         \morpheus{}. Shown are histograms of the fraction of sources with
         a non-zero probability of belonging to the {\it spheroid} (upper
         left), {\it disk} (upper right), {\it irregular} (lower left), or {\it
         point source/compact} classes, as determined visual classification by
         expert astronomers. The histograms have been normalized to show the
         distribution of classification probabilities for each class, and
         consist of ${\approx}7,600$ sources.}
\label{fig:confidence}
\end{figure}

Galaxies in the K15 training set have been visually classified by multiple
experts, providing a distribution of possible classifications for each object in
the sample. Figure \ref{fig:confidence} presents histograms of the fraction of
K15 classifiers recording votes for {\it spheroid}, {\it disk}, {\it irregular},
and {\it point source/compact} classes for each object. Only classes with more
than one vote are plotted.

\subsubsection{Classification Agreement in Training Sample}

To aid these comparisons, we introduce the {\it agreement} statistic
\begin{equation}
\label{eqn:agreement}
a(\bm{p}) = 1 - \frac{H(\bm{p})}{\log(n_c)}
\end{equation}
\noindent
where $\bm{p}$ is the distribution of classifications and $n_c$ is the number of
classes. The quantity
\begin{equation}
H(\bm{p})\equiv -\sum_{k=1}^{n_c} p_{k} \log p_k
\end{equation}
\noindent
is the self entropy. According to these definitions, $H(\bm{p}){\in}$ $[0,\log
n_c]$ and $a(\bm{p}){\in}[0,1]$. The agreement $a(\bm{p})\to1$ when the
distribution of classifications $\bm{p}$ is concentrated in a single class, and
$a(\bm{p})\to0$ when the classifications are equally distributed. For reference,
$a(\bm{p})\approx0.57$ for two equal classes and $a(\bm{p})\approx0.8$ for a
90\% / 10\% split between two classes for $n_c=5$ possible classes.

\subsubsection{Training and Test Set Statistics}

\begin{figure}
\centering
\includegraphics[width=3.5in]{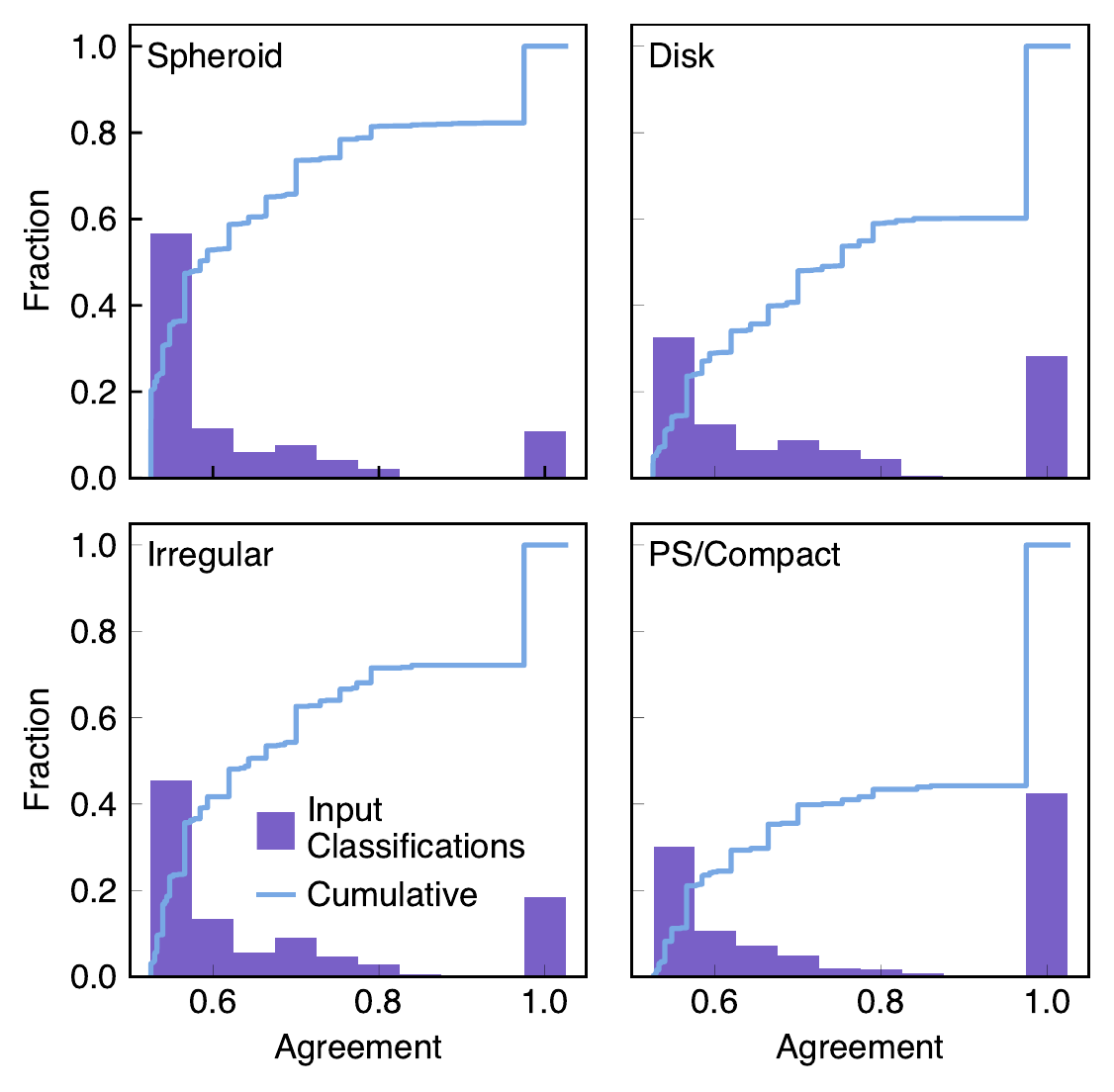}
\caption{Histograms (purple) and cumulative distribution (blue lines) of
         agreement $a(\bm{p})$ for the \citet[][K15]{kartaltepe2015a} visual
         morphological classifications, for objects with {\it spheroid} (upper
         left),  {\it disk} (upper right), {\it irregular} (lower left), and
         {\it point source/compact} (lower right) as their dominant
         classification. Agreement $a(\bm{p})$ (see Equation \ref{eqn:agreement}
         for a definition) characterizes the breadth of the distribution of
         morphological classes assigned by the K15 classifiers for each object,
         with $a(\bm{p}){=}1$ indicating perfect agreement of a single class and
         $a(\bm{p}){=}0$ corresponding to perfect disagreement with equal
         probability among classes. The distribution of agreement in the K15
         training classifications is roughly bimodal, with a strong peak
         near-perfect agreement and a broader peak near
         $a(\bm{p}){\approx}0.5$, close to the agreement value for an even split
         between two classes.}
\label{fig:cumulative_agreement}
\end{figure}

The K15 classifications have substantial variation in their agreement
$a(\bm{p})$. Figure \ref{fig:cumulative_agreement} shows histograms and the
cumulative distribution of $a(\bm{p})$ for objects with {\it spheroid}, {\it
disk}, {\it irregular}, and {\it point source/compact} dominant classes. These
distributions of $a(\bm{p})$ are roughly bimodal, consisting of a single peak
near $a(\bm{p}){=}1$ and a broader peak near $a(\bm{p}){\approx}0.5$ with a tail
to larger $a(\bm{p})$. As the cumulative distributions indicate, roughly
20\%-60\% of objects in the K15 sample had perfect agreement in their
morphological classification, with {\it disk} and {\it point source/compact}
being the most distinctive classes.

\begin{figure}
\centering
\includegraphics[width=\columnwidth]{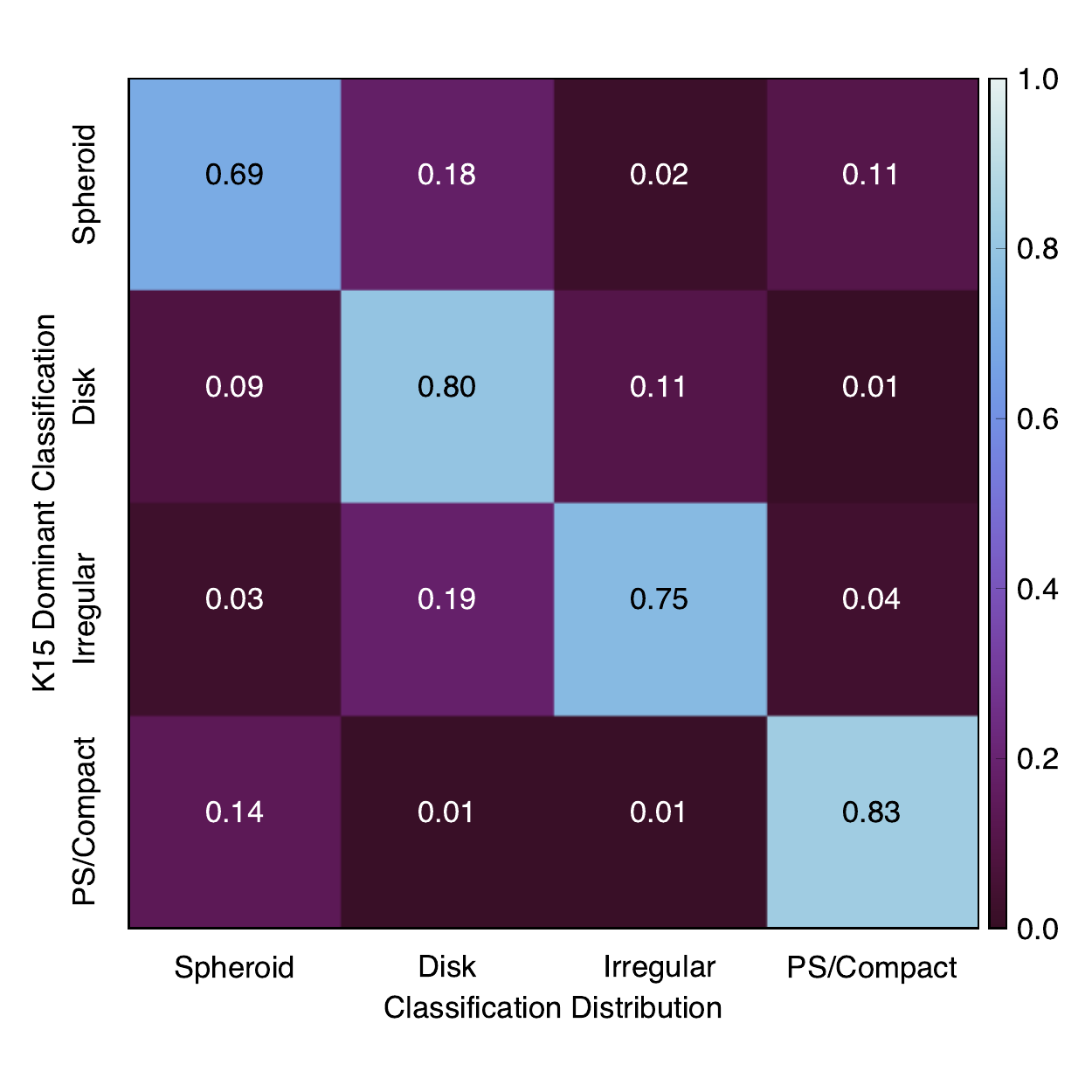}
\caption{Confusion matrix for the distribution of K15 morphological
         classifications. Shown is the distribution of
         morphologies assigned by K15 visual classifiers for objects of a given
         dominant classification. Objects with a dominant {\it spheroid} class
         show the most variation, with frequent additional {\it disk} and {\it
         point source/compact} morphologies assigned. The most distinctive
         dominant class is {\it point source/compact}, which also receives a
         {\it spheroid} classification in 14\% of objects. The off-diagonal
         components of the confusion matrix indicate imperfect agreement among
         the K15 classifiers, consistent with the distributions of the agreement
         statistic shown in Figure \ref{fig:cumulative_agreement}.}
\label{fig:label_confusion}
\end{figure}

\begin{figure}
\centering
\includegraphics[width=\columnwidth]{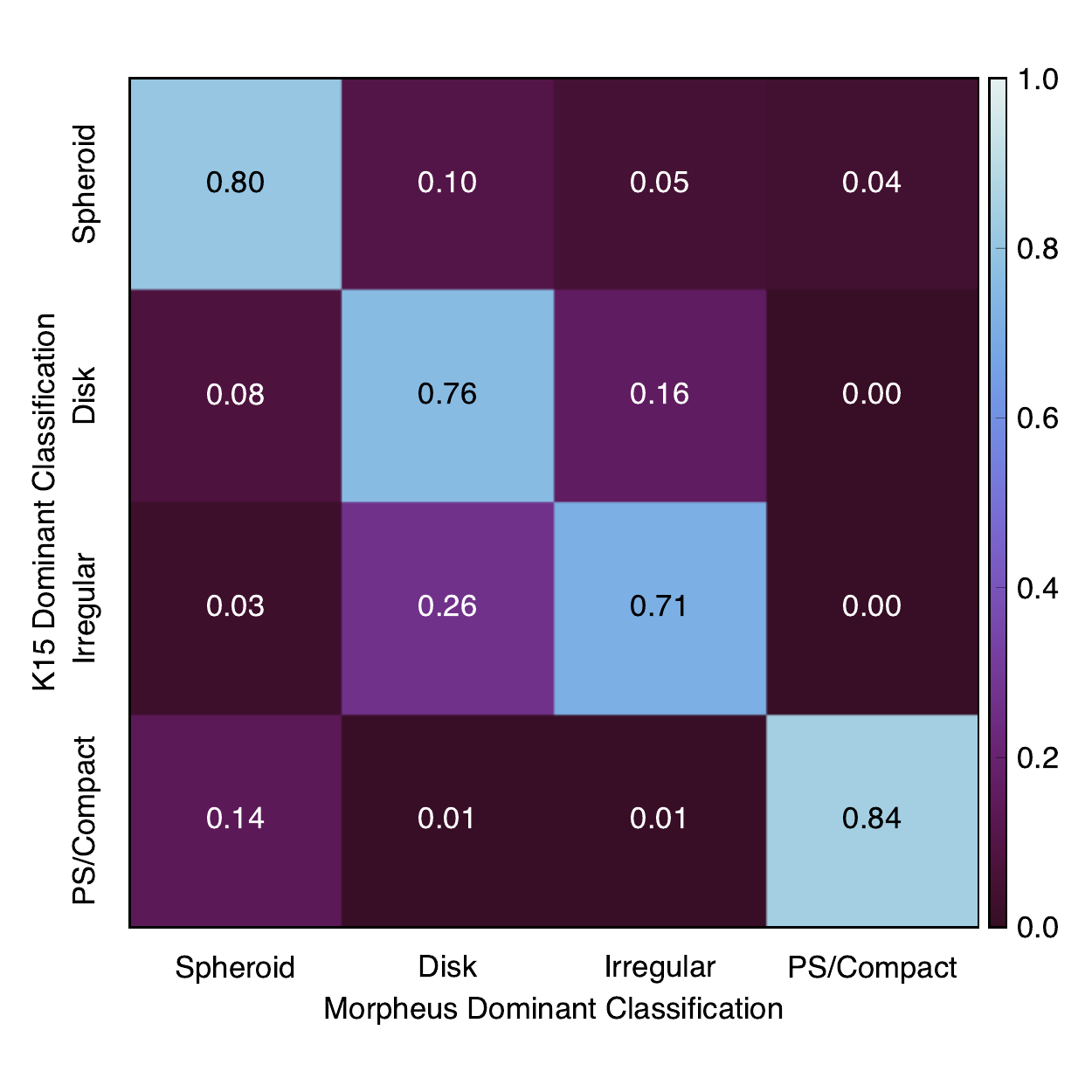}
\caption{Confusion matrix showing the spread in \morpheus{} dominant
         classifications for objects with a given K15 dominant classifications.
         The \morpheus{} framework is trained to reproduce the input K15
         distributions, and this confusion matrix should,
         therefore, largely match Figure \ref{fig:label_confusion}. The
         relative agreement between the two confusion matrices demonstrates that
         the \morpheus{} output can approximate the input K15 classification
         distributions.}
\label{fig:confusion_all}
\end{figure}

\begin{figure}
\centering
\includegraphics[width=\columnwidth]{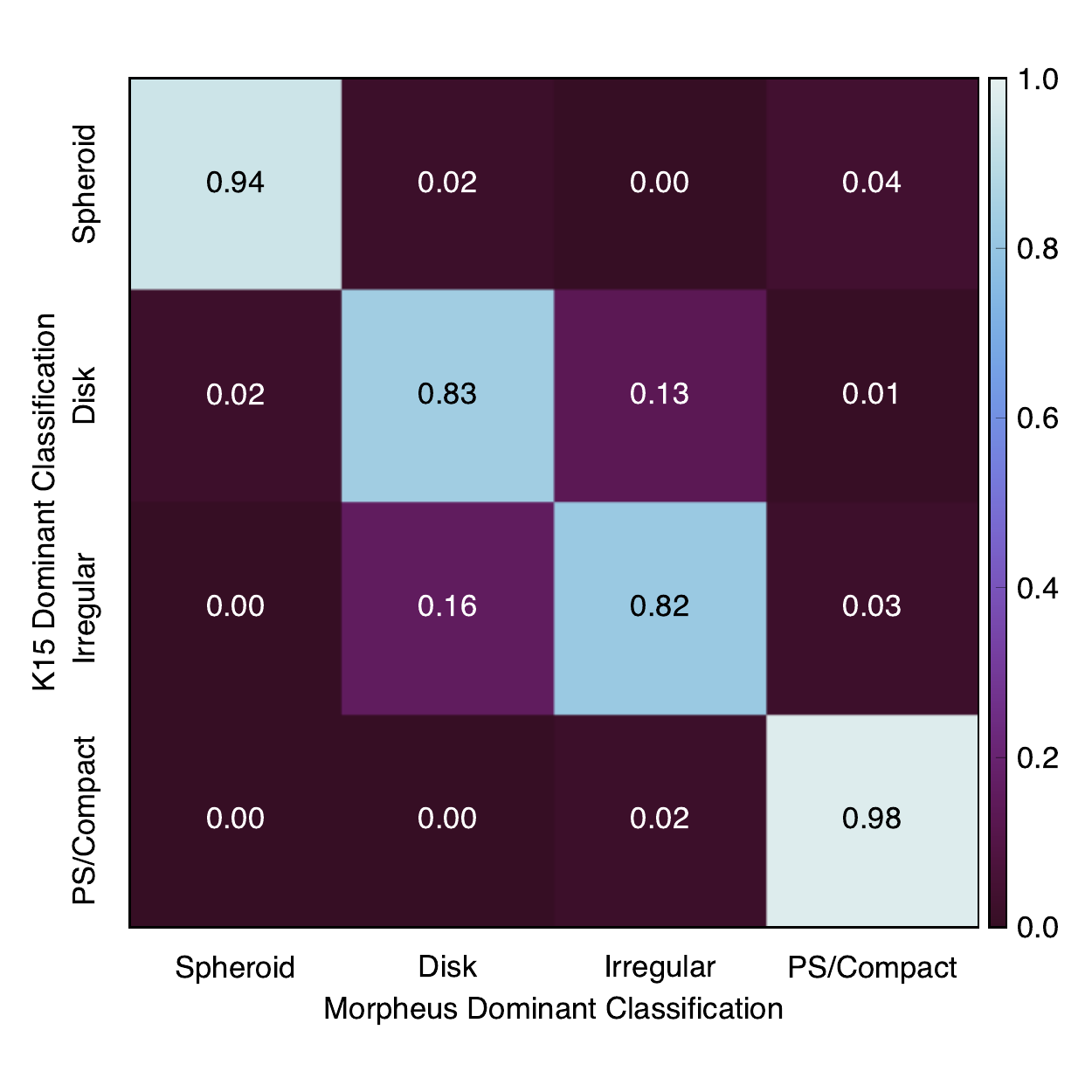}
\caption{Confusion matrix quantifying the spread in \morpheus{} dominant
         classifications for K15 objects with a distinctive morphology. Shown
         are the output \morpheus{} classification distributions for K15 objects
         where all visual classifiers agreed on the input classification. The
         \morpheus{} pixel-by-pixel classifications computed for the HLF GOODS
         South images were aggregated into flux-weighted object-by-object
         classifications following Equation \ref{eqn:label_aggregation} using
         the K15 segmentation maps. The results demonstrate that \morpheus{} can
         reproduce the results of the dominant K15 visual classifications for
         objects with distinct morphologies, even as the \morpheus{}
         classifications were computed from per-pixel classifications using
         different FITS images of the same region of the sky.}
\label{fig:morph_confusion_matrix}
\end{figure}

The breadth in the agreement statistic for the input K15 data indicates
substantial variation in how expert astronomers would visually classify
individual objects. As these data are used to train \morpheus{}, understanding
exactly what \morpheus{} should reproduce requires further analysis of the K15
data. An important characterization of the input K15 data is the confusion
matrix of object classifications. This matrix describes the typical
classification distribution for objects of a given dominant class. Figure
\ref{fig:label_confusion} presents the confusion matrix for the K15
classifications, showing the typical spread in classifications for objects
assigned {\it spheroid}, {\it disk}, {\it irregular}, or {\it point
source/compact} dominant morphologies. For reference, a confusion matrix for a
distribution with perfect agreement is the identity matrix. Figure
\ref{fig:label_confusion} provides some insight into the natural degeneracies
present in visually-classified morphologies. Objects with a dominant {\it disk}
classification are partially classified as {\it spheroid} (10\%) and {\it
irregular} (11\%). The {\it irregular} objects frequently receive an alternative
{\it disk} classification (19\%). The {\it point source/compact} objects also
are assigned {\it spheroid} classifications (14\%). Objects with a dominant {\it
spheroid} class have the highest variation and receive substantial
{\it disk} (18\%) and {\it point source/compact} (11\%) classifications. This
result is consistent with Figure \ref{fig:cumulative_agreement}, which shows a
relatively large disagreement for objects with a dominant {\it spheroid}
classification.

Since \morpheus{} is trained to reproduce the distribution of K15
classifications, the confusion matrix between the dominant \morpheus{}
classifications and the K15 classification distributions should be similar to
Figure \ref{fig:label_confusion}. Indeed, Figure \ref{fig:confusion_all} shows
the distribution of K15 classifications for objects with a given dominant
\morpheus{} classification agrees well with the input K15 distributions shown in
Figure \ref{fig:label_confusion}. This result demonstrates \morpheus{}
reproduces well the intrinsic uncertainty in the K15 classifications, as
measured by the distribution of morphologies, recovered for a given K15
dominant classification.

The ability of \morpheus{} to reproduce the distribution of K15 classifications
is not the only metric of interest, as it does not indicate whether the
object-by-object \morpheus{} classifications agree with the K15 classifications
for objects with distinctive morphologies. Figure \ref{fig:cumulative_agreement}
shows that 20-60\% of objects in the K15 classifications have an agreement
$a(\bm{p}){=}1$, meaning that all K15 visual classifiers agreed on the object
morphology. The confusion matrix for these distinctive objects constructed from
the K15 data is diagonal, and the confusion matrix for these objects constructed
from the \morpheus{} output should also be diagonal if \morpheus{} perfectly
reproduced the object-by-object K15 classifications. Further, to ensure
that \morpheus{} captures the distribution of the K15 morphologies, the
cumulative distribution of dominant K15 morphologies and dominant \morpheus{}
morphologies as a function of color were compared using a two-sample
Kolmogorov–Smirnov test. For each morphology, the p-values ($p=0.3-0.99$)
indicate consistency between the Morpheus and K15 distributions as a function of
color. These results suggest that \morpheus{} accurately captures the K15
representation of morphology without significant color bias.

To characterize the performance of \morpheus{} for the $a(\bm{p}){=}1$ K15
subsample, we used the \morpheus{} output classification images computed from
the HLF GOODS South images. The flux-weighted \morpheus{} morphological
classifications were computed following Equation
\ref{eqn:label_aggregation} and using the K15 segmentation maps to
ensure the same pixels were being evaluated. Figure
\ref{fig:morph_confusion_matrix} presents the resulting confusion matrix showing
the \morpheus{} dominant classification for each object's dominant
classification determined by K15. As Figure \ref{fig:morph_confusion_matrix}
demonstrates, \morpheus{} achieves extremely high agreement ($\ge90\%$) with K15
for {\it spheroid} and {\it point source/compact} objects, and good agreement
($\ge80\%$) for {\it disk} and {\it irregular} objects with some mixing
${\sim}15\%$ between them. This performance is comparable to other
object-by-object morphological classifications in the literature
\citep[e.g.,][]{huertas-company2015a}, but is constructed directly from a
flux-weighted average of pixel-by-pixel classifications by \morpheus{} using
real FITS image data of differing formats and depth.

\subsubsection{Redshift Evolution of Morphology in CANDELS Galaxies}
\label{sec:catalog_comparison}

\begin{figure}
\centering
\includegraphics[width=\columnwidth]{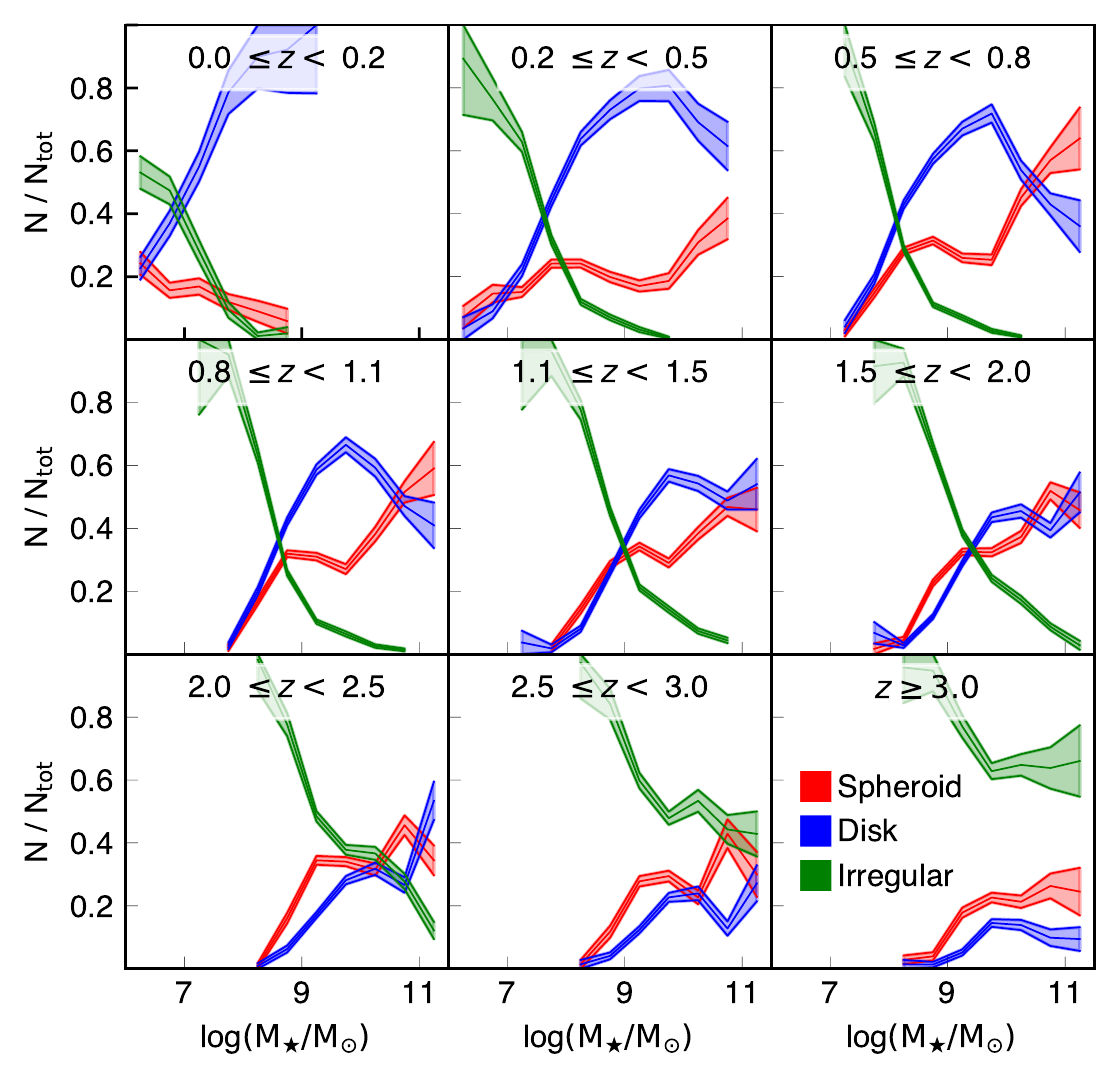}
\caption{Morphology as a function of stellar mass and redshift for ~54,000
        sources in the five CANDELS fields. Sources included in the plot are
        those where $H<24.5$AB and the \morpheus{} confidence for {\it spheroid},
        {\it disk},
        or {\it irregular} is greater than 0.7. See Section \ref{sec:catalog_comparison}.}
\label{fig:stellar_mass_morph}
\end{figure}

To illustrate the scientific applications of \morpheus{,
we examine the morphological distribution of $\sim$54,000
3D-HST sources in the five CANDELS fields as a function of redshift and stellar
mass (Figure \ref{fig:stellar_mass_morph}). We combine together the
flux-weighted \morpheus{} classifications of galaxies identified in CANDELS with
the 3D-HST stellar masses and
redshift, dividing the sample into coarse redshift bins. The fraction of objects
$N/N_{tot}$ with a flux-weighted classification of {\it spheroid} (red),
{\it disk} (blue), or {\it irregular} (green) are shown as a function of
stellar mass for each redshift bin, along with Poisson uncertainties on the
binned values. The well-established trends of increasing fractions of irregular
objects at small masses and high redshifts are correctly reproduced by
\morpheus{}, as well
as the growth of the disk population at low redshifts.
These results can be compared
with the results
reported in Figure 3 of \citet[][HC16]{huertas-company2016a}.
To ensure comparable samples
between HC16 and this work, the \morpheus{}-classified samples in
Figure \ref{fig:stellar_mass_morph} are limited to
objects with $H<$24.5AB. Since HC16 and Morpheus use similar
but not
identical morphological classifications, we adapt the sample definitions
used by HC16 to the \morpheus{} classification scheme. To be counted as a
part of a morphological class, each galaxy's flux-weighted confidence value assigned by
\morpheus{} must be
greater than 0.7. This threshold ensures each classification is
mutually exclusive, but
low enough to ensure a comparable sample size to HC16.}

The trends in Figure \ref{fig:stellar_mass_morph} agree with those found by
HC16 in two important aspects. First, at lower redshifts, disks tend to dominate
spheroids, and as redshift increases, spheroids tend to dominate disks. Second,
irregular sources are a larger portion of the population than spheroids and
disks at lower stellar masses and more become less abundant than spheroids and
disks as stellar mass increases. The agreement between \morpheus{} and the
results of HC16, which were based on object-level classifications, confirms
the ability of \morpheus{} to capture source-level
morphologies by aggregating pixel-level classifications.

\subsection{Simulated Detection Tests}
\label{section:detection_limit_tests}

The \morpheus{} framework enables the detection of astronomical objects by
producing a {\it background} classification image, with source locations
corresponding to regions where {\it background}${<}1$. If generating points in
the form of a source catalog are not supplied, the
segmentation algorithm of \morpheus{} uses an even more restrictive condition
that regions near sources must contain pixels with {\it background}${=}0$. Given
that the semantic segmentation algorithm of \morpheus{} was trained on the K15
sample that has a completeness limit, whether the regions identified by
\morpheus{} to have {\it background}${=}0$ correspond to an approximate flux
limit should be tested. Similarly, whether noise fluctuations lead to regions
assigned {\it background}${\approx}0$ in error should also be evaluated.

Below, we summarize detection tests for \morpheus{} using simulated images. For
these tests, a simulated sky background was generated using
Gaussian random noise with RMS
scatter measured in $0.5''$ apertures after convolving with a model HST PSF and
scaled to that measured from the K15 training images. The Tiny Tim software
\citep{krist2011a} software was used to produce the PSF models appropriate for
each band.

\subsubsection{Simulated False Positive Test}
\label{section:false_positives}

Provided a large enough image of the sky, random sampling of the noise could
produce regions with local fluctuation some factor $f$ above the
RMS background $\sigma$ and lead to a false positive
detection. A classical extraction technique using aperture flux thresholds would
typically identify such regions as a $SNR=f$ source. Here, we evaluate whether
\morpheus{} behaves similarly.

Using the Gaussian random noise field,
single-pixel fluctuations were added to the $H$-band only such that the
local flux measured in a $0.5''$ aperture after convolving with Tiny Tim
corresponded to $SNR{=}[0.5, 1, 2, 3, 4, 5, 6, 7, 10]$. The false signals were
inserted at well-separated locations such that \morpheus{} evaluated them
independently. The $V$, $z$, and $J$ images were left as blank noise, and then
all four images were supplied to \morpheus{}. We find that \morpheus{} assigns
none of these fake signals pixels with {\it background}${=}0$. However, the
$SNR{=}7$ and $SNR{=}10$ regions have some {\it background}${<}1$ pixels, and
while in the default algorithm, \morpheus{} would not assign these
regions segmentation maps, a more permissive version of the
algorithm could. An alternative test was performed by replacing the $SNR=10$
noise fluctuation in the $H$-band image with a Tiny Tim $H$-band PSF, added
after the convolution step with an amplitude corresponding to $SNR=10$ measured
in a $0.5''$ aperture. This test evaluates whether the shape of flux
distribution influences the detection of single-band noise fluctuations.
In this case, the minimum pixel values decreased to {\it
background}${\approx}0.05$ for a single band $SNR{=}10$ fluctuation shaped like
an $H$-band PSF, but did not lead to a detection. We conclude that \morpheus{}
is robust to false positives arising from relatively large ($SNR\lesssim7$)
noise fluctuations.

\subsubsection{False Negative Test}
\label{section:false_negatives}

Given that \morpheus{} seems insensitive to false positives from noise
fluctuations, it may also miss real but low $SNR$ sources. By performing a
similar test to that presented in Section \ref{section:false_positives} but with
sources inserted in all bands rather than noise fluctuations inserted in a
single band, the typical $SNR$ where \morpheus{} becomes incomplete for real
objects can be estimated.

\begin{figure}
    \centering
    \includegraphics[width=\columnwidth]{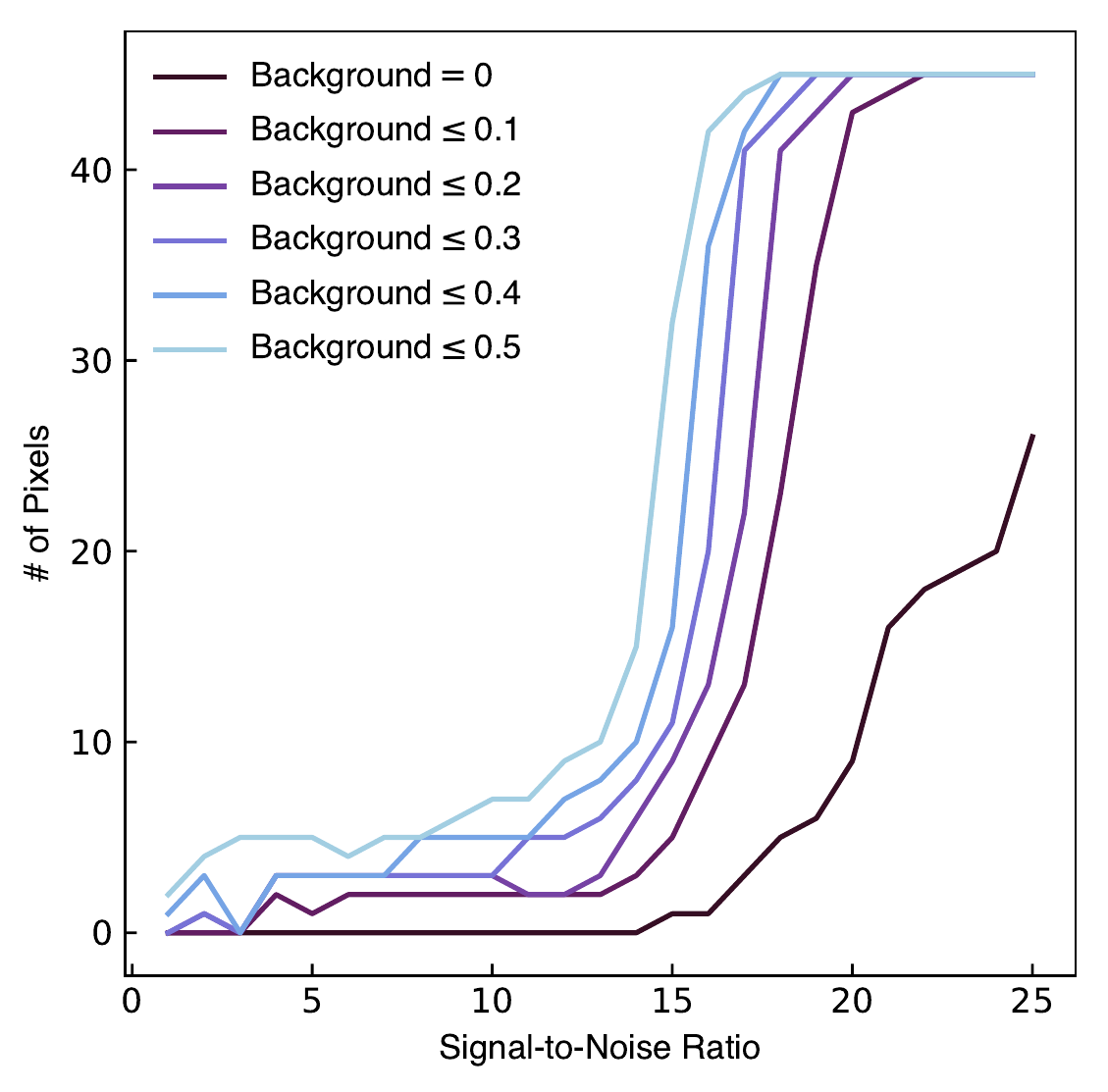}
    \caption{False negative test for the \morpheus{} source detection scheme.
             Simulated sources with different signal-to-noise ratios ($SNRs$)
             were inserted into a noise image and then recovered by \morpheus,
             which assigns a low {\it background} value to regions it identifies
             as containing source flux (see Section
             \ref{section:false_negatives}). Shown are lines corresponding to
             the number of pixels assigned to sources of different $SNR$, as a
             function of the {\it background} threshold. As trained on the K15
             sample, \morpheus{} becomes incomplete for objects with
             $SNR\lesssim15$, and is more complete if the threshold for
             identifying sources is made more permissive (i.e., at a higher {\it
             background} value).}
    \label{fig:false_negatives}
\end{figure}

Noise images were generated to have the same RMS noise as
the K15 images by convolving Gaussian random variates
with the Tiny Tim \citep{krist2011a} model for the HST PSF. An array of
well-separated point sources modeled by the PSF were then inserted with a range
of $SNR{\in}[1,25]$ into all four input band images. The \morpheus{} model was
then applied to the images, and the output {\it background} image analyzed to
find regions with {\it background} below some threshold value. Figure
\ref{fig:false_negatives} shows the number of pixels below various {\it
background} threshold values assigned to objects with different $SNR$. Below
about $SNR\sim15$, the number of pixels identified as low {\it background}
begins to decline rapidly. We therefore expect \morpheus{} to show
incompleteness in real data for $SNR\lesssim15$ sources. However, we emphasize
that this limitation likely depends on the training sample used. Indeed, the K15
training data set is complete to $H=24.5$AB in images with $5\sigma$ source
sensitivities of $H\approx27$AB. If trained on deeper samples, \morpheus{} may
prove more complete to fainter magnitudes. We revisit this issue in Section
\ref{section:detection} below, but will explore training \morpheus{} on deeper
training sets in future work.

\subsection{Morphological Classification vs. Surface Brightness Profile}
\label{section:sersic}

In this paper, the \morpheus{} framework is trained on the K15 visual
classifications to provide pixel-level morphologies for galaxies. The K15
galaxies are real astronomical objects with a range of surface brightness
profiles for a given dominant morphology. Correspondingly, the typical
classification that \morpheus{} would assign to idealized objects
with a specified surface brightness profile is difficult to anticipate without
computing it directly. Understanding how \morpheus{} would classify idealized
galaxy models can provide some intuition about how the deep learning framework
operates and what image features are related to output \morpheus{}
classifications.

\begin{figure}
\centering
\includegraphics[width=\linewidth]{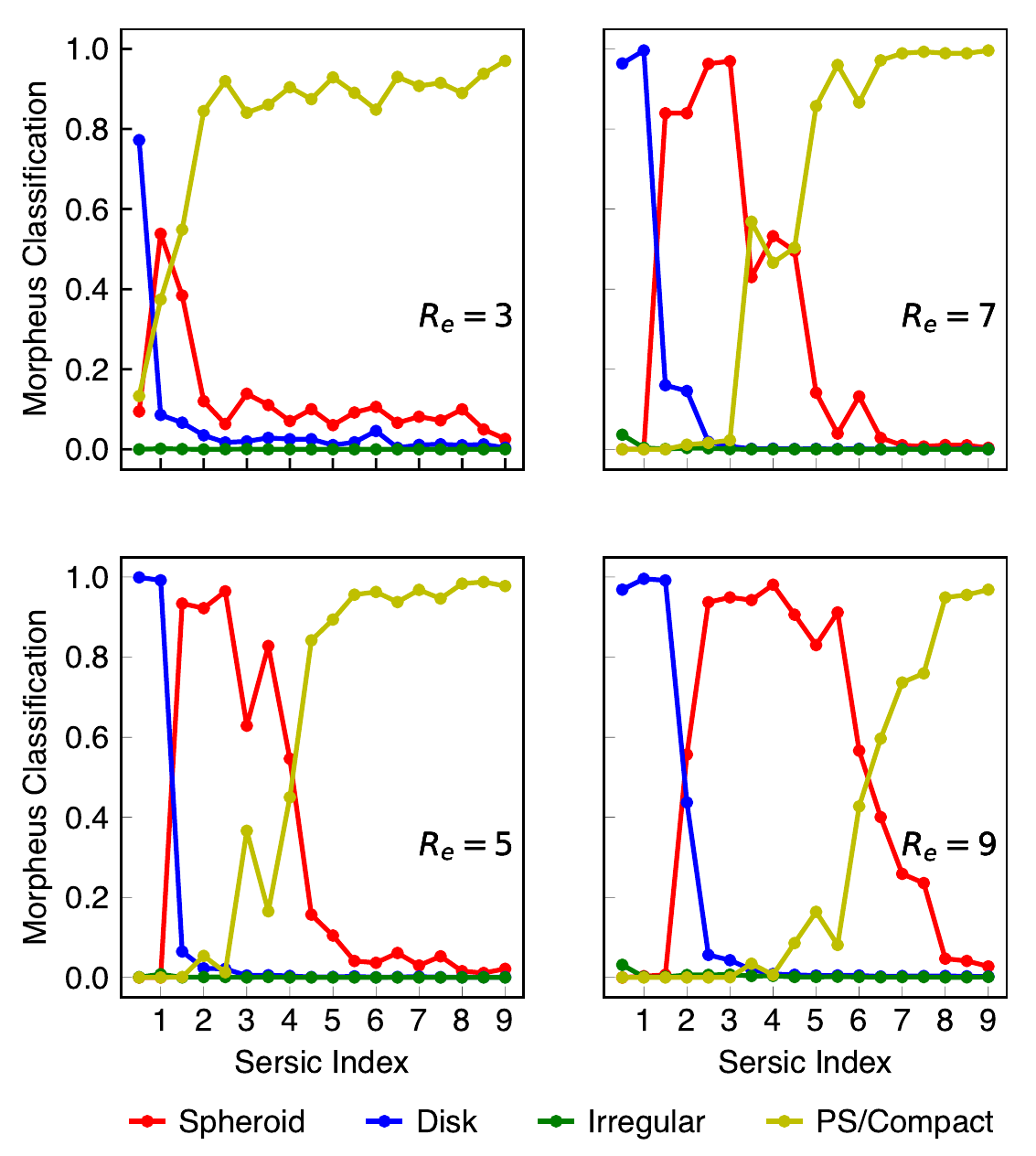}
\caption{Morphological classifications as a function of simulated source surface
         brightness profile Sersic index. Shown are the \morpheus{}
         classification distributions for simulated $SNR=20$ objects with
         circular \citet{sersic1968a} profiles, as a function of the Sersic
         index $\eta{\in}[1,9]$. The experiment was repeated on objects with
         effective radii of three (upper left panel), five (upper right panel),
         seven (lower left panel), and nine (lower right panel) pixels. Objects
         with $\eta=1$ were dominantly classified as {\it disk} or {\it
         spheroid}. Intermediate Sersic profiles ($\eta\sim2-3$) were mostly
         classified as {\it spheroid}. Objects with high Sersic index
         ($\eta{\ge}4$) were classified as {\it point source/compact}. These
         simulated objects with azimuthally symmetrical surface brightness
         profiles were assigned almost no {\it irregular} classifications by
         \morpheus{}.}
\label{fig:sersic_n_test_lines}
\end{figure}

Figure \ref{fig:sersic_n_test_lines} shows the output \morpheus{} classification
distribution for simulated objects with circular \citet{sersic1968a} surface
brightness profiles, for objects with $SNR=20$, Sersic indices $\eta{\in}[1,9]$,
and effective radii ranging from three to nine pixels. Synthetic FITS images for
each object in each band were constructed by assuming zero color gradients and a
flat $f_{\nu}$ spectrum, populating the image with a Sersic profile object and
noise consistent with the K15 images, and then convolving the images with a Tiny
Tim point spread function model appropriate for each input HST filter.

\begin{figure}
\centering
\includegraphics[width=\linewidth]{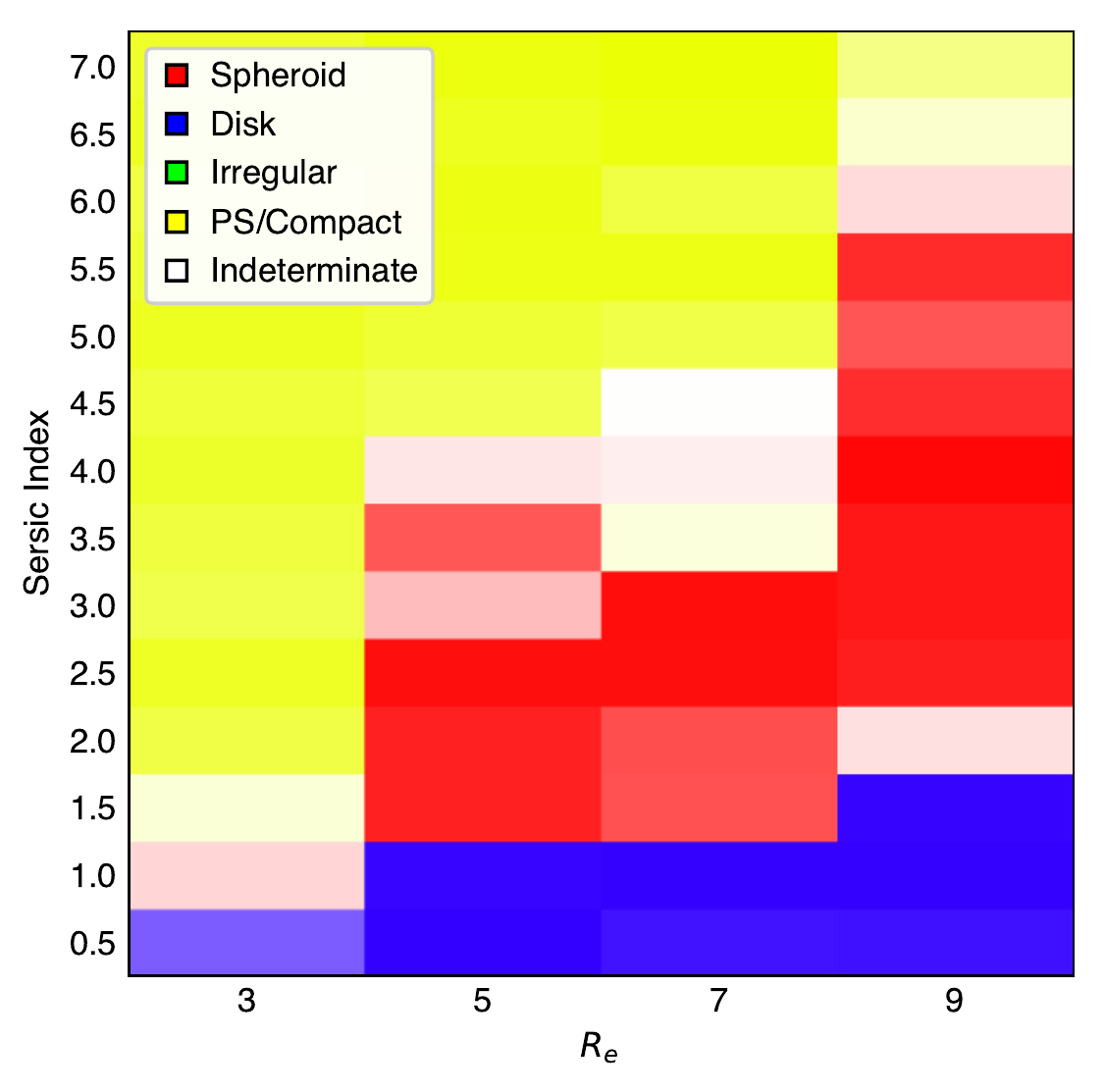}
\caption{Dominant morphological classification as a function of simulated source
         surface brightness profile Sersic index $\eta$ and effective radius
         $R_e$ in pixels. Each element of the matrix is color-coded to
         indicate the dominant \morpheus{} classification assigned for each
         $[\eta, R_e]$ pair, with the saturation of the color corresponding to
         the difference between the dominant and second \morpheus{}
         classification values. Large objects with low Sersic index are
         classified as {\it disk} (blue). Large objects with high Sersic index
         are classified as {\it spheroid} (red). Small objects with high Sersic
         index are classified as {\it point source/compact} (yellow). None of
         the symmetrical objects in the test were classified as {\it irregular}
         (green).}
\label{fig:sersic_n_test_matrix}
\end{figure}

The results from \morpheus{} reflect common expectations for the typical Sersic
profile of morphological classes. Objects with $\eta=1$ were typically
classified as {\it disk} or {\it spheroid}, while intermediate Sersic index
objects (e.g., $\eta{\approx}2-3$) were classified as {\it spheroid}. More
compact objects, with Sersic indices $\eta{\ge}4$, were dominantly classified as
{\it point source/compact}. Also, as expected for azimuthally-symmetric
surface brightness profiles, \morpheus{} did not significantly classify any
objects as {\it irregular}. Figure \ref{fig:sersic_n_test_matrix} provides a
complementary summary of the \morpheus{} classification of Sersic profile
objects, showing a matrix indicating the dominant classification assigned for
each pair of $[\eta, R_e]$ values. The \morpheus{} model classifies large
objects with low $\eta$ as {\it disk}, large objects with high $\eta$ as {\it
spheroid}, and small objects with high $\eta$ as {\it point source/compact}.

Overall, this test indicates that for objects with circular Sersic profiles,
\morpheus{} reproduces the expected morphological classifications and that
asymmetries in the surface brightness are needed for \morpheus{} to return an
{\it irregular} morphological classification.

\begin{figure}
\centering
\includegraphics[width=\columnwidth]{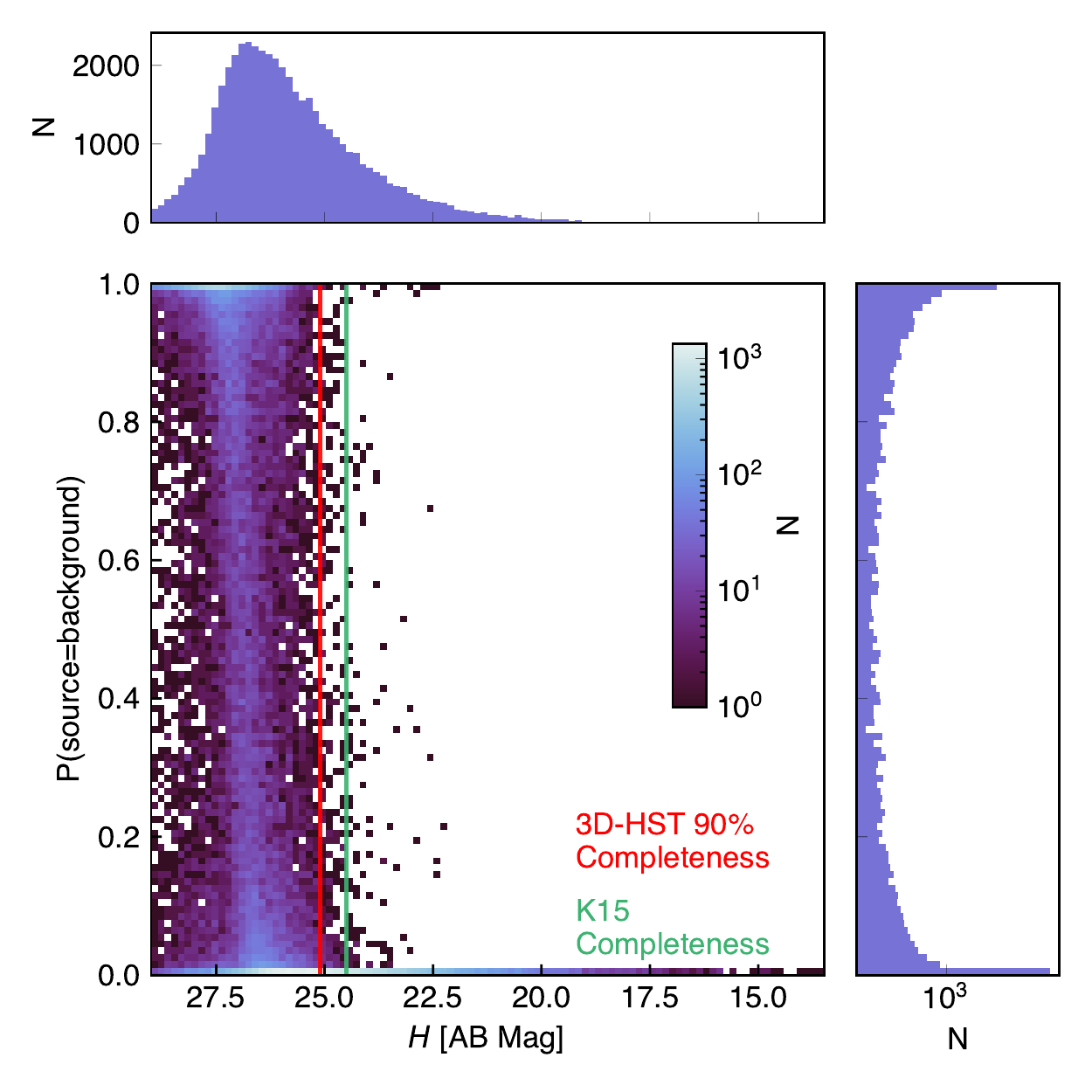}
\caption{Two-dimensional histogram of \morpheus{} {\it background} values and
         3D-HST source flux in GOODS South. Shown is the distribution of
         {\it background} at the location of 3D-HST sources
         \citep{skelton2014a,momcheva2016a} in GOODS South of various
         $H$-band magnitudes, along with the marginal histograms for both
         quantities (side panels). For reference, the K15 completeness (green
         line) and 3D-HST 90\% completeness (red line) flux limits are also
         shown. The 3D-HST sources most frequently have {\it background}${=}0$,
         and the majority of 3D-HST sources of any flux $H<29$ have {\it
         background}${<}0.5$. The {\it background} values for objects where K15
         and 3D-HST are complete is frequently zero. The \morpheus{} {\it
         background} values increase for many objects at flux levels $H>26$AB.}
\label{fig:bkg_pixel_hist}
\end{figure}

\subsection{Source Detection and Completeness}
\label{section:detection}

The semantic segmentation capability of \morpheus{} allows for the detection of
astronomical objects directly from the pixel classifications. In its simplest
form, this object detection corresponds to regions of the output \morpheus{}
classification images with low {\it background} class values. However,
the \morpheus{} object detection capability raises several
questions. The model was trained on the K15 sample, which has a reported
completeness of $H=24.5$AB, and given the pixel-by-pixel {\it background}
classifications computed by \morpheus{}, it is unclear whether the
object-level detection of sources in images would match the K15 completeness. In
regions of low {\it background}, the transition to regions of high {\it
background} likely depends on the individual pixel fluxes, but this transition
should be characterized.

In what follows below, we provide some quantification of the \morpheus{}
performance for identifying objects with different fluxes. To do this, we use
results from the 3D-HST catalog of sources for the GOODS South
\citep{skelton2014a,momcheva2016a}. Given the output \morpheus{}
{\it background} classification images computed from the HLF GOODS South FITS
images in $F606W$, $F850LP$, $F125W$, and $F160W$, we can report the
pixel-by-pixel {\it background} values and typical {\it background} values
aggregated for objects. These measurements can be compared directly with sources
in the \citet{momcheva2016a} catalog to characterize how \morpheus{} detects
objects and the corresponding completeness relative to 3D-HST.

\begin{figure}
\centering
\includegraphics[width=\columnwidth]{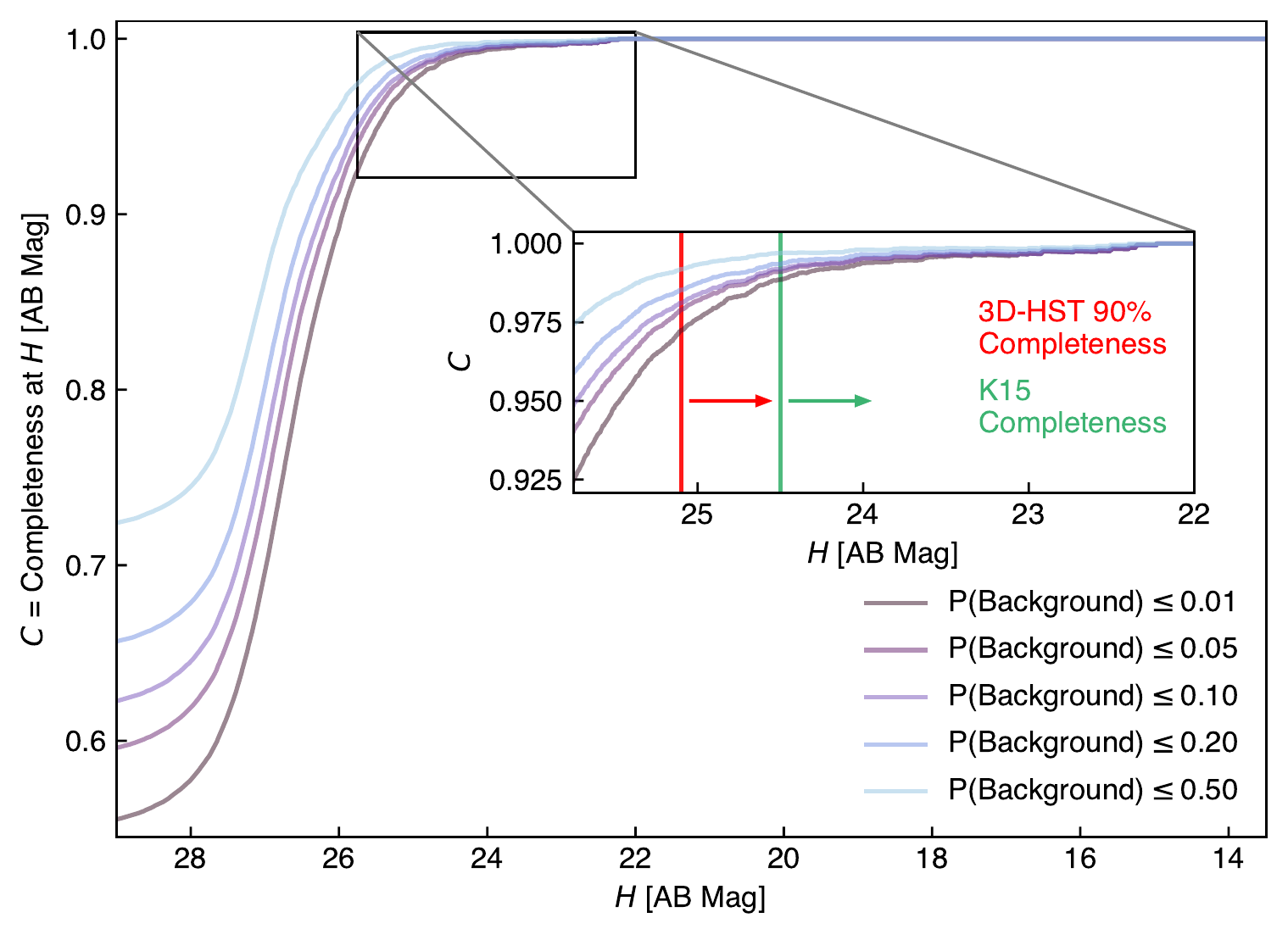}
\caption{Completeness of \morpheus{} in source detection relative to 3D-HST
         \citep{skelton2014a,momcheva2016a} in GOODS South. Shown
         is the fraction of 3D-HST sources in GOODS
         South detected by \morpheus{} brighter than some $H$-band source
         magnitude, for different {\it background} thresholds defining a
         detection (purple lines). The inset shows the \morpheus{} completeness
         for the brightest objects where 3D-HST (red line and arrow) and K15
         (green line and arrow) are both highly complete. The completeness of
         \morpheus{} relative to 3D-HST is ${>}90\%$ where 3D-HST is highly
         complete. The completeness of \morpheus{} declines rapidly at faint
         magnitudes ($H\gtrsim26.5$), but some objects are detected to
         $H\sim29$, about 100$\times$ fainter than objects in the training set.}
\label{fig:bkg_pixel_line}
\end{figure}

In a first test, we can locate the \citet{momcheva2016a} catalog objects based
on their reported coordinates in the \morpheus{} {\it background} image, and
then record the {\it background} pixel values at those locations. Figure
\ref{fig:bkg_pixel_hist} shows the two-dimensional histogram of
\morpheus{} {\it background} value and 3D-HST source $H$-band AB magnitude,
along with the marginal distributions of both quantities. The figure also
indicates the reported K15 sample and 3D-HST 90\% completeness flux levels. The
results demonstrate that for the majority of 3D-HST sources and for the vast
majority of bright 3D-HST sources with $H<25$, the local \morpheus{} {\it
background}${=}0$. The low {\it background} values computed by \morpheus{}
extend to extremely faint magnitudes (e.g., $H\approx29$), indicating that for
some faint sources, \morpheus{} reports {\it background}${=}0$ and that
{\it background} is not a simple function of the local $SNR$ of an object. For
many objects with fluxes below the 3D-HST completeness, the \morpheus{} {\it
background} value does increase with decreasing flux, and there is a
rapid transition between detected sources at $H{\approx}26.5$ to undetected
sources at $H{\le}27.5$.

Owing to this transition in {\it background} with decreasing flux, the
completeness of \morpheus{} relative to 3D-HST will depend on a threshold in
{\it background} used to define a detection. Figure \ref{fig:bkg_pixel_line}
shows the completeness of \morpheus{} in recovering 3D-HST objects as a function
of $H$-band source flux for different {\it background} levels
defining a \morpheus{} detection. The completeness flux limits for K15 and
3D-HST are indicated for reference. For magnitudes $H<25$AB, where 3D-HST and
K15 are complete, \morpheus{} is highly complete and recovers more than $99\%$
of all 3D-HST sources. The \morpheus{} completeness declines rapidly at fluxes
$H>26.5$AB, where \morpheus{} is 90\% relative to 3D-HST for {\it background}
thresholds of $P{\le}0.5$. Perhaps remarkably, for all {\it background}
thresholds $P{\le}0.01-0.5$ \morpheus{} detects some objects as faint as
$H{\approx}29$, about $100\times$ fainter in flux than the training set objects.

\begin{figure}
    \centering
    \includegraphics[width=\linewidth]{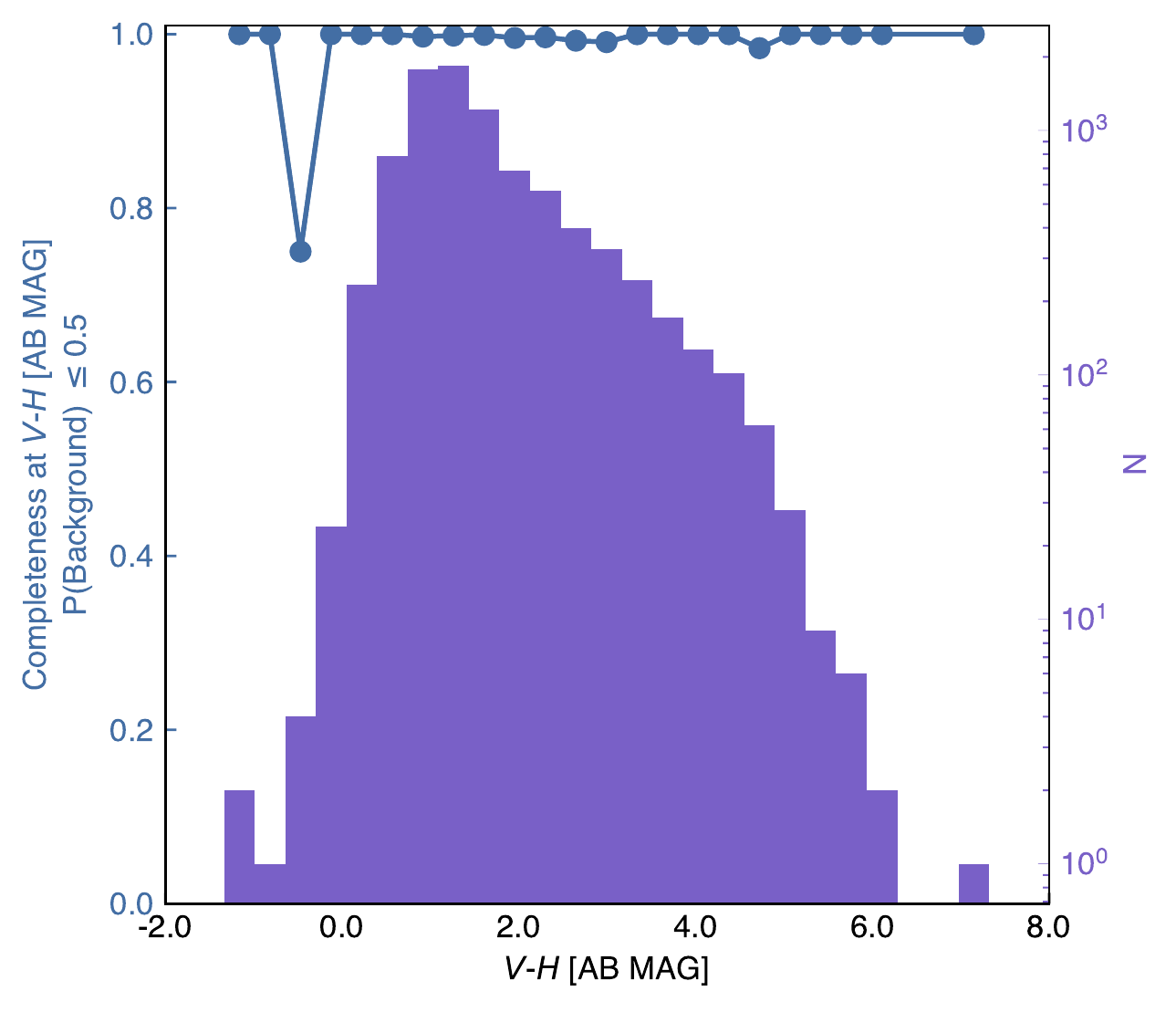}
    \caption{Source detection completeness as a function of color for
             sources with an $H$-band (F160W) AB magnitude of $H<$24.5. Sources
             that had a $V$ band flux less than the $V$ band
             error, had their flux replaced with three times the error value to
             limit unrealistically large $V$-$H$ values. \morpheus{} does not show
             bias in the detection of objects with respect to color. There is a
             dip in completeness at $V-H\sim0.2$, where the completeness is
             $\sim$75\%. However, this bin only has four sources, indicating
             \morpheus{} only missed one source at this color.}
    \label{fig:mag_vs_color}
\end{figure}

We further examined the detection of 3D-HST sources as a function of color
($V$-$H$) to evaluate bias that may have been inherited as result of the training
dataset. In our tests, we found that Morpheus is not biased with respect to
color for those sources which are brighter than the K15 magnitude limit
(Figure \ref{fig:mag_vs_color}). When considering all sources
within the 3D-HST catalog, Morpheus detects sources well, with a slight bias for
bluer sources, but performs less well for very red ($(V-H) \geq 9$) and
($(V-H) < 0$) sources. However, it should be noted that there are very few
such sources in the training set, and with a more extensive training sample,
Morpheus could be more complete.

\subsection{Morphological Classification vs. Source Magnitude}
\label{section:morph_vs_mag}

The tests of \morpheus{} on simulated Sersic objects of different effective
radii and the completeness study suggest that the ability of \morpheus{} to
provide informative morphological information about astronomical sources will
depend on the size and signal-to-noise of the object. While these are intuitive
limitations on any morphological classification method, the distribution of
morphological classifications with source flux determined by \morpheus{} should
be quantified.

\begin{figure}
\centering
\includegraphics[width=\linewidth]{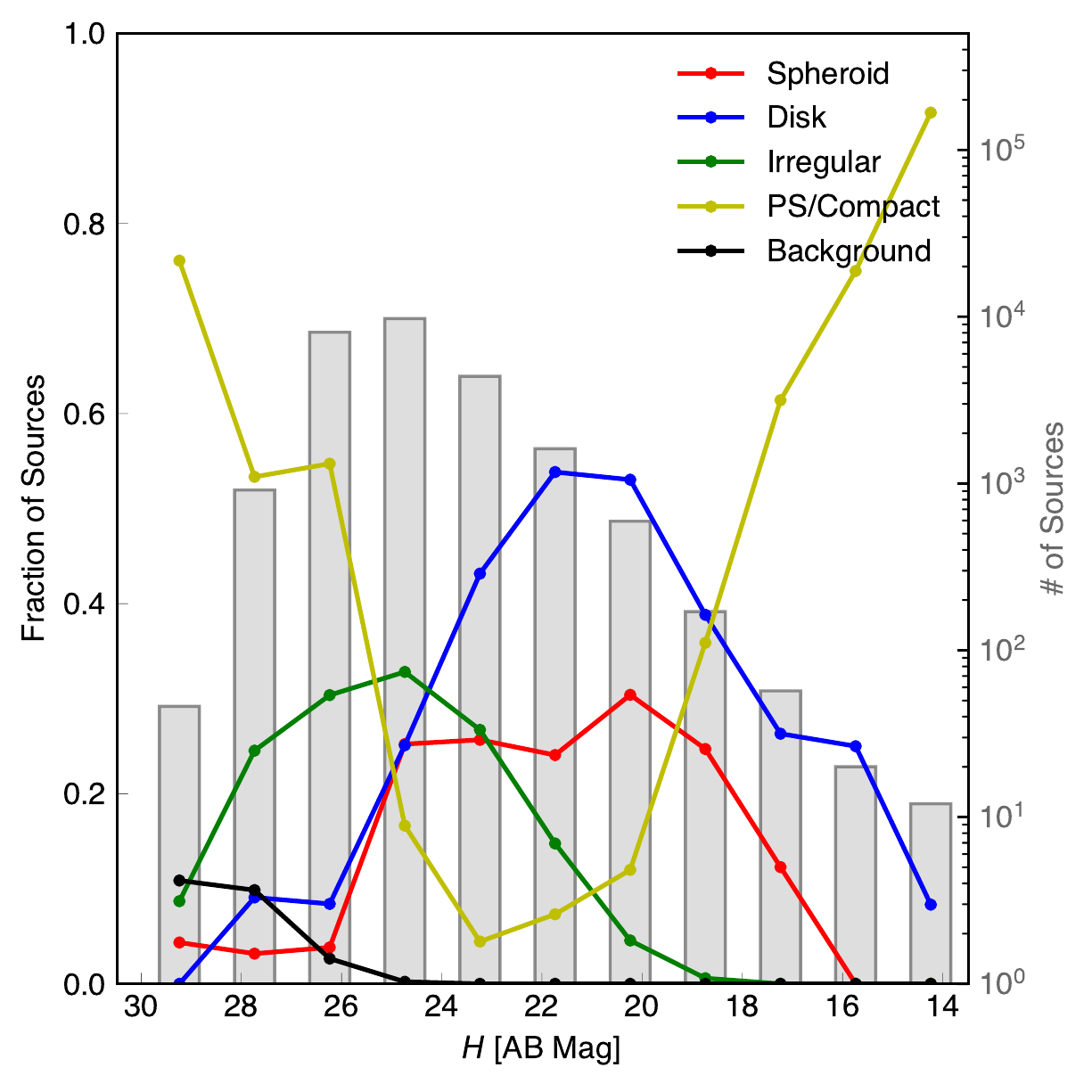}
\caption{Morphological classification as a function of object flux in
         GOODS South. Shown are the fraction of 3D-HST objects (see left axis)
         with \morpheus{} dominant, flux-weighted classifications of {\it
         spheroid} (red line), {\it disk} (blue line), {\it irregular} (green
         line), and {\it point source/compact} (yellow line), each as a function
         of their $H$-band ($F160W$) AB magnitude. The brightest objects in the
         image are stars that are classified as {\it point source/compact}. The
         faintest objects in the image are compact faint galaxies classified as
         {\it point source/compact} or irregular. At intermediate fluxes, the
         objects are primarily classified as {\it disk} and {\it spheroid}. Also
         shown as a gray histogram (see right axis) is the
         number of 3D-HST objects detected and classified by \morpheus{} with
         source magnitude.}
\label{fig:morph_vs_mag}
\end{figure}

Figure \ref{fig:morph_vs_mag} shows the fraction of 3D-HST objects detected and
classified by \morpheus{} as {\it spheroid}, {\it disk}, {\it irregular}, and
{\it point source/compact} as a function of their $H$-band magnitude. Most of
the brightest objects in the image are nearby stars, classified as {\it point
source/compact}. At intermediate magnitudes, \morpheus{} classifies the objects
as primarily a mix of {\it disk} (${\sim}50\%$) and {\it spheroid}
(${\sim}30\%$), with contributions from {\it irregular} (${\sim}10-30\%$) and
{\it point source/compact} (${\sim}5-15\%$). For fainter objects, below the
completeness limit of the K15 training sample, \morpheus{} increasingly
classifies objects as {\it irregular} and {\it point source/compact}. This
behavior is in part physical, in that many low mass galaxies are irregular and
distant galaxies are physically compact. In part, it reflects a natural
bias in how the morphologies are defined during training. In K15, the class
{\it point/source compact} can describe bright stars and compact unresolved
sources (see Section 3.1 of K15). However, the trend also reflects how \morpheus{}
becomes less effective at distinguishing morphologies in small, faint objects
and returns either {\it point source/compact} and {\it irregular} for low $SNR$
and compact sources. While training \morpheus{} on fainter objects with
well-defined morphologies could enhance the ability of \morpheus{} to
distinguish the features of faint sources, the results of this test make sense
in the context of the completeness limit of the K15 training sample used.

\subsection{False Positives in GOODS South}
\label{sec:deblend_fp}

The segmentation and deblending of real astronomical datasets
are challenging tasks.
An important test of the efficacy of the \morpheus{} segmentation and deblending algorithms
is to examine false positives generated when \morpheus{} is applied
to a real image. To quantify the
propensity for \morpheus{} to
generate false positives, the segmentation and deblending algorithms were run on
the HLF GOODS South image without the specified marker set parameter
\textit{p} (See Algorithms \ref{alg:segmentation} and \ref{alg:deblending}). For the purposes of this test, a
false positive is then defined as a set of pixels classified by the \morpheus{}
segmentation and deblending algorithms as a source but that
does not contain a source
from the 3DHST and CANDELS \citep{guo2013a} catalogs.
Additionally, since the edges of the GOODS South
classified image are a frayed mix of pixels, to
minimize the effects of data artifacts sources
less than 20 pixels from the edge of the classified area were excluded from the
analysis. Further, we conservatively use the ``default''
\morpheus{} algorithms that identify sources with
$background=0$, i.e., when \morpheus{} indicates a source
detection with high confidence.
With these choices, the sample used for the
false positive analysis was a total of 19,481 sources.

Among the objects classified by the segmentation and deblending
algorithms, 123 sources were not present in the CANDELS or 3D-HST catalogs. Upon
visual inspection of these sources, each can be categorized as an image
artifact, a poor deblend, a missed source, or an actual false positive. We list
the number of sources in each category in Table \ref{tbl:deblend_fp}.

Sources in the \textit{image artifiact category are false positives
caused by image artifacts. The \textit{poor deblend} category represents false
positives caused by the \morpheus{} deblending algorithm, where single sources
in the CANDELS or 3D-HST catalogs were shredded into multiple \morpheus{}
sources. The \textit{missed sources} are \morpheus{} sources that upon visual
inspection correspond to real objects missed by the 3D-HST or CANDELS catalogs.
Sources in the \textit{actual false positive} category are false positives not
associated with any image artifact or real source after visual inspection.}

As Table \ref{tbl:deblend_fp} shows, \morpheus{}
can identify real sources that other methods used
to generate catalogs can miss, although
the algorithms used by \morpheus{} can very rarely
cause actual false positives (at roughly the $0.1\%$ rate).
Given the delicate nature of deblending, this analysis
suggests that the \morpheus{} deblending algorithm could be integrated with
other methods to generate more robust segmentation maps.

\begin{table}
    \centering
    \begin{tabular}{c c p{1.4cm} p{1.3cm}}
        \multicolumn{4}{c}{False Positives in GOODS South} \\
        \toprule
        Category & Count & \% of False Positives & \% of All Sources\\
        \midrule
        Image Artifact          & $27$           & $21.95\%$         & $0.139\%$       \\
        Poor Deblend            & $31$           & $25.20\%$         & $0.159\%$        \\
        Missed Source           & $47$           & $38.21\%$         & $0.241\%$        \\
        Actual False Positive   & $18$           & $14.64\%$         & $0.092\%$        \\
        \textbf{Total} & $\textbf{123}$    & $\textbf{100\%}$ & $\textbf{0.631\%}$ \\
        \bottomrule
    \end{tabular}
    \caption{Summary of sources identified by \morpheus{ in GOODS-S that
    were absent in the CANDELS or 3D-HST catalogs. Of the $19,481$ sources identified
    by \morpheus{} in a subregion of GOODS-S, $123$ sources did not have CANDELS
    or 3D-HST counterparts. Upon visual inspection, these objects could be
    categorized as either {\it image artifacts}, {\it poor deblends} where \morpheus{}
    had shredded sources, {\it missed sources} corresponding to real objects missed
    by CANDELS and 3D-HST, or {\it actual false positives} incorrectly identified
    as \morpheus{} as real sources. The false positive rate for the \morpheus{}
    algorithm is only roughly $0.09\%$, defined relative to the CANDELS and 3D-HST
    catalogs. See Section
    \ref{sec:deblend_fp} for more discussion.}}
    \label{tbl:deblend_fp}
\end{table}

\section{Value Added Catalog for 3D-HST Sources with \morpheus{} Morphologies}
\label{section:vac}

The \morpheus{} framework provides a system for performing the pixel-level
analysis of astronomical images and has been engineered to allow
for the processing of large-format scientific FITS data. As described in Section
\ref{section:hfl_morphology}, \morpheus{} was applied to the Hubble Legacy
Fields \citep[HLF;][]{illingworth2016a} reduction of HST imaging in GOODS
South\footnote{Some bright pixels in the released HLF images are censored with
zeros. For the purpose of computing the segmentation maps only, we replaced
these censored pixels with nearby flux values.} and a suite of morphological
classification images produced. Using the \morpheus{} {\it background} in GOODS
South, the detection efficiency of \morpheus{} relative to the
\citet{momcheva2016a} 3D-HST catalog was computed (see Section
\ref{section:detection}) and a high level of completeness demonstrated for
objects comparably bright to the \citet{kartaltepe2015a} galaxy sample used to
train the model. By segmenting and deblending the HLF images, \morpheus{} can
then compute flux-weighted morphologies for all the 3D-HST sources.

Table \ref{table:vac} provides the \morpheus{} morphological classifications for
$50,506$ sources from the 3D-HST catalog of \citet{momcheva2016a}. This
value-added catalog lists the 3D-HST ID, the source right ascension and
declination, the $F160W$-band AB magnitude (or $-1$ for negative flux objects),
and properties for the sources computed by \morpheus{}. The value-added
properties include a flag denoting whether and how \morpheus{} detected the
object, the area in pixels assigned to each source, and the {\it spheroid}, {\it
disk}, {\it irregular}, {\it point source/compact}, and {\it background}
flux-weighted classifications determined by \morpheus{}. The size of the
segmentation regions assigned to each 3D-HST object following Algorithms
\ref{alg:segmentation} and \ref{alg:deblending} is reported for all objects. If
the segmentation region assigned to an object was smaller than a circle with a
0.36'' radius, or the object was undetected, instead, use a 0.36'' radius
aperture (about 109 pixels) to measure flux-weighted quantities. Only objects
with joint coverage in the HLF $V$, $z$, $J$, and $H$ FITS images are classified
and receive an assigned pixel area. The full results for the \morpheus{}
morphological classifications of 3D-HST objects are released as a
machine-readable table accompanying this paper. Appendix
\ref{appendix:data_release} describes the \morpheus{} Data Release associated
with this paper, including FITS images of the classification images, the
value-added catalog, and segmentation maps generated by \morpheus{} for
the 3D-HST sources used to compute flux-weighted morphologies. Additionally, we
release an interactive online map at
\href{https://morpheus-project.github.io/morpheus/}{https://morpheus-project.github.io/morpheus/}, which provides an interface to examine the data and overlay the 3D-HST
catalog on the \morpheus{} classification images, morphological color images,
and segmentation maps.

\begin{table*}
\centering
\caption{\morpheus{} + 3D-HST Value Added Catalog for GOODS South}
\label{table:vac}
\renewcommand{\arraystretch}{0.85}
\begin{tabular}{@{}cccccccccccc@{}}
\hline
\hline
ID & RA    & Dec   & $H160$   & Detection & Area     & {\it spheroid} & {\it disk}  & {\it irregular} & {\it ps/compact} & {\it background} & min(\it background)\\
   & [deg] & [deg] & [AB mag] & Flag      & [pixels] & & & & & & \\
\hline
     1 & 53.093012 & -27.954546 &  19.54 & 1 &   4408 & 0.092 & 0.797 & 0.106 & 0.003 & 0.003 & 0.000 \\
     2 & 53.089613 & -27.959742 &  25.49 & 0 &     -- & -- & -- & -- & --& -- & -- \\
     3 & 53.102913 & -27.959642 &  25.37 & 1 &    121 & 0.013 & 0.033 & 0.894 & 0.025 & 0.034 & 0.000 \\
     4 & 53.101709 & -27.958481 &  21.41 & 1 &    725 & 0.001 & 0.874 & 0.120 & 0.004 & 0.001 & 0.000 \\
     5 & 53.102277 & -27.958683 &  24.62 & 1 &    144 & 0.098 & 0.003 & 0.020 & 0.746 & 0.133 & 0.000 \\
     6 & 53.090577 & -27.958515 &  25.07 & 2 &    109 & 0.000 & 0.831 & 0.034 & 0.000 & 0.134 & 0.001 \\
     7 & 53.099964 & -27.958278 &  23.73 & 1 &    266 & 0.000 & 0.712 & 0.284 & 0.000 & 0.003 & 0.000 \\
     8 & 53.096144 & -27.957583 &  21.41 & 1 &   1322 & 0.001 & 0.752 & 0.238 & 0.003 & 0.006 & 0.000 \\
     9 & 53.091572 & -27.958367 &  25.90 & 2 &    109 & 0.000 & 0.044 & 0.083 & 0.081 & 0.792 & 0.431 \\
    10 & 53.091852 & -27.958181 &  25.88 & 2 &    109 & 0.000 & 0.000 & 0.038 & 0.186 & 0.776 & 0.570 \\
\hline
\end{tabular}
\\
\vspace{-1mm}
\begin{flushleft}
Column 1 provides the 3D-HST source ID. Columns 2 and 3 list the right ascension
and declination in degrees. Column 4 shows the $F160W$ AB magnitude of the
3D-HST source, with $-1$ indicating a negative flux reported by 3D-HST. Column 5
lists the detection flag, with 0 indicating the object was not within the region
of GOODS South classified by \morpheus{}, 1 indicating a detection with {\it
background}${=}0$ at the source location, 2 indicating a possible detection with
$0{<}${\it background}${<}1$ at the source location, and 3 indicating a
non-detection with {\it background}${=}1$ at the source location. Column 6
reports the area in pixels for the object determined by the \morpheus{}
segmentation algorithm. For non-detections and objects with very small
segmentation regions, we instead use a 0.36'' radius circle (about 109 pixels)
for their segmentation region. Columns 7-11 list the flux-weighted \morpheus{}
morphological classifications of the objects within their assigned area. These
columns are normalized such that the classifications sum to one for objects
where the detection flag $!=2$. Column 12 reports the minimum {\it background}
value within the segmentation region. Table \ref{table:vac} is published in its
entirety in the machine-readable format along with comparable tables for
the other CANDELS fields. A portion is shown here for guidance regarding its
form and content.
\end{flushleft}
\end{table*}

\section{Discussion}
\label{section:discussion}

The analysis of astronomical imagery necessarily involves pixel-level
information to be used to characterize sources. The semantic segmentation
approach of \morpheus{} delivers pixel-level separation between sources and the
background sky, and provides an automated classification of the source pixels.
In this paper, we trained \morpheus{} with the visual morphological
classifications from \citet{kartaltepe2015a}. We then characterized the
performance of \morpheus{} in reproducing the object-level classifications of
K15 after aggregating the pixel information through flux-weighted averages of
pixels in \morpheus{}-derived segmentation maps, and in detecting objects via
completeness measured relative to the 3D-HST catalog \citep{momcheva2016a}. The
potential applications of \morpheus{} extend well beyond object-level
morphological classification. Below, we discuss some applications of the
pixel-level information to understanding the complexities of galaxy morphology
and future applications of the semantic segmentation approach of \morpheus{} in
areas besides morphological classification. We also comment on some features of
\morpheus{} specific to its application on astronomical images.

\subsection{Pixel-Level Morphology}
\label{section:pixel_morphology}

The complex morphologies of astronomical objects have been described by both
visual classification schemes and quantitative morphological measures for many
years. Both \citet{hubble1926a} and \citet{devaucouleurs1959a} sought to
subdivide broad morphological classifications into more descriptive categories.
Quantitative morphological decompositions of galaxies \citep[e.g.,][]{peng2010a}
also characterize the relative strength of bulge and disk components in
galaxies, and quantitative morphological classifications often measure the
degree of object asymmetry
\citep[e.g.,][]{abraham1994a,conselice2000a,lotz2004a}.

The object-level classifications computed by \morpheus{} provide a mixture of
the pixel-level morphologies from the \morpheus{} classification images. The
classification distributions reported in the \morpheus{} value-added catalog in
GOODS South provide many examples of flux-weighted measures of morphological
type. However, more information is available in the pixel-level classifications
than flux-weighted summaries provide.

\begin{figure*}
\centering
\includegraphics[width=\linewidth]{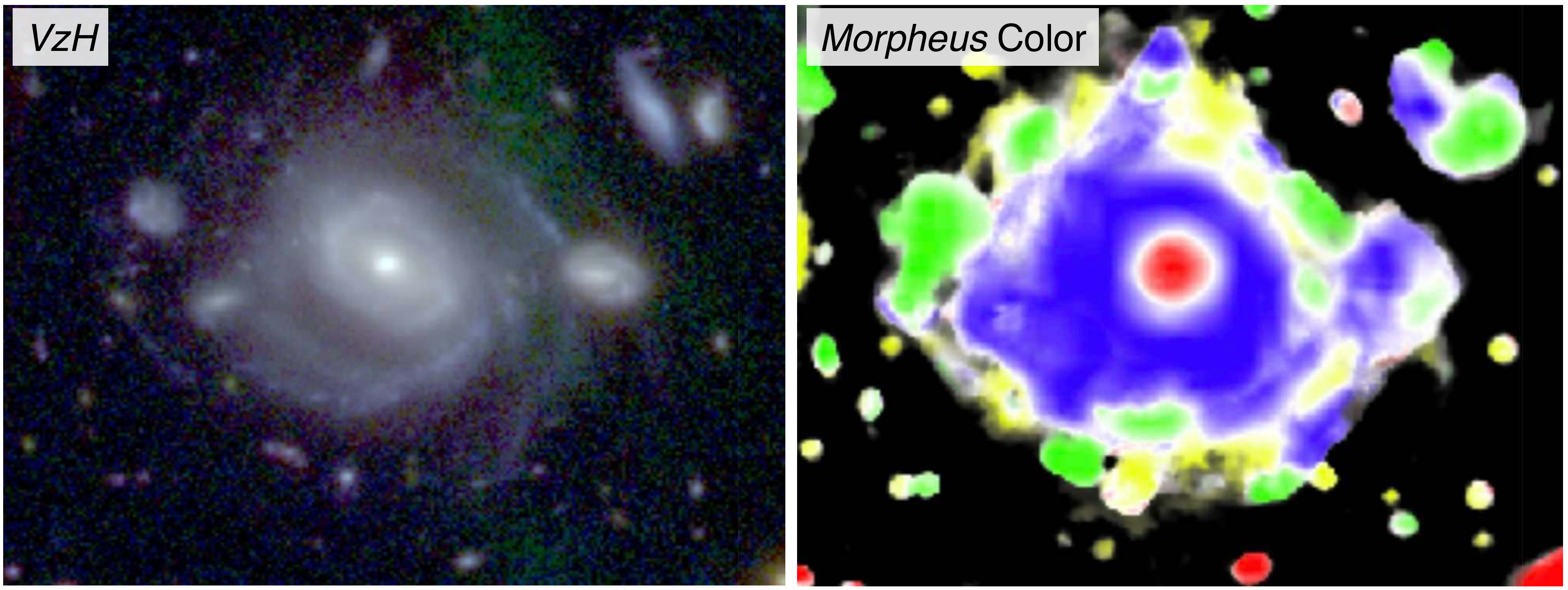}
\caption{Example automated morphological decomposition by \morpheus{}. The left
         panel shows the $VzH$ multicolor image of a galaxy in GOODS South from
         the Hubble Legacy Fields. The disk galaxy, 3D-HST ID 46386, has a
         prominent central bulge. The right panel shows the \morpheus{}
         classification color image, with pixels displaying {\it spheroid}, {\it
         disk}, {\it irregular}, or {\it point source/compact} dominant
         morphologies shown in red, blue, green, and yellow, respectively. The
         figure demonstrates that \morpheus{} correctly classifies the spheroid
         and disk structural components of the galaxy correctly, even though the
         training process for \morpheus{} does not involve spatially-varying
         morphologies for galaxy interiors. We note that there is a large-scale
         image artifact in $F850LP$ that appears as green in the left image, but
         does not strongly affect the \morpheus{} pixel-level classifications.}
\label{fig:bulge_disk_decomposition}
\end{figure*}

Figure \ref{fig:bulge_disk_decomposition} shows an example object for which the
\morpheus{} pixel-level classifications provide direct information about its
complex morphology. The figure shows a disk galaxy with a prominent central
bulge. The pixel-level classifications capture both the central bulge and the
extended disk, with the pixels in each structural component receiving dominant
bulge or disk classifications from \morpheus{}. Note that \morpheus{} was not
trained to perform this automated bulge--disk decomposition, as in the training
process, all pixels in a given object are assigned the same distribution
of classifications as determined by the K15 visual classifiers. As the use
of pixel-level morphological classifications becomes wide-spread, the
development of standard datasets that include labels at the pixel-level will be
needed to evaluate the efficacy of classifiers. Simulations of galaxy formation
may be useful for generating such training datasets
\citep[e.g.,][]{huertas-company2019a}. We leave a more thorough analysis of
automated morphological decompositions with \morpheus{} to future work.


\begin{figure*}
\centering
\includegraphics[width=\linewidth]{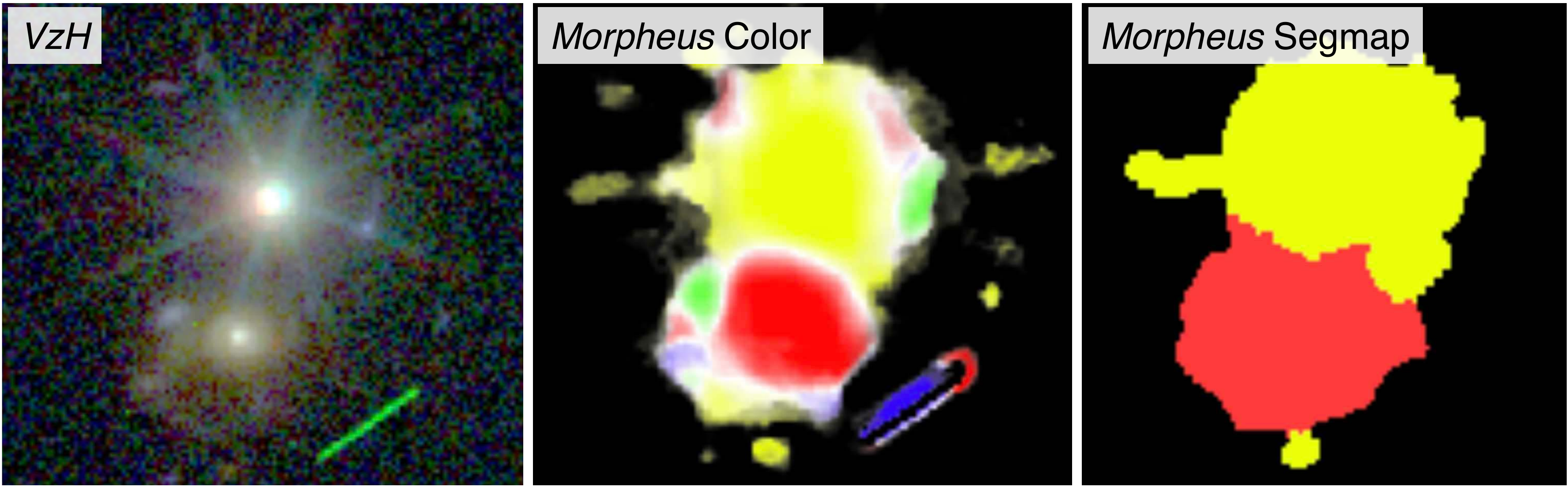}
\caption{Example of morphological deblending by \morpheus{}. The leftmost
         panel shows the $VzH$ image of a star-galaxy blend in GOODS South from
         the Hubble Legacy Fields. The star, 3D-HST ID 601, overlaps with a
         spheroidal galaxy 3D-HST ID 543. The center panel shows the \morpheus{}
         classification color image, with pixels displaying {\it spheroid}, {\it
         disk}, {\it irregular}, or {\it point source/compact} dominant
         morphologies shown in red, blue, green, and yellow, respectively. The
         pixel regions dominated by the star or spheroid are correctly
         classified by \morpheus{}. The right panel shows the resulting
         \morpheus{} segmentation map, illustrating that the dominant object
         classification in each segmentation region is also correct. The
         pixel-level classifications could be used to refine the segmentation to
         more precisely include only pixels that contained a single dominant
         class. The green feature in the left panel is an image artifact in
         $F850LP$.}
\label{fig:star_galaxy_separation}
\end{figure*}

\subsection{Morphological Deblending}
\label{section:morphological_deblending}

The ability of \morpheus{} to provide pixel-level morphological classifications
has applications beyond the bulk categorization of objects. One potential
additional application is the morphological deblending of overlapping objects,
where the pixel-level classifications are used to augment the deblending
process. Figure \ref{fig:star_galaxy_separation} shows an example of two blended
objects, 3D-HST IDs 543 and 601, where the \morpheus{} pixel-level
classifications could be used to perform or augment star-galaxy separation. As
the figure makes clear, when \morpheus{} correctly assigns dominant
classifications to pixels, there exists an interface region between regions with
distinctive morphologies (in this case, {\it spheroid} and {\it point
source/compact}) that could serve as an interface between segmented regions in
the image. The deblending algorithm used in this work could include
other forms of machine learning
\citep[e.g.,][]{masters2015a,hemmati2019a}
information in the deblending process. If \morpheus{} was trained on information
other than morphology, such as photometric redshift, those pixel-level
classifications could be used in the deblending process as well. We plan to
explore this idea in future applications of \morpheus{}.

\subsection{Classifications Beyond Morphology}
\label{section:other_classifications}

The semantic segmentation approach of \morpheus{} allows for complex features of
astronomical objects to be learned from the data, as long as those features can
be spatially localized by other means. In this paper, we used the segmentation
maps of K15 to separate source pixels from the sky, and then assigned
pixels within the segmentation maps the morphological classification determined
by K15 on an object-by-object basis. In principle, this approach can be extended
to identify regions of pixels that contain a wide variety of features. For
instance, \morpheus{} could be trained to identify image artifacts, spurious
cosmic rays, or other instrumental or data effects that lead to distinctive
pixel-level features in images. Of course, real features in images could also be
identified, such as the pixels containing arcs in gravitational lenses, or
perhaps low-surface brightness features in interacting systems and stellar
halos. These pixel-level applications of \morpheus{} complement machine
learning-based methods already deployed, such as those that discover and model
gravitational lenses
\citep{agnello2015a,hezaveh2017a,morningstar2018a,morningstar2019a}. Pixel-level
photometric redshift estimates could also be adopted by \morpheus{} and compared
with existing methods based on SED fitting or other forms of machine learning
\citep[e.g.,][]{masters2015a,hemmati2019a}.

\subsection{Deep Learning and Astronomical Imagery}
\label{section:deep_astronomy}

An important difference in the approach of \morpheus{}, where a purpose-built
framework was constructed from TensorFlow primitives, compared with the
adaptation and retraining of existing frameworks like Inception
\citep[e.g.,][]{szegedy2016a} is the use of astronomical FITS images as
training, test, and input data rather than preprocessed PNG or JPG files. The
incorporation of deep learning into astronomical pipelines will benefit from the
consistency of the data format. The output data of \morpheus{} are also FITS
classification images, allowing pixel-by-pixel information to be easily
referenced between the astronomical science images and the \morpheus{} model
images. As indicated in Section \ref{section:nn}, the \morpheus{} framework is
extensible and allows for any number of astronomical filter images to be used,
as opposed to a fixed red-blue-green set of layers in PNG or JPG files. The
\morpheus{} framework has been engineered to allow for the classification of
arbitrarily-sized astronomical images. The same approach also provides
\morpheus{} a measure of the dispersion of the classifications of individual
pixels, allowing the user to choose a metric for the ``best'' pixel-by-pixel
classification. The combination of these features allows for immense flexibility
in adapting the \morpheus{} framework to problems in astronomical image
classification.

\section{Summary and Conclusions}
\label{section:summary}

In this paper, we presented \morpheus{}, a deep learning framework for
the pixel-level analysis of astronomical images. The architecture of
\morpheus{} consists of our original implementation of a U-Net
\citep{ronneberger2015a} convolutional neural network. \morpheus{} applies the
semantic segmentation technique adopted from computer vision to enable
pixel-by-pixel classifications, and by separately identifying background and
source pixels \morpheus{} combines object detection and classification into a
single analysis. \morpheus{} represents a new approach to astronomical data
analysis, with wide applicability in enabling per-pixel classification of images
where suitable training datasets exist. Important results from this
paper include:

\begin{itemize}
	\item \morpheus{} provides pixel-level classifications of astronomical FITS
	images. By using user-supplied segmentation maps during training, the model
	learns to distinguish {\it background} pixels from pixels containing source
	flux. The pixels associated with astronomical objects are then classified
	according to the classification scheme of the training data set. The entire
	\morpheus{} source code has been publicly released, and a Python package
    installer for \morpheus{} provided. Further, we have a
    citable ``frozen''
	version of code available through Zenodo \citep{hausen2020}.
    \item As a salient application, we trained \morpheus{} to provide
    pixel-level classifications of galaxy morphology by using the
    \citet{kartaltepe2015a} visual morphological classifications of galaxies in
    the CANDELS dataset \citep{grogin2011a,koekemoer2011a} as our training
    sample.
	\item Applying \morpheus{} to the Hubble Legacy Fields
	\citep{illingworth2016a} v2.0 reduction of the CANDELS data in GOODS South
	and the v1.0 data \citep{grogin2011a,koekemoer2011a} for COSMOS, EGS,
	GOODS North and UDS, we generated morphological classifications for every
	pixel in the HLF mosaics.
	The resulting \morpheus{} morphological
	classification images have been publicly released.
	\item The pixel-level morphological classifications in GOODS South were then
	used to compute and publicly release a ``value-added'' catalog of
	morphologies for all objects in the public 3D-HST source catalog
	\citep{skelton2014a,momcheva2016a}.
	\item The CANDELS HLF and 3D-HST data were used to quantify the
	performance of \morpheus{}, both for morphological classification and its
	completeness in object detection. As trained, the \morpheus{} code shows
	high completeness at magnitudes $H\lesssim26.5$AB. We demonstrate that
	\morpheus{} can detect objects in astronomical images at flux levels up to
	$100\times$ fainter than the completeness limit of its training sample
	($H\sim29$AB).
	\item Tutorials for using the \morpheus{} deep learning framework have been
	created and publicly released as Jupyter notebooks.
	\item An interactive visualization of the \morpheus{} model results for
	GOODS South, including the \morpheus{} segmentation maps and pixel-level
	morphological classifications of 3D-HST sources, has been publicly released.
\end{itemize}

We expect that semantic segmentation will be increasingly used in astronomical
applications of deep learning, and \morpheus{} serves as an example framework
that leverages this technique to identify and classify objects in astronomical
images. We caution that \morpheus{ may be most effective at wavelengths
similar to the data
on which the model was trained (i.e., the $F606W$, $F850LP$, $F125W$, and $F160W$
bands).
However, \citet{dominguez_sanchez2019a} have shown recent success in applying
transfer learning on astronomical datasets with morphological labels.} With the
advent of large imaging data sets such those provided by Dark Energy Survey
\citep{des2016a} and Hyper Suprime-Cam \citep{aihara2018a,aihara2018b}, and
next-generation surveys to be conducted by Large Synoptic Survey Telescope
\citep{ivezic2019a,robertson2019a}, Euclid \citep{laureijs2011a,rhodes2017a},
and the Wide Field Infrared Survey Telescope \citep{akeson2019a}, pixel-level
analysis of massive imaging data sets with deep learning will find many
applications. While the details of the \morpheus{} neural network architecture
will likely change and possibly improve, we expect the approach of using
semantic segmentation to provide pixel-level analyses of astronomical images
with deep learning models will be broadly useful. The public release of the
\morpheus{} code, tutorials, and example data products should provide a basis
for future applications of deep learning for astronomical datasets.

\acknowledgements

B.E.R. acknowledges a Maureen and John Hendricks Visiting Professorship at the
Institute for Advanced Study, and NASA contract NNG16PJ25C.
R.D.H acknowledges the Eugene V. Cota-Robles Fellowship.
The authors acknowledge use of the lux supercomputer at UC Santa Cruz,
funded by NSF MRI grant AST 1828315. This work is based on observations taken by
the CANDELS Multi-Cycle Treasury Program with the NASA/ESA HST, which is
operated by the Association of Universities for Research in Astronomy, Inc.,
under NASA contract NAS5-26555. This work is based on observations taken by the
3D-HST Treasury Program (GO 12177 and 12328) with the NASA/ESA HST, which is
operated by the Association of Universities for Research in Astronomy, Inc.,
under NASA contract NAS5-26555.

\software{
    Python \citep{rossum1995},
    NumPy \citep{vanderwal2011a},
    astropy \citep{price2018},
    scikit-learn \citep{pedregosa2011a},
    matplotlib \citep{hunter2007},
    Docker \citep{merkel2014a},
    TensorFlow \citep{tensorflow2015}
}

\appendix

\section{Deep Learning}
\label{appendix:dl}

The \morpheus{} deep learning framework incorporates a variety of technologies
developed for machine learning applications. The following descriptions of deep
learning techniques complement the overview of \morpheus{} provided in Section
\ref{section:framework}, and are useful for understanding optional
configurations of the model.

\subsection{Artificial Neuron}
\label{section:an}

The basic unit of the \morpheus{} neural network is the artificial neuron (AN),
which transforms an input vector $\mathbf{x}$ to a single output
$AN(\mathbf{x})$. The AN is designed to mimic the activation of a neuron,
producing a nonlinear response to an input stimulus value when it exceeds a
rough threshold.

The first stage of an AN consists of a function
\begin{equation}
\label{eq:z}
    z(\mathbf{x})  = \sum_{i=1}^n w_i x_i + b
\end{equation}
\noindent
that adds the dot product of the $n$-element vector $\mathbf{x}$ with a vector
of weights $\mathbf{w}$ to a bias $b$. The values of the $\mathbf{w}$ elements
and $b$ are parameters of the model that are set during optimization. The
function $z(\mathbf{x})$ is equivalent to a  linear
transformation on input data $\mathbf{x}$.

In the second stage, a nonlinear function $a$ is applied to the output of
$z(\mathbf{x})$. We write
\begin{equation}
\label{eq:an}
AN(\mathbf{x}) \equiv a(z(\mathbf{x})),
\end{equation}
\noindent
where $a(z)$ is called the activation function. The \morpheus{} framework allows
the user to specify the activation function, including the sigmoid
\begin{equation}
    \mathrm{sigmoid}(z) = \frac{1}{1 + e^{-z}},
    \label{eq:sigmoid}
\end{equation}
\noindent
the hyperbolic tangent
\begin{equation}
    \tanh(z) = \frac{e^z - e^{-z}}{e^z + e^{-z}},
    \label{eq:tanh}
\end{equation}
\noindent
and the rectified linear unit
\begin{equation}
    \mathrm{relu}(z) = \max(0, z).
    \label{eq:relu}
\end{equation}
\noindent
These functions share a thresholding behavior, such that the function activates
a nonlinear behavior at a characteristic value of $z$, but the domain of these
functions differ. For the morphological classification problem presented in this
paper, the rectified linear unit (Equation \ref{eq:relu}) was used as the
activation function.

\subsection{Neural Networks}
\label{section:nnet}
\begin{figure}
\centering
\includegraphics[width=\linewidth]{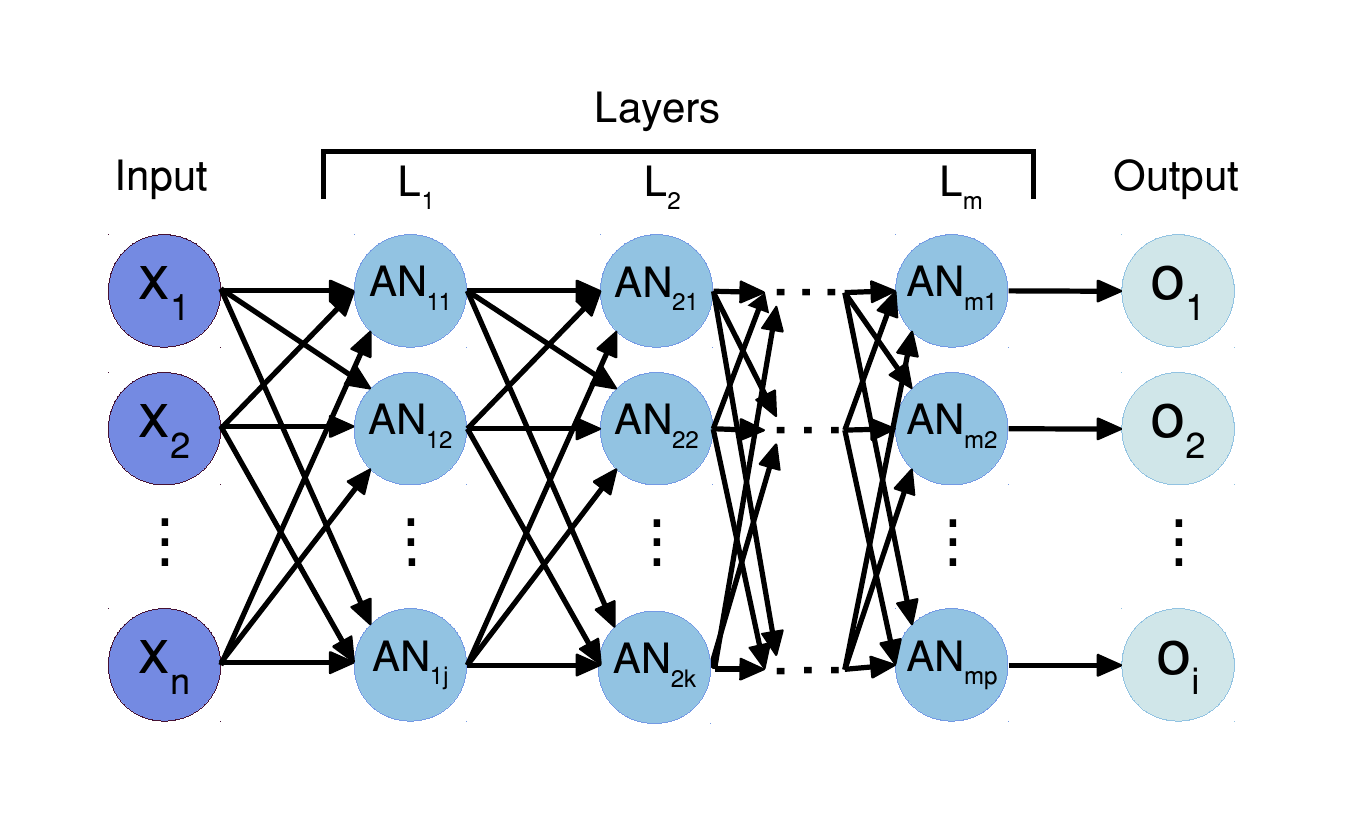}
\caption{Schematic of a simple neural network. Given an input vector
         $\mathbf{x}$, the neural network applies a series of reductions and
         nonlinear transformations through a collection of layers $\mathbf{L}$
         to produce an output $\mathbf{o}$. Each layer $L$ consists of a set of
         artificial neurons $\mathbf{AN}$ that perform a linear rescaling of
         their input data, followed by a nonlinear transformation via the
         application of an activation function (see Equation \ref{eq:an}). The
         activation function may vary across layers.}
\label{fig:nn_layers}
\end{figure}

Increasingly complex computational structures can be constructed from ANs.
Single ANs are combined into layers, which are collections of distinct ANs that
process the same input vector $\mathrm{x}$. A collection of layers forms a
neural network (NN), with the layers ordered such that the outputs from one
layer provide the inputs to the neurons in the subsequent layer. Figure
\ref{fig:nn_layers} shows a schematic of a NN and how the initial input vector
$\mathrm{x}$ is processed by multiple layers. As shown, these layers are
commonly called fully-connected since each neuron in a given layer receives the
outputs $\mathrm{z}$ from all neurons in the previous layer.

\subsection{Convolutional Neural Networks}
\label{section:convolution}
The \morpheus{} framework operates on image data with a convolutional neural
network (CNN). A CNN includes at least one layer of ANs whose \textit{z}
function uses a discrete cross-correlation (convolution) in place of the dot
product in Equation \ref{eq:z}. For a convolutional artificial neuron (CAN), we
write
\begin{equation}
\label{eq:can_z}
z(\mathbf{X}) = (\mathbf{X} * \mathbf{W}) + b\mathbf{J},
\end{equation}
\noindent
where $\mathbf{X} * \mathbf{W}$ represents the convolution of an input image
$\mathbf{X}$ and a kernel $\mathbf{W}$. The elements of the kernel $\mathbf{W}$
are parameters of the model, and $\mathbf{W}$ may differ in dimensions
from $\mathbf{X}$. In \morpheus{}, the dimensions of $\mathbf{W}$ are set to be
$3 \times 3$ throughout. The bias $b$ is a scalar as before, and $\mathbf{J}$
represents a matrix of $1$s with the same dimensions as the result of the
convolution. In \morpheus{}, the convolution is zero-padded to maintain the
dimensions of the input data.

The activation function of the neuron is computed element-wise after the
convolution and bias have been applied to the input. We write
\begin{equation}
\label{eq:can}
CAN(\mathbf{X}) \equiv a(z(\mathbf{X})).
\end{equation}
\noindent
We refer to the output from a CAN as a feature map.

\begin{figure*}
\centering
\includegraphics[width=\textwidth]{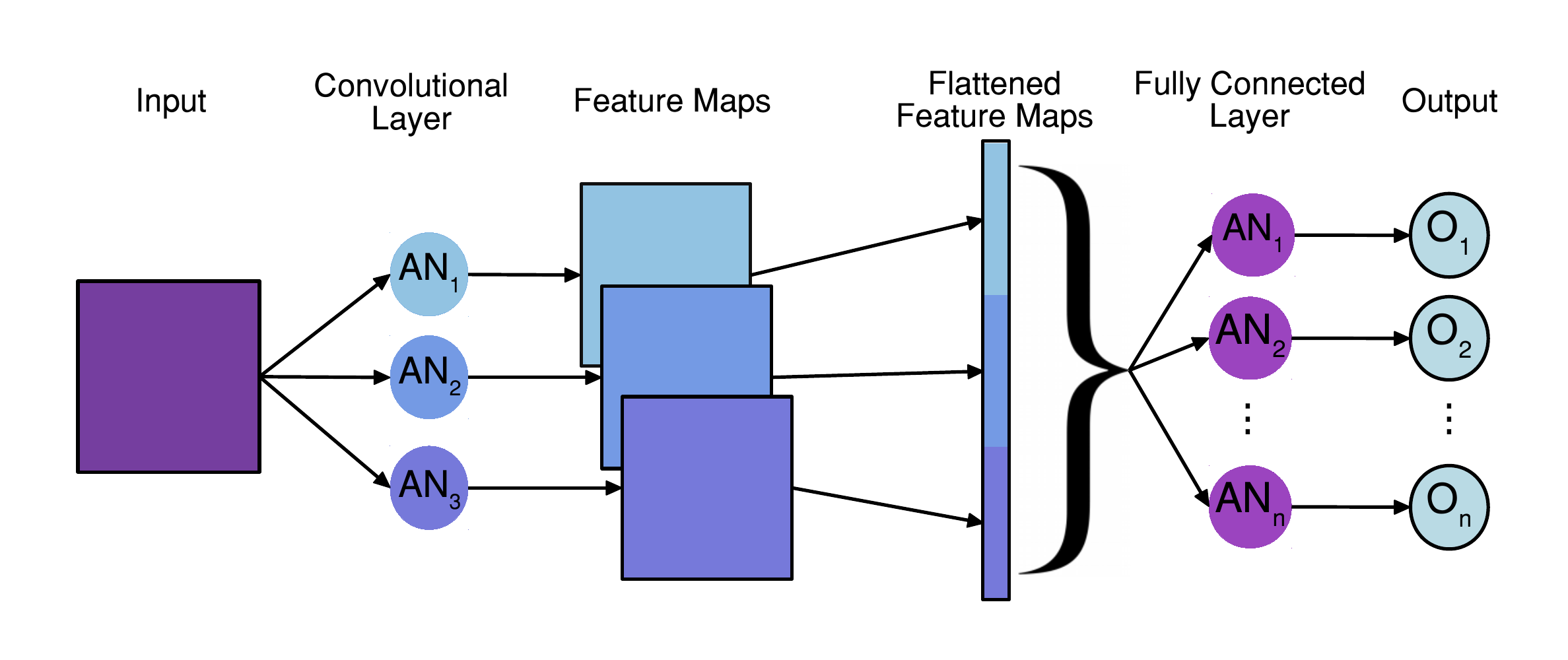}
\caption{Schematic of a convolutional neural network (CNN). Shown is a
         simplified CNN consisting of a convolutional layer feeding a fully
         connected layer. Each artificial neuron (AN) in the convolutional layer
         outputs a feature map as described by Equation \ref{eq:can}. Each
         output feature map is flattened and concatenated
         into a single vector. This vector is processed by each AN in the fully
         connected layer (see Equation \ref{eq:an}). The curly
         brace represents connections from all elements of
         the vector input.}
\label{fig:cnn}
\end{figure*}

As with fully-connected layers, convolutional layers consist of a group of CANs
that process the same input data $\mathbf{X}$. Convolutional layers can also be
arranged sequentially such that the output from one convolutional layer serves
as input to the next. \morpheus{}' neural network architecture,
being U-Net based, is comprised of CANs (see Figure \ref{fig:cnn} for a
schematic). In typical convolutional neural network topologies, CANs are
used to extract features from input images. The resulting feature maps are
eventually flattened into a single vector and processed by a fully connected
layer to produce the output classification values.

\subsection{Other Functions in Neural Networks}

The primary computational elements of \morpheus{} are a convolutional neural
network (Section \ref{section:convolution}) and a fully connected layer (Section
\ref{section:nnet}). In detail, other layers are used to reformat or summarize
the data, renormalize it, or combine data from different stages in the network.

\subsubsection{Pooling}
\label{section:pooling}

\begin{figure}
\centering
\includegraphics[width=\linewidth]{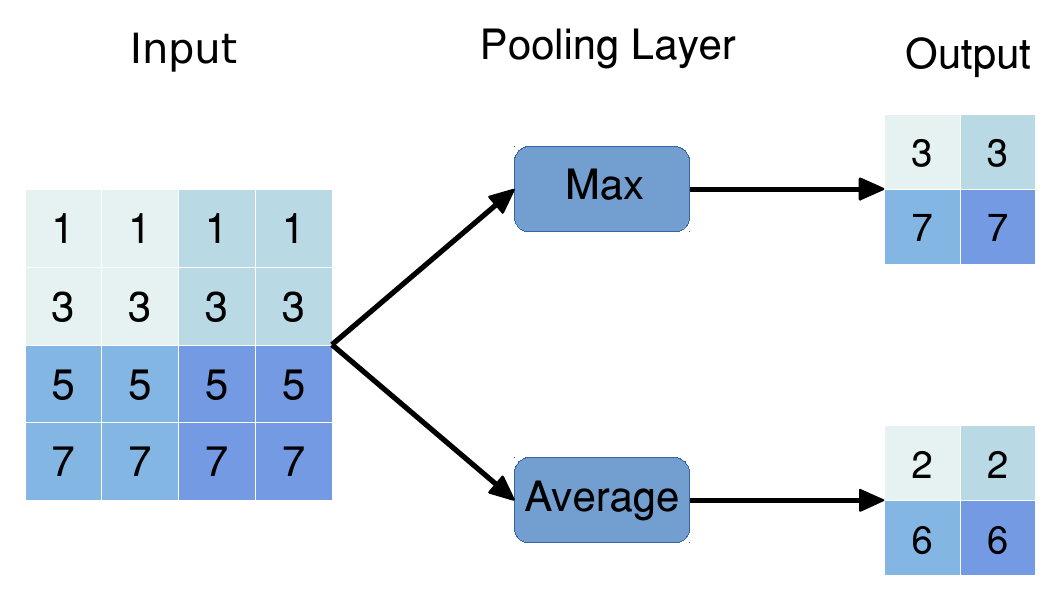}
\caption{Comparison of max and average pooling layers. Pooling layers perform
         reductions on subsets of feature maps, providing a local average or
         maximum of data elements in a window ($2\times2$ in this schematic).
         Shown are cells of an input feature map (left), color-coded
         within a window to match the corresponding regions of the output
         feature map (right). The pooling layers perform a simple reduction with
         these windows, taking either a maximum (upper branch) or average (lower
         branch).}
\label{fig:pooling}
\end{figure}

Pooling layers (Figure \ref{fig:pooling}) are composed of functions that
summarize their input data to reduce its size while preserving some information.
These layers perform a moving average (average pooling) or maximum (max pooling)
over a window of data elements, repeating these reductions as the window scans
through the input image with a stride equal to the window size. In the
morphological classification tasks described in this paper, \morpheus{} uses
$2\times2$ windows and max pooling.

\subsubsection{Up-sampling}

Up-sampling layers expand the size of feature maps by a specified factor through
an interpolation between input data elements. The up-sampling layers operate in
the image dimensions of the feature map and typically employ
bicubic and bilinear interpolation. In the morphological classification
application explored in this paper, \morpheus{} used $2\times2$ up-sampling and
bicubic interpolation.

\subsubsection{Concatenation}
\label{sec:concatenation}

Concatenation layers combine multiple feature maps by appending them without
changing their contents. For instance, the concatenation of red, green, and blue
(RGB) channels into a three-color image would append three $N\times M$ images
into an RGB image with dimensions $N\times M \times 3$. This operation is used
in \morpheus{} to combine together data from the contraction phase with the
output from bicubic interpolations in the expansion phase (see Figure
\ref{fig:architecture}).

\subsubsection{Batch Normalization}
\label{section:batch_normalization}
A common preprocessing step for neural network architectures is to normalize the
input data $x$ using, e.g., the operation
\begin{equation}
\label{eq:simple_normalization}
    \hat{x} = (x-\mu) / \sqrt{\sigma^2}
\end{equation}
\noindent
where $\hat{x}$ is the normalized data, and $\mu$ and $\sigma$ are parameters of
the model. \citet{ioffe2015a} extended this normalization step to apply to the
inputs of layers within the network, such that activations (AN) and feature maps
(CAN) are normalized over each batch. A batch consists of a subset of the
training examples used during the training process. Simple normalization
operations like Equation \ref{eq:simple_normalization} can reduce the range of
values represented in the data provided to a layer, which can inhibit learning.
\citet{ioffe2015a} addressed this issue by providing an alternative
normalization operation that introduces additional parameters to be learned
during training. The input data elements $x_i$ are first rescaled as
\begin{equation}
    \hat{x_i} = \frac{x_i - \mu_{\mathbf{x}}}{\sqrt{\sigma^2_{\mathbf{x}}+\epsilon}}.
\end{equation}
\noindent
Here, $x_i$ is a single element from the data output by a single AN or CAN over
a batch, $\mu_\mathbf{x}$ is their mean, and $\sigma^2_{\mathbf{x}}$ is their
variance. The parameter $\epsilon$ is learned during optimization. The new
normalization $BN_{\hat{x}_i}$ is then taken to be a linear transformation
\begin{equation}
\label{eq:bn}
    BN_{\hat{x}_i} = \gamma_x\hat{x_i} + \beta_x.
\end{equation}
\noindent
The parameters $\gamma_x$ and $\beta_x$ are also learned during optimization.
\citet{ioffe2015a} demonstrated that batch normalization, in the form of
Equation \ref{eq:bn}, can increase overall accuracy and decrease training
time, and we adopt this approach in the \morpheus{} framework.

\subsection{U-Net Architecture} \label{section:unet}

The \morpheus{} framework uses a U-Net architecture, first introduced by
\citet{ronneberger2015a}. The U-Net architecture was originally designed for
segmentation of medical imagery, but has enjoyed success in other fields. The
U-Net takes as input a set of images and outputs a classification image
of pixel-level probability distributions. The architecture begins with a
contraction phase composed of a series of convolutional and pooling layers,
followed by an expansion phase composed of a series of
convolutional and up-sampling layers. Each of the outputs from the down-sampling
layers is concatenated with the output of an up-sampling layer when the height
and width dimensions of the feature maps match. These concatenations help
preserve the locality of learned features in the output of the NN.

\section{code release}
\label{appendix:code_release}

The code for \morpheus{} has been release via GitHub
(\href{https://github.com/morpheus-project/morpheus}{https://github.com/morpheus-project/morpheus}).
\morpheus{} is also available as a python package installable via pip
(\href{https://pypi.org/project/morpheus-astro/}{https://pypi.org/project/morpheus-astro/})
and as Docker images available via Docker Hub
(\href{https://hub.docker.com/r/morpheusastro/morpheus}{https://hub.docker.com/r/morpheusastro/morpheus}).
\morpheus{} includes both a Python API and a command-line
interface, the documentation of which can be found online at
\href{https://morpheus-astro.readthedocs.io/en/latest/}{https://morpheus-astro.readthedocs.io/en/latest/}.

\section{code tutorial}
\label{appendix:tutorial}

An online tutorial demonstrating the \morpheus{} Python API in the form of a Jupyter
notebook can be found at\\
\href{https://github.com/morpheus-project/morpheus/blob/master/examples/example_array.ipynb}{https://github.com/morpheus-project/morpheus/blob/master/examples/example\_array.ipynb}.
The tutorial walks through the classification of an example image. Additionally,
the tutorial explores other features of \morpheus{}, including generating
segmentation maps and morphological catalogs.

\section{Data Release}
\label{appendix:data_release}

The data release associated with this work consists of multiple data
products. For each field in the CANDELS survey, we provide the following data
products: pixel-level morphological classifications, segmentation maps, and
value-added catalogs(see also Section \ref{section:vac}) for the 3D-HST
catalogs. Tables \ref{table:cosmos}-\ref{table:uds} provide the URLs for each of
the data products. Each of the fields has two types of segmentation maps, a
segmentation map informed by the 3D-HST survey and a segmentation map informed
only by the background values provided by \morpheus{} (see Algorithm
\ref{alg:segmentation}). The classifications for the EGS and UDS fields may vary
as a result of using the $F814W$ band in place of the $F850LP$ due to availability.

An interactive online visualization of the HST images, \morpheus{} classification
images, and 3D-HST sources is available at
\href{https://morpheus-project.github.io/morpheus/}{https://morpheus-project.github.io/morpheus/}.

\begin{figure}
    \centering
    \includegraphics[height=\textheight]{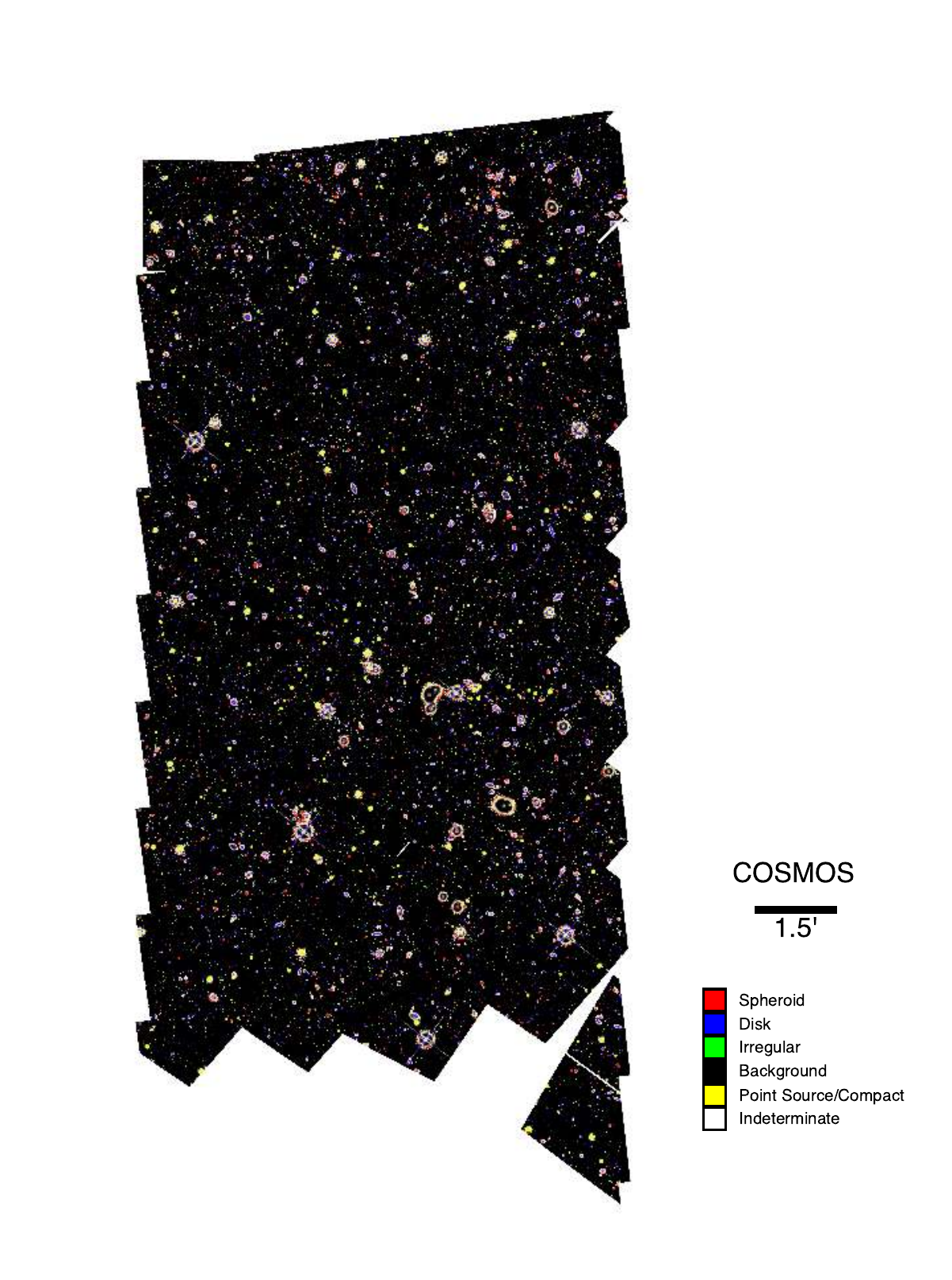}
    \caption{Color composite of the \morpheus{} morphological classifications
             for the COSMOS field from the CANDELS survey
             \citep{grogin2011a, koekemoer2011a}.}
    \label{fig:color_COSMOS}
\end{figure}

\begin{figure}
    \centering
    \includegraphics[height=\textheight]{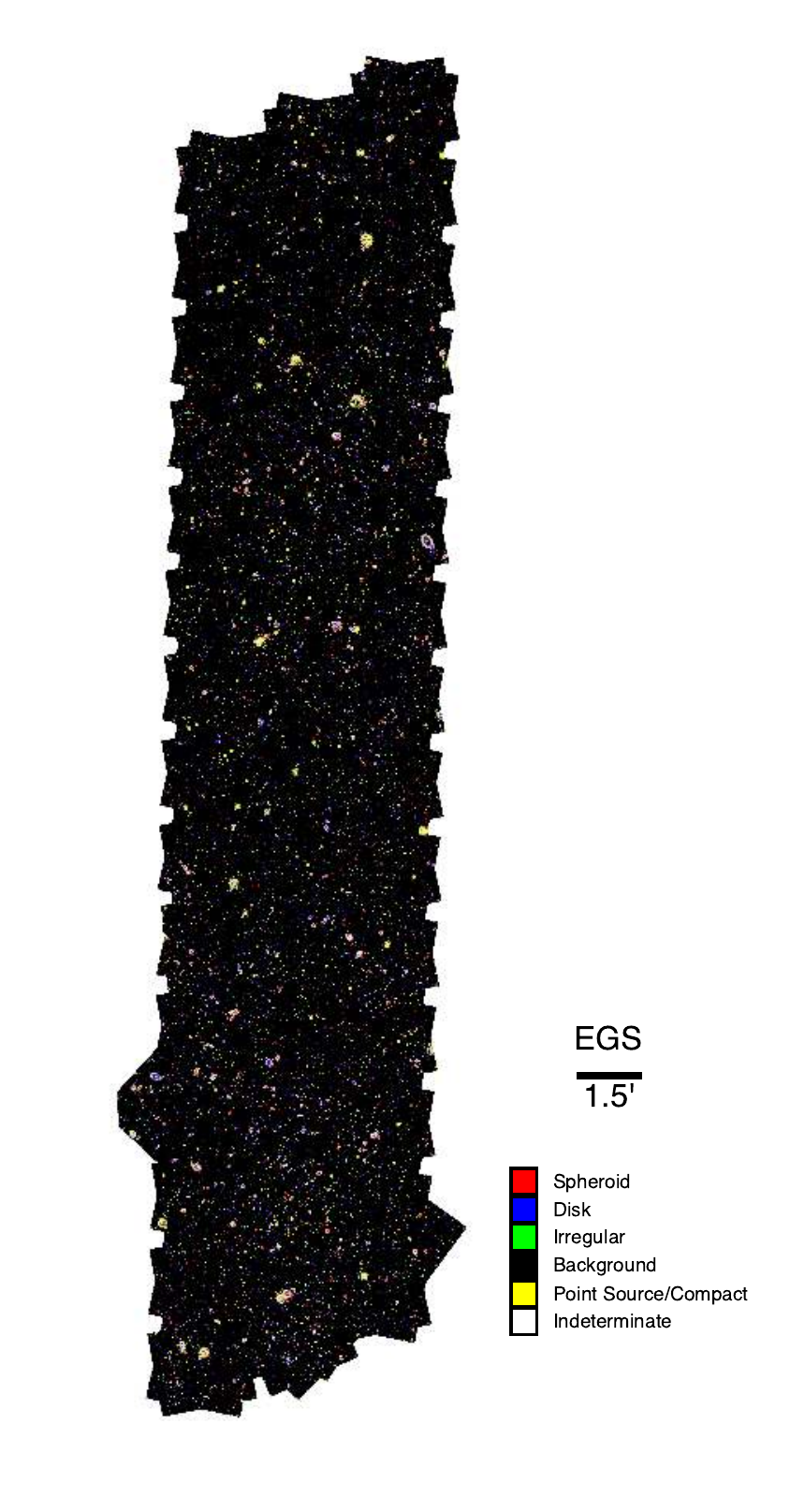}
    \caption{Color composite of the \morpheus{} morphological classifications
             for the EGS field from the CANDELS survey
             \citep{grogin2011a, koekemoer2011a}.}
    \label{fig:color_COSMOS}
\end{figure}

\begin{figure}
    \centering
    \includegraphics[width=\textwidth]{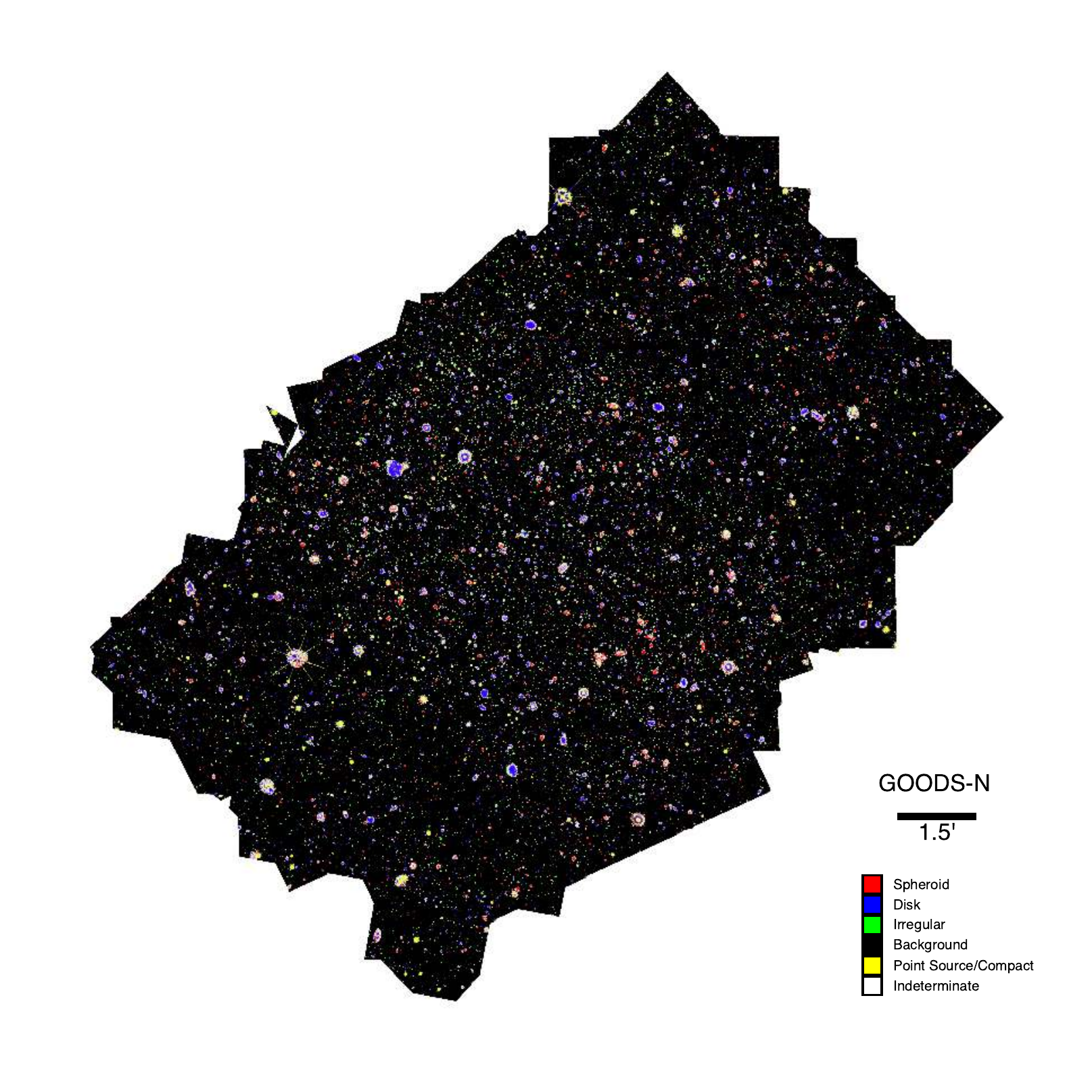}
    \caption{Color composite of the \morpheus{} morphological classifications
             for the GOODS North field from the CANDELS survey
             \citep{grogin2011a, koekemoer2011a}.}
    \label{fig:color_COSMOS}
\end{figure}

\begin{figure}
    \centering
    \includegraphics[height=\textheight]{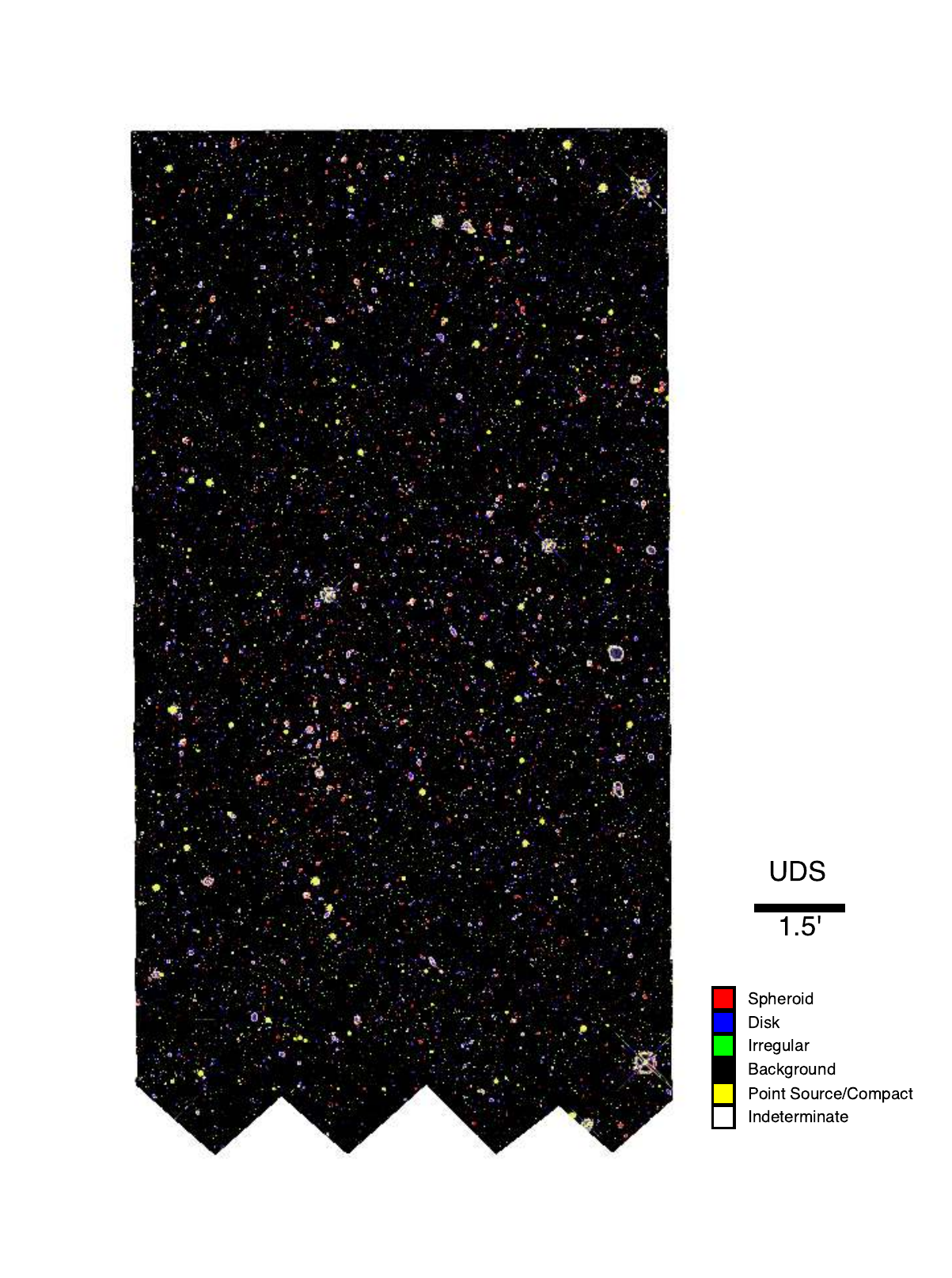}
    \caption{Color composite of the \morpheus{} morphological classifications
             for the UDS field from the CANDELS survey
             \citep{grogin2011a, koekemoer2011a}.}
    \label{fig:color_COSMOS}
\end{figure}

\begin{table}{\morpheus{} Data Products for the COSMOS Field}
    \centering
    \begin{tabular}{l l}
    \toprule
    File Name & URL \\
    \midrule
    \textbf{Pixel-level Morphological Classifications} & \\
    morpheus\_COSMOS\_spheroid.v1.0.fits    & \href{http://morpheus-project.github.io/morpheus/data-release/cosmos/spheroid.html}   {morpheus-project.github.io/morpheus/data-release/cosmos/spheroid.html} \\
    morpheus\_COSMOS\_disk.v1.0.fits        & \href{http://morpheus-project.github.io/morpheus/data-release/cosmos/disk.html}       {morpheus-project.github.io/morpheus/data-release/cosmos/disk.html} \\
    morpheus\_COSMOS\_irregular.v1.0.fits   & \href{http://morpheus-project.github.io/morpheus/data-release/cosmos/irregular.html}  {morpheus-project.github.io/morpheus/data-release/cosmos/irregular.html} \\
    morpheus\_COSMOS\_ps\_compact.v1.0.fits & \href{http://morpheus-project.github.io/morpheus/data-release/cosmos/ps-compact.html} {morpheus-project.github.io/morpheus/data-release/cosmos/ps\_compact.html} \\
    morpheus\_COSMOS\_background.v1.0.fits  & \href{http://morpheus-project.github.io/morpheus/data-release/cosmos/background.html} {morpheus-project.github.io/morpheus/data-release/cosmos/background.html} \\
    morpheus\_COSMOS\_mask.v1.0.fits        & \href{http://morpheus-project.github.io/morpheus/data-release/cosmos/mask.html}       {morpheus-project.github.io/morpheus/data-release/cosmos/mask.html} \\

    \textbf{Segmentation Maps} & \\
    morpheus\_COSMOS\_segmap.v1.0.fits  & \href{http://morpheus-project.github.io/morpheus/data-release/cosmos/segmap.html}        {morpheus-project.github.io/morpheus/data-release/cosmos/segmap.html} \\
    morpheus\_COSMOS\_3dhst-segmap.v1.0.fits & \href{http://morpheus-project.github.io/morpheus/data-release/cosmos/3dhst-segmap.html} {morpheus-project.github.io/morpheus/data-release/cosmos/3dhst-segmap.html} \\

    \textbf{3D-HST Value Added Catalog} & \\
    morpheus\_COSOMS\_3dhst\_catalog.v1.0.csv & \href{http://morpheus-project.github.io/morpheus/data-release/cosmos/value-added-catalog.html}{morpheus-project.github.io/morpheus/data-release/cosmos/value-added-catalog.html} \\
    morpheus\_COSOMS\_3dhst\_catalog.v1.0.txt & \href{http://morpheus-project.github.io/morpheus/data-release/cosmos/value-added-catalog.html}{morpheus-project.github.io/morpheus/data-release/cosmos/value-added-catalog-mrt.html} \\

    \textbf{All Files} & \\
    morpheus\_COSMOS\_all.v1.0.tar.gz  &  \href{http://morpheus-project.github.io/morpheus/data-release/cosmos/all.html}  {morpheus-project.github.io/morpheus/data-release/cosmos/all.html} \\

    \bottomrule
    \end{tabular}
    \caption{Data release files generated by Morpheus and associated URLs for
             the COSMOS CANDELS field. The data release files for each field are
             organized into three groups: \textit{pixel-level morphological
             classifications}, \textit{segmentation maps}, and \textit{3D-HST
             value-added catalogs}. The \textit{pixel-level morphological
             classification} files are named according to the following scheme
             \Verb|morpheus\_COSMOS\_[morphology].v1.0.fits|, where
             \Verb|[morphology]| can be one of the morphological classes
             (\textit{spheroid}, \textit{disk}, \textit{irregular},
             \textit{ps\_compact}, \textit{background}) or \textit{mask}, a
             binary image mask indicating which pixels in the image we're
             classified by \morpheus{}. The \textit{segmentation map} files are
             named according to the following scheme
             \Verb|morpheus\_COSMOS\_[segmap\_type].v1.0.fits| , where
             \Verb|[segmap\_type]| can be \textit{3dhst-segmap} (indicating the
             3D-HST informed segmap) or \textit{segmap} (indicating a segmap
             based only on background class/flux values). Finally, the 3D-HST
             value-added catalog files are named according to the following
             scheme \Verb|morpheus\_COSMOS\_3dhst-catalog.v1.0.[file\_type]|,
             where \Verb|[file\_type]| can be \textit{csv} for a
             comma-separated-value version of the value-added catalog and
             \textit{txt} for the machine-readable table version described in
             Table \ref{table:vac}. Additionally, a link to an archive
             containing all of the files associated with the COSMOS field is
             available in an additional section called \textit{All Files}. See
             Appendix \ref{appendix:data_release} for details.}
    \label{table:cosmos}
\end{table}

\begin{table}{\morpheus{} Data Products for the EGS Field}
    \centering
    \begin{tabular}{l l}
    \toprule
    File Name & URL \\
    \midrule
    \textbf{Pixel-level Morphological Classifications} & \\
    morpheus\_EGS\_spheroid.v1.0.fits    & \href{http://morpheus-project.github.io/morpheus/data-release/egs/spheroid.html}   {morpheus-project.github.io/morpheus/data-release/egs/spheroid.html} \\
    morpheus\_EGS\_disk.v1.0.fits        & \href{http://morpheus-project.github.io/morpheus/data-release/egs/disk.html}       {morpheus-project.github.io/morpheus/data-release/egs/disk.html} \\
    morpheus\_EGS\_irregular.v1.0.fits   & \href{http://morpheus-project.github.io/morpheus/data-release/egs/irregular.html}  {morpheus-project.github.io/morpheus/data-release/egs/irregular.html} \\
    morpheus\_EGS\_ps\_compact.v1.0.fits & \href{http://morpheus-project.github.io/morpheus/data-release/egs/ps-compact.html} {morpheus-project.github.io/morpheus/data-release/egs/ps\_compact.html} \\
    morpheus\_EGS\_background.v1.0.fits  & \href{http://morpheus-project.github.io/morpheus/data-release/egs/background.html} {morpheus-project.github.io/morpheus/data-release/egs/background.html} \\
    morpheus\_EGS\_mask.v1.0.fits        & \href{http://morpheus-project.github.io/morpheus/data-release/egs/mask.html}       {morpheus-project.github.io/morpheus/data-release/egs/mask.html} \\

    \textbf{Segmentation Maps} & \\
    morpheus\_EGS\_segmap.v1.0.fits     & \href{http://morpheus-project.github.io/morpheus/data-release/egs/segmap.html}           {morpheus-project.github.io/morpheus/data-release/egs/segmap.html} \\
    morpheus\_EGS\_3dhst-segmap.v1.0.fits & \href{http://morpheus-project.github.io/morpheus/data-release/uds/3dhst-segmap.html} {morpheus-project.github.io/morpheus/data-release/egs/3dhst-segmap.html} \\

    \textbf{3D-HST Value Added Catalogs} & \\
    morpheus\_EGS\_3dhst\_catalog.v1.0.csv & \href{http://morpheus-project.github.io/morpheus/data-release/egs/value-added-catalog.html}{morpheus-project.github.io/morpheus/data-release/egs/value-added-catalog.html} \\
    morpheus\_EGS\_3dhst\_catalog.v1.0.txt & \href{http://morpheus-project.github.io/morpheus/data-release/egs/value-added-catalog.html}{morpheus-project.github.io/morpheus/data-release/egs/value-added-catalog-mrt.html} \\

    \textbf{All Files} & \\
    morpheus\_EGS\_all.v1.0.tar.gz     &  \href{http://morpheus-project.github.io/morpheus/data-release/egs/all.html}     {morpheus-project.github.io/morpheus/data-release/egs/all.html} \\

    \bottomrule
    \end{tabular}
    \caption{Data release files generated by Morpheus and associated URLs for
             the EGS CANDELS field. The data release files for each field are
             organized into three groups: \textit{pixel-level morphological
             classifications}, \textit{segmentation maps}, and \textit{3D-HST
             value-added catalogs}. The \textit{pixel-level morphological
             classification} files are named according to the following scheme
             \Verb|morpheus\_EGS\_[morphology].v1.0.fits|, where
             \Verb|[morphology]| can be one of the morphological classes
             (\textit{spheroid}, \textit{disk}, \textit{irregular},
             \textit{ps\_compact}, \textit{background}) or \textit{mask}, a
             binary image mask indicating which pixels in the image we're
             classified by \morpheus{}. The \textit{segmentation map} files are
             named according to the following scheme
             \Verb|morpheus\_EGS\_[segmap\_type].v1.0.fits| , where
             \Verb|[segmap\_type]| can be \textit{3dhst-segmap} (indicating the
             3D-HST informed segmap) or \textit{segmap} (indicating a segmap
             based only on background class/flux values). Finally, the 3D-HST
             value-added catalog files are named according to the following
             scheme \Verb|morpheus\_EGS\_3dhst-catalog.v1.0.[file\_type]|,
             where \Verb|[file\_type]| can be \textit{csv} for a
             comma-separated-value version of the value-added catalog and
             \textit{txt} for the machine-readable table version described in
             Table \ref{table:vac}. Additionally, a link to an archive
             containing all of the files associated with the EGS field is
             available in an additional section called \textit{All Files}. See
             Appendix \ref{appendix:data_release} for details.}
    \label{table:egs}
\end{table}

\begin{table}{\morpheus{} Data Products for the GOODS North Field}
    \centering
    \begin{tabular}{l l}
    \toprule
    File Name & URL \\
    \midrule
    \textbf{Pixel-level Morphological Classifications} & \\
    morpheus\_GOODS-N\_spheroid.v1.0.fits    & \href{http://morpheus-project.github.io/morpheus/data-release/goods-n/spheroid.html}   {morpheus-project.github.io/morpheus/data-release/goods-n/spheroid.html} \\
    morpheus\_GOODS-N\_disk.v1.0.fits        & \href{http://morpheus-project.github.io/morpheus/data-release/goods-n/disk.html}       {morpheus-project.github.io/morpheus/data-release/goods-n/disk.html} \\
    morpheus\_GOODS-N\_irregular.v1.0.fits   & \href{http://morpheus-project.github.io/morpheus/data-release/goods-n/irregular.html}  {morpheus-project.github.io/morpheus/data-release/goods-n/irregular.html} \\
    morpheus\_GOODS-N\_ps\_compact.v1.0.fits & \href{http://morpheus-project.github.io/morpheus/data-release/goods-n/ps-compact.html} {morpheus-project.github.io/morpheus/data-release/goods-n/ps\_compact.html} \\
    morpheus\_GOODS-N\_background.v1.0.fits  & \href{http://morpheus-project.github.io/morpheus/data-release/goods-n/background.html} {morpheus-project.github.io/morpheus/data-release/goods-n/background.html} \\
    morpheus\_GOODS-N\_mask.v1.0.fits        & \href{http://morpheus-project.github.io/morpheus/data-release/goods-n/mask.html}       {morpheus-project.github.io/morpheus/data-release/goods-n/mask.html} \\

    \textbf{Segmentation Maps} & \\
    morpheus\_GOODS-N\_segmap.v1.0.fits       & \href{http://morpheus-project.github.io/morpheus/data-release/goods-n/segmap.html}       {morpheus-project.github.io/morpheus/data-release/goods-n/segmap.html} \\
    morpheus\_GOODS-N\_3dhst-segmap.v1.0.fits & \href{http://morpheus-project.github.io/morpheus/data-release/goods-n/3dhst-segmap.html} {morpheus-project.github.io/morpheus/data-release/goods-n/3dhst-segmap.html} \\

    \textbf{3D-HST Value Added Catalogs} & \\
    morpheus\_GOODS-N\_3dhst\_catalog.v1.0.csv & \href{http://morpheus-project.github.io/morpheus/data-release/goods-n/value-added-catalog.html}{morpheus-project.github.io/morpheus/data-release/goods-n/value-added-catalog.html} \\
    morpheus\_GOODS-N\_3dhst\_catalog.v1.0.txt & \href{http://morpheus-project.github.io/morpheus/data-release/goods-n/value-added-catalog.html}{morpheus-project.github.io/morpheus/data-release/goods-n/value-added-catalog-mrt.html} \\

    \textbf{All Files} & \\
    morpheus\_GOODS-N\_all.v1.0.tar.gz &  \href{http://morpheus-project.github.io/morpheus/data-release/goods-n/all.html} {morpheus-project.github.io/morpheus/data-release/goods-n/all.html} \\

    \bottomrule
    \end{tabular}
    \caption{Data release files generated by Morpheus and associated URLs for
             the GOODS North CANDELS field. The data release files for each
             field are organized into three groups: \textit{pixel-level
             morphological classifications}, \textit{segmentation maps}, and
             \textit{3D-HST value-added catalogs}. The \textit{pixel-level
             morphological classification} files are named according to the
             following scheme \Verb|morpheus\_GOODS-N\_[morphology].v1.0.fits|,
             where \Verb|[morphology]| can be one of the morphological classes
             (\textit{spheroid}, \textit{disk}, \textit{irregular},
             \textit{ps\_compact}, \textit{background}) or \textit{mask}, a
             binary image mask indicating which pixels in the image we're
             classified by \morpheus{}. The \textit{segmentation map} files are
             named according to the following scheme
             \Verb|morpheus\_GOODS-N\_[segmap\_type].v1.0.fits| , where
             \Verb|[segmap\_type]| can be \textit{3dhst-segmap} (indicating the
             3D-HST informed segmap) or \textit{segmap} (indicating a segmap
             based only on background class/flux values). Finally, the 3D-HST
             value-added catalog files are named according to the following
             scheme \Verb|morpheus\_GOODS-N\_3dhst-catalog.v1.0.[file\_type]|,
             where \Verb|[file\_type]| can be \textit{csv} for a
             comma-separated-value version of the value-added catalog and
             \textit{txt} for the machine-readable table version described in
             Table \ref{table:vac}. Additionally, a link to an archive
             containing all of the files associated with the GOODS North field
             is available in an additional section called \textit{All Files}.
             See Appendix \ref{appendix:data_release} for details.}
    \label{table:goods-n}
\end{table}

\begin{table}{\morpheus{} Data Products for the GOODS South Field}
    \centering
    \begin{tabular}{l l}
    \toprule
    File Name & URL \\
    \midrule
    \textbf{Pixel-level Morphological Classifications} & \\
    morpheus\_GOODS-S\_spheroid.v1.0.fits    & \href{http://morpheus-project.github.io/morpheus/data-release/goods-s/spheroid.html}   {morpheus-project.github.io/morpheus/data-release/goods-s/spheroid.html} \\
    morpheus\_GOODS-S\_disk.v1.0.fits        & \href{http://morpheus-project.github.io/morpheus/data-release/goods-s/disk.html}       {morpheus-project.github.io/morpheus/data-release/goods-s/disk.html} \\
    morpheus\_GOODS-S\_irregular.v1.0.fits   & \href{http://morpheus-project.github.io/morpheus/data-release/goods-s/irregular.html}  {morpheus-project.github.io/morpheus/data-release/goods-s/irregular.html} \\
    morpheus\_GOODS-S\_ps\_compact.v1.0.fits & \href{http://morpheus-project.github.io/morpheus/data-release/goods-s/ps-compact.html} {morpheus-project.github.io/morpheus/data-release/goods-s/ps\_compact.html} \\
    morpheus\_GOODS-S\_background.v1.0.fits  & \href{http://morpheus-project.github.io/morpheus/data-release/goods-s/background.html} {morpheus-project.github.io/morpheus/data-release/goods-s/background.html} \\
    morpheus\_GOODS-S\_mask.v1.0.fits        & \href{http://morpheus-project.github.io/morpheus/data-release/goods-s/mask.html}       {morpheus-project.github.io/morpheus/data-release/goods-s/mask.html} \\
    morpheus\_GOODS-S\_spheroid.v1.0.fits    & \href{http://morpheus-project.github.io/morpheus/data-release/goods-s/spheroid.html}   {morpheus-project.github.io/morpheus/data-release/goods-s/spheroid.html} \\

    \textbf{Segmentation Maps} & \\
    morpheus\_GOODS-S\_segmap.v1.0.fits     & \href{http://morpheus-project.github.io/morpheus/data-release/goods-s/segmap.html}           {morpheus-project.github.io/morpheus/data-release/goods-s/segmap.html} \\
    morpheus\_GOODS-S\_3dhst\_segmap.v1.0.fits & \href{http://morpheus-project.github.io/morpheus/data-release/goods-s/3dhst-segmap.html} {morpheus-project.github.io/morpheus/data-release/goods-s/3dhst-segmap.html} \\

    \textbf{3D-HST Value Added Catalogs} & \\
    morpheus\_GOODS-S\_3dhst\_catalog.v1.0.csv & \href{http://morpheus-project.github.io/morpheus/data-release/goods-s/value-added-catalog.html}{morpheus-project.github.io/morpheus/data-release/goods-s/value-added-catalog.html} \\
    morpheus\_GOODS-S\_3dhst\_catalog.v1.0.txt & \href{http://morpheus-project.github.io/morpheus/data-release/goods-s/value-added-catalog.html}{morpheus-project.github.io/morpheus/data-release/goods-s/value-added-catalog-mrt.html} \\

    \textbf{All Files} & \\
    morpheus\_GOODS-S\_all.v1.0.tar.gz        &  \href{http://morpheus-project.github.io/morpheus/data-release/goods-s/all.html} {morpheus-project.github.io/morpheus/data-release/goods-s/all.html} \\

    \bottomrule
    \end{tabular}
    \caption{Data release files generated by Morpheus and associated URLs for
             the GOODS South CANDELS field. The data release files for each
             field are organized into three groups: \textit{pixel-level
             morphological classifications}, \textit{segmentation maps}, and
             \textit{3D-HST value-added catalogs}. The \textit{pixel-level
             morphological classification} files are named according to the
             following scheme \Verb|morpheus\_GOODS-S\_[morphology].v1.0.fits|,
             where \Verb|[morphology]| can be one of the morphological classes
             (\textit{spheroid}, \textit{disk}, \textit{irregular},
             \textit{ps\_compact}, \textit{background}) or \textit{mask}, a
             binary image mask indicating which pixels in the image we're
             classified by \morpheus{}. The \textit{segmentation map} files are
             named according to the following scheme
             \Verb|morpheus\_GOODS-S\_[segmap\_type].v1.0.fits| , where
             \Verb|[segmap\_type]| can be \textit{3dhst-segmap} (indicating the
             3D-HST informed segmap) or \textit{segmap} (indicating a segmap
             based only on background class/flux values). Finally, the 3D-HST
             value-added catalog files are named according to the following
             scheme \Verb|morpheus\_GOODS-S\_3dhst-catalog.v1.0.[file\_type]|,
             where \Verb|[file\_type]| can be \textit{csv} for a
             comma-separated-value version of the value-added catalog and
             \textit{txt} for the machine-readable table version described in
             Table \ref{table:vac}. Additionally, a link to an archive
             containing all of the files associated with the GOODS South field
             is available in an additional section called \textit{All Files}.
             See Appendix \ref{appendix:data_release} for details.}
    \label{table:goods-s}
\end{table}

\begin{table}{\morpheus{} Data Products for the UDS Field}
    \centering
    \begin{tabular}{l l}
    \toprule
    File Name & URL \\
    \midrule
    \textbf{Pixel-level Morphological Classifications} & \\
    morpheus\_UDS\_spheroid.v1.0.fits    & \href{http://morpheus-project.github.io/morpheus/data-release/uds/spheroid.html}   {morpheus-project.github.io/morpheus/data-release/uds/spheroid.html} \\
    morpheus\_UDS\_disk.v1.0.fits        & \href{http://morpheus-project.github.io/morpheus/data-release/uds/disk.html}       {morpheus-project.github.io/morpheus/data-release/uds/disk.html} \\
    morpheus\_UDS\_irregular.v1.0.fits   & \href{http://morpheus-project.github.io/morpheus/data-release/uds/irregular.html}  {morpheus-project.github.io/morpheus/data-release/uds/irregular.html} \\
    morpheus\_UDS\_ps\_compact.v1.0.fits & \href{http://morpheus-project.github.io/morpheus/data-release/uds/ps-compact.html} {morpheus-project.github.io/morpheus/data-release/uds/ps\_compact.html} \\
    morpheus\_UDS\_background.v1.0.fits  & \href{http://morpheus-project.github.io/morpheus/data-release/uds/background.html} {morpheus-project.github.io/morpheus/data-release/uds/background.html} \\
    morpheus\_UDS\_mask.v1.0.fits        & \href{http://morpheus-project.github.io/morpheus/data-release/uds/mask.html}       {morpheus-project.github.io/morpheus/data-release/uds/mask.html} \\

    \textbf{Segmentation Maps} & \\
    morpheus\_UDS\_segmap.v1.0.fits     & \href{http://morpheus-project.github.io/morpheus/data-release/uds/segmap.html}           {morpheus-project.github.io/morpheus/data-release/uds/segmap.html} \\
    morpheus\_UDS\_3dhst-segmap.v1.0.fits & \href{http://morpheus-project.github.io/morpheus/data-release/uds/3dhst-segmap.html} {morpheus-project.github.io/morpheus/data-release/uds/3dhst-segmap.html} \\

    \textbf{3D-HST Value Added Catalogs} & \\
    morpheus\_UDS\_3dhst\_catalog.v1.0.csv & \href{http://morpheus-project.github.io/morpheus/data-release/uds/value-added-catalog.html}{morpheus-project.github.io/morpheus/data-release/uds/value-added-catalog.html} \\
    morpheus\_UDS\_3dhst\_catalog.v1.0.txt & \href{http://morpheus-project.github.io/morpheus/data-release/uds/value-added-catalog.html}{morpheus-project.github.io/morpheus/data-release/uds/value-added-catalog-mrt.html} \\

    \textbf{All Files} & \\
    morpheus\_UDS\_all.v1.0.tar.gz     &  \href{http://morpheus-project.github.io/morpheus/data-release/uds/all.html}     {morpheus-project.github.io/morpheus/data-release/uds/all.html} \\

    \bottomrule
    \end{tabular}
    \caption{Data release files generated by Morpheus and associated URLs for
             the UDS CANDELS field. The data release files for each field are
             organized into three groups: \textit{pixel-level morphological
             classifications}, \textit{segmentation maps}, and \textit{3D-HST
             value-added catalogs}. The \textit{pixel-level morphological
             classification} files are named according to the following scheme
             \Verb|morpheus\_UDS\_[morphology].v1.0.fits|, where
             \Verb|[morphology]| can be one of the morphological classes
             (\textit{spheroid}, \textit{disk}, \textit{irregular},
             \textit{ps\_compact}, \textit{background}) or \textit{mask}, a
             binary image mask indicating which pixels in the image we're
             classified by \morpheus{}. The \textit{segmentation map} files are
             named according to the following scheme
             \Verb|morpheus\_UDS\_[segmap\_type].v1.0.fits| , where
             \Verb|[segmap\_type]| can be \textit{3dhst-segmap} (indicating the
             3D-HST informed segmap) or \textit{segmap} (indicating a segmap
             based only on background class/flux values). Finally, the 3D-HST
             value-added catalog files are named according to the following
             scheme \Verb|morpheus\_UDS\_3dhst-catalog.v1.0.[file\_type]|, where
             \Verb|[file\_type]| can be \textit{csv} for a comma-separated-value
             version of the value-added catalog and \textit{txt} for the
             machine-readable table version described in Table \ref{table:vac}.
             Additionally, a link to an archive containing all of the files
             associated with the UDS field is available in an additional section
             called \textit{All Files}. See Appendix \ref{appendix:data_release}
             for details.}
    \label{table:uds}
\end{table}

\bibliography{all}

\begin{thebibliography}{}
\expandafter\ifx\csname natexlab\endcsname\relax\def\natexlab#1{#1}\fi
\providecommand{\url}[1]{\href{#1}{#1}}

\bibitem[{{Abadi} {et~al.}(2016){Abadi}, {Agarwal}, {Barham}, {Brevdo}, {Chen},
  {Citro}, {Corrado}, {Davis}, {Dean}, {Devin}, {Ghemawat}, {Goodfellow},
  {Harp}, {Irving}, {Isard}, {Jia}, {Jozefowicz}, {Kaiser}, {Kudlur},
  {Levenberg}, {Mane}, {Monga}, {Moore}, {Murray}, {Olah}, {Schuster},
  {Shlens}, {Steiner}, {Sutskever}, {Talwar}, {Tucker}, {Vanhoucke},
  {Vasudevan}, {Viegas}, {Vinyals}, {Warden}, {Wattenberg}, {Wicke}, {Yu}, \&
  {Zheng}}]{tensorflow2015}
{Abadi}, M., {Agarwal}, A., {Barham}, P., {et~al.} 2016, arXiv e-prints,
  arXiv:1603.04467

\bibitem[{{Abraham} {et~al.}(1996){Abraham}, {Tanvir}, {Santiago}, {Ellis},
  {Glazebrook}, \& {van den Bergh}}]{abraham1996a}
{Abraham}, R.~G., {Tanvir}, N.~R., {Santiago}, B.~X., {et~al.} 1996, \mnras,
  279, L47

\bibitem[{{Abraham} {et~al.}(1994){Abraham}, {Valdes}, {Yee}, \& {van den
  Bergh}}]{abraham1994a}
{Abraham}, R.~G., {Valdes}, F., {Yee}, H.~K.~C., \& {van den Bergh}, S. 1994,
  \apj, 432, 75

\bibitem[{Abraham \& van~den Bergh(2001)}]{abraham2001a}
Abraham, R.~G., \& van~den Bergh, S. 2001, Science, 293, 1273.
\newblock \url{http://science.sciencemag.org/content/293/5533/1273}

\bibitem[{{Agnello} {et~al.}(2015){Agnello}, {Kelly}, {Treu}, \&
  {Marshall}}]{agnello2015a}
{Agnello}, A., {Kelly}, B.~C., {Treu}, T., \& {Marshall}, P.~J. 2015, \mnras,
  448, 1446

\bibitem[{{Aihara} {et~al.}(2018{\natexlab{a}}){Aihara}, {Arimoto},
  {Armstrong}, {Arnouts}, {Bahcall}, {Bickerton}, {Bosch}, {Bundy}, {Capak}, \&
  {Chan}}]{aihara2018a}
{Aihara}, H., {Arimoto}, N., {Armstrong}, R., {et~al.} 2018{\natexlab{a}},
  \pasj, 70, S4

\bibitem[{{Aihara} {et~al.}(2018{\natexlab{b}}){Aihara}, {Armstrong},
  {Bickerton}, {Bosch}, {Coupon}, {Furusawa}, {Hayashi}, {Ikeda}, {Kamata}, \&
  {Karoji}}]{aihara2018b}
{Aihara}, H., {Armstrong}, R., {Bickerton}, S., {et~al.} 2018{\natexlab{b}},
  \pasj, 70, S8

\bibitem[{{Akeson} {et~al.}(2019){Akeson}, {Armus}, {Bachelet}, {Bailey},
  {Bartusek}, {Bellini}, {Benford}, {Bennett}, {Bhattacharya}, \&
  {Bohlin}}]{akeson2019a}
{Akeson}, R., {Armus}, L., {Bachelet}, E., {et~al.} 2019, arXiv e-prints,
  arXiv:1902.05569

\bibitem[{{Allen} {et~al.}(2017){Allen}, {Kacprzak}, {Glazebrook}, {Labb{\'e}},
  {Tran}, {Spitler}, {Cowley}, {Nanayakkara}, {Papovich}, {Quadri},
  {Straatman}, {Tilvi}, \& {van Dokkum}}]{allen2017a}
{Allen}, R.~J., {Kacprzak}, G.~G., {Glazebrook}, K., {et~al.} 2017, \apjl, 834,
  L11

\bibitem[{{Beck} {et~al.}(2018){Beck}, {Scarlata}, {Fortson}, {Lintott},
  {Simmons}, {Galloway}, {Willett}, {Dickinson}, {Masters}, \&
  {Marshall}}]{beck2018a}
{Beck}, M.~R., {Scarlata}, C., {Fortson}, L.~F., {et~al.} 2018, \mnras, 476,
  5516

\bibitem[{{Bell} {et~al.}(2012){Bell}, {van der Wel}, {Papovich}, {Kocevski},
  {Lotz}, {McIntosh}, {Kartaltepe}, {Faber}, {Ferguson}, {Koekemoer}, {Grogin},
  {Wuyts}, {Cheung}, {Conselice}, {Dekel}, {Dunlop}, {Giavalisco},
  {Herrington}, {Koo}, {McGrath}, {de Mello}, {Rix}, {Robaina}, \&
  {Williams}}]{bell2012a}
{Bell}, E.~F., {van der Wel}, A., {Papovich}, C., {et~al.} 2012, \apj, 753, 167

\bibitem[{{Bender} {et~al.}(1992){Bender}, {Burstein}, \&
  {Faber}}]{bender1992a}
{Bender}, R., {Burstein}, D., \& {Faber}, S.~M. 1992, \apj, 399, 462

\bibitem[{{Bertin} \& {Arnouts}(1996)}]{bertin1996a}
{Bertin}, E., \& {Arnouts}, S. 1996, \aaps, 117, 393

\bibitem[{{Bezanson} {et~al.}(2013){Bezanson}, {van Dokkum}, {van de Sande},
  {Franx}, {Leja}, \& {Kriek}}]{bezanson2013a}
{Bezanson}, R., {van Dokkum}, P.~G., {van de Sande}, J., {et~al.} 2013, \apjl,
  779, L21

\bibitem[{{Binney}(1978)}]{binney1978a}
{Binney}, J. 1978, \mnras, 183, 501

\bibitem[{{Binney} \& {Tremaine}(1987)}]{binney1987a}
{Binney}, J., \& {Tremaine}, S. 1987, {Galactic dynamics} (Princeton, NJ:
  Princeton University Press)

\bibitem[{{Boucaud} {et~al.}(2019){Boucaud}, {Huertas-Company}, {Heneka},
  {Ishida}, {Sedaghat}, {de Souza}, {Moews}, {Dole}, {Castellano}, \&
  {Merlin}}]{boucard2019a}
{Boucaud}, A., {Huertas-Company}, M., {Heneka}, C., {et~al.} 2019, arXiv
  e-prints, arXiv:1905.01324

\bibitem[{{Bruce} {et~al.}(2016){Bruce}, {Dunlop}, {Mortlock}, {Kocevski},
  {McGrath}, \& {Rosario}}]{bruce2016a}
{Bruce}, V.~A., {Dunlop}, J.~S., {Mortlock}, A., {et~al.} 2016, \mnras, 458,
  2391

\bibitem[{{Bruce} {et~al.}(2012){Bruce}, {Dunlop}, {Cirasuolo}, {McLure},
  {Targett}, {Bell}, {Croton}, {Dekel}, {Faber}, {Ferguson}, {Grogin},
  {Kocevski}, {Koekemoer}, {Koo}, {Lai}, {Lotz}, {McGrath}, {Newman}, \& {van
  der Wel}}]{bruce2012a}
{Bruce}, V.~A., {Dunlop}, J.~S., {Cirasuolo}, M., {et~al.} 2012, \mnras, 427,
  1666

\bibitem[{{Cire{\c s}an} {et~al.}(2012){Cire{\c s}an}, {Meier}, \&
  {Schmidhuber}}]{ciresan2012a}
{Cire{\c s}an}, D., {Meier}, U., \& {Schmidhuber}, J. 2012, arXiv e-prints,
  arXiv:1202.2745

\bibitem[{{Conselice}(2003)}]{conselice2003a}
{Conselice}, C.~J. 2003, \apjs, 147, 1

\bibitem[{{Conselice} {et~al.}(2000){Conselice}, {Bershady}, \&
  {Jangren}}]{conselice2000a}
{Conselice}, C.~J., {Bershady}, M.~A., \& {Jangren}, A. 2000, \apj, 529, 886

\bibitem[{{Conselice} {et~al.}(2005){Conselice}, {Blackburne}, \&
  {Papovich}}]{conselice2005a}
{Conselice}, C.~J., {Blackburne}, J.~A., \& {Papovich}, C. 2005, \apj, 620, 564

\bibitem[{{Cooper} {et~al.}(2012){Cooper}, {Griffith}, {Newman}, {Coil},
  {Davis}, {Dutton}, {Faber}, {Guhathakurta}, {Koo}, {Lotz}, {Weiner},
  {Willmer}, \& {Yan}}]{cooper2012a}
{Cooper}, M.~C., {Griffith}, R.~L., {Newman}, J.~A., {et~al.} 2012, \mnras,
  419, 3018

\bibitem[{{Couprie} \& {Bertrand}(1997)}]{couprie1997a}
{Couprie}, M., \& {Bertrand}, G. 1997, in Society of Photo-Optical
  Instrumentation Engineers (SPIE) Conference Series, Vol. 3168, Vision
  Geometry VI, ed. R.~A. {Melter}, A.~Y. {Wu}, \& L.~J. {Latecki}, 136--146

\bibitem[{{Dai} \& {Tong}(2018)}]{dai2018b}
{Dai}, J.-M., \& {Tong}, J. 2018, ArXiv e-prints, arXiv:1807.10406

\bibitem[{{Dark Energy Survey Collaboration} {et~al.}(2016){Dark Energy Survey
  Collaboration}, {Abbott}, {Abdalla}, {Aleksi{\'c}}, {Allam}, {Amara},
  {Bacon}, {Balbinot}, {Banerji}, \& {Bechtol}}]{des2016a}
{Dark Energy Survey Collaboration}, {Abbott}, T., {Abdalla}, F.~B., {et~al.}
  2016, \mnras, 460, 1270

\bibitem[{Deng {et~al.}(2009)Deng, Dong, Socher, Li, Li, \&
  Fei-Fei}]{deng2009a}
Deng, J., Dong, W., Socher, R., {et~al.} 2009, in CVPR09

\bibitem[{{Dieleman} {et~al.}(2015){Dieleman}, {Willett}, \&
  {Dambre}}]{dielman2015a}
{Dieleman}, S., {Willett}, K.~W., \& {Dambre}, J. 2015, \mnras, 450, 1441

\bibitem[{{Dimauro} {et~al.}(2018){Dimauro}, {Huertas-Company}, {Daddi},
  {P{\'e}rez-Gonz{\'a}lez}, {Bernardi}, {Barro}, {Buitrago}, {Caro},
  {Cattaneo}, {Dominguez-S{\'a}nchez}, {Faber}, {H{\"a}u{\ss}ler}, {Kocevski},
  {Koekemoer}, {Koo}, {Lee}, {Mei}, {Margalef-Bentabol}, {Primack},
  {Rodriguez-Puebla}, {Salvato}, {Shankar}, \& {Tuccillo}}]{dimauro2018a}
{Dimauro}, P., {Huertas-Company}, M., {Daddi}, E., {et~al.} 2018, \mnras, 478,
  5410

\bibitem[{{Djorgovski} \& {Davis}(1987)}]{djorgovski1987a}
{Djorgovski}, S., \& {Davis}, M. 1987, \apj, 313, 59

\bibitem[{{Dom{\'{\i}}nguez S{\'a}nchez} {et~al.}(2019){Dom{\'{\i}}nguez
  S{\'a}nchez}, {Huertas-Company}, {Bernardi}, {Kaviraj}, {Fischer}, {Abbott},
  {Abdalla}, {Annis}, {Avila}, {Brooks}, {Buckley-Geer}, {Carnero Rosell},
  {Carrasco Kind}, {Carretero}, {Cunha}, {D'Andrea}, {da Costa}, {Davis}, {De
  Vicente}, {Doel}, {Evrard}, {Fosalba}, {Frieman}, {Garc{\'{\i}}a-Bellido},
  {Gaztanaga}, {Gerdes}, {Gruen}, {Gruendl}, {Gschwend}, {Gutierrez},
  {Hartley}, {Hollowood}, {Honscheid}, {Hoyle}, {James}, {Kuehn}, {Kuropatkin},
  {Lahav}, {Maia}, {March}, {Melchior}, {Menanteau}, {Miquel}, {Nord},
  {Plazas}, {Sanchez}, {Scarpine}, {Schindler}, {Schubnell}, {Smith}, {Smith},
  {Soares-Santos}, {Sobreira}, {Suchyta}, {Swanson}, {Tarle}, {Thomas},
  {Walker}, \& {Zuntz}}]{dominguez_sanchez2019a}
{Dom{\'{\i}}nguez S{\'a}nchez}, H., {Huertas-Company}, M., {Bernardi}, M.,
  {et~al.} 2019, \mnras, 484, 93

\bibitem[{{Dressler}(1980)}]{dressler1980a}
{Dressler}, A. 1980, \apj, 236, 351

\bibitem[{{Dressler} {et~al.}(1987){Dressler}, {Lynden-Bell}, {Burstein},
  {Davies}, {Faber}, {Terlevich}, \& {Wegner}}]{dressler1987a}
{Dressler}, A., {Lynden-Bell}, D., {Burstein}, D., {et~al.} 1987, \apj, 313, 42

\bibitem[{{Dressler} {et~al.}(1997){Dressler}, {Oemler}, {Couch}, {Smail},
  {Ellis}, {Barger}, {Butcher}, {Poggianti}, \& {Sharples}}]{dressler1997a}
{Dressler}, A., {Oemler}, Jr., A., {Couch}, W.~J., {et~al.} 1997, \apj, 490,
  577

\bibitem[{{Elmegreen} {et~al.}(2005){Elmegreen}, {Elmegreen}, {Rubin}, \&
  {Schaffer}}]{elmegreen2005a}
{Elmegreen}, D.~M., {Elmegreen}, B.~G., {Rubin}, D.~S., \& {Schaffer}, M.~A.
  2005, \apj, 631, 85

\bibitem[{{Franx} {et~al.}(2008){Franx}, {van Dokkum}, {F{\"o}rster Schreiber},
  {Wuyts}, {Labb{\'e}}, \& {Toft}}]{franx2008a}
{Franx}, M., {van Dokkum}, P.~G., {F{\"o}rster Schreiber}, N.~M., {et~al.}
  2008, \apj, 688, 770

\bibitem[{{Gardner} {et~al.}(2006){Gardner}, {Mather}, {Clampin}, {Doyon},
  {Greenhouse}, {Hammel}, {Hutchings}, {Jakobsen}, {Lilly}, {Long}, {Lunine},
  {McCaughrean}, {Mountain}, {Nella}, {Rieke}, {Rieke}, {Rix}, {Smith},
  {Sonneborn}, {Stiavelli}, {Stockman}, {Windhorst}, \&
  {Wright}}]{gardner2006a}
{Gardner}, J.~P., {Mather}, J.~C., {Clampin}, M., {et~al.} 2006, \ssr, 123, 485

\bibitem[{{Gonz{\'a}lez} {et~al.}(2018){Gonz{\'a}lez}, {Mu{\~n}oz}, \&
  {Hern{\'a}ndez}}]{gonzales2018a}
{Gonz{\'a}lez}, R.~E., {Mu{\~n}oz}, R.~P., \& {Hern{\'a}ndez}, C.~A. 2018,
  ArXiv e-prints, arXiv:1809.01691

\bibitem[{{Grogin} {et~al.}(2011){Grogin}, {Kocevski}, {Faber}, {Ferguson},
  {Koekemoer}, {Riess}, {Acquaviva}, {Alexander}, {Almaini}, {Ashby}, {Barden},
  {Bell}, {Bournaud}, {Brown}, {Caputi}, {Casertano}, {Cassata}, {Castellano},
  {Challis}, {Chary}, {Cheung}, {Cirasuolo}, {Conselice}, {Roshan Cooray},
  {Croton}, {Daddi}, {Dahlen}, {Dav{\'e}}, {de Mello}, {Dekel}, {Dickinson},
  {Dolch}, {Donley}, {Dunlop}, {Dutton}, {Elbaz}, {Fazio}, {Filippenko},
  {Finkelstein}, {Fontana}, {Gardner}, {Garnavich}, {Gawiser}, {Giavalisco},
  {Grazian}, {Guo}, {Hathi}, {H{\"a}ussler}, {Hopkins}, {Huang}, {Huang},
  {Jha}, {Kartaltepe}, {Kirshner}, {Koo}, {Lai}, {Lee}, {Li}, {Lotz}, {Lucas},
  {Madau}, {McCarthy}, {McGrath}, {McIntosh}, {McLure}, {Mobasher},
  {Moustakas}, {Mozena}, {Nandra}, {Newman}, {Niemi}, {Noeske}, {Papovich},
  {Pentericci}, {Pope}, {Primack}, {Rajan}, {Ravindranath}, {Reddy}, {Renzini},
  {Rix}, {Robaina}, {Rodney}, {Rosario}, {Rosati}, {Salimbeni}, {Scarlata},
  {Siana}, {Simard}, {Smidt}, {Somerville}, {Spinrad}, {Straughn}, {Strolger},
  {Telford}, {Teplitz}, {Trump}, {van der Wel}, {Villforth}, {Wechsler},
  {Weiner}, {Wiklind}, {Wild}, {Wilson}, {Wuyts}, {Yan}, \&
  {Yun}}]{grogin2011a}
{Grogin}, N.~A., {Kocevski}, D.~D., {Faber}, S.~M., {et~al.} 2011, \apjs, 197,
  35

\bibitem[{{Guo} {et~al.}(2013){Guo}, {Ferguson}, {Giavalisco}, {Barro},
  {Willner}, {Ashby}, {Dahlen}, {Donley}, {Faber}, {Fontana}, {Galametz},
  {Grazian}, {Huang}, {Kocevski}, {Koekemoer}, {Koo}, {McGrath}, {Peth},
  {Salvato}, {Wuyts}, {Castellano}, {Cooray}, {Dickinson}, {Dunlop}, {Fazio},
  {Gardner}, {Gawiser}, {Grogin}, {Hathi}, {Hsu}, {Lee}, {Lucas}, {Mobasher},
  {Nand ra}, {Newman}, \& {van der Wel}}]{guo2013a}
{Guo}, Y., {Ferguson}, H.~C., {Giavalisco}, M., {et~al.} 2013, \apjs, 207, 24

\bibitem[{{Hahnloser} {et~al.}(2000){Hahnloser}, {Sarpeshkar}, {Mahowald},
  {Douglas}, \& {Seung}}]{hahnloser2000a}
{Hahnloser}, R.~H.~R., {Sarpeshkar}, R., {Mahowald}, M.~A., {Douglas}, R.~J.,
  \& {Seung}, H.~S. 2000, \nat, 405, 947

\bibitem[{Hausen(2020)}]{hausen2020}
Hausen, R. 2020, morpheus-project/morpheus: Zenodo Release, vv0.4.6,  Zenodo,
  doi:10.5281/zenodo.3601680.
\newblock \url{https://doi.org/10.5281/zenodo.3601680}

\bibitem[{{Hemmati} {et~al.}(2019){Hemmati}, {Capak}, {Masters}, {Davidzon},
  {Dor{\`e}}, {Kruk}, {Mobasher}, {Rhodes}, {Scolnic}, \&
  {Stern}}]{hemmati2019a}
{Hemmati}, S., {Capak}, P., {Masters}, D., {et~al.} 2019, \apj, 877, 117

\bibitem[{{Hezaveh} {et~al.}(2017){Hezaveh}, {Perreault Levasseur}, \&
  {Marshall}}]{hezaveh2017a}
{Hezaveh}, Y.~D., {Perreault Levasseur}, L., \& {Marshall}, P.~J. 2017, \nat,
  548, 555

\bibitem[{{Hocking} {et~al.}(2018){Hocking}, {Geach}, {Sun}, \&
  {Davey}}]{hocking2017a}
{Hocking}, A., {Geach}, J.~E., {Sun}, Y., \& {Davey}, N. 2018, \mnras, 473,
  1108

\bibitem[{{Hubble}(1926)}]{hubble1926a}
{Hubble}, E.~P. 1926, \apj, 64, doi:10.1086/143018

\bibitem[{{Huertas-Company} {et~al.}(2015){Huertas-Company}, {Gravet},
  {Cabrera-Vives}, {P{\'e}rez-Gonz{\'a}lez}, {Kartaltepe}, {Barro}, {Bernardi},
  {Mei}, {Shankar}, {Dimauro}, {Bell}, {Kocevski}, {Koo}, {Faber}, \&
  {Mcintosh}}]{huertas-company2015a}
{Huertas-Company}, M., {Gravet}, R., {Cabrera-Vives}, G., {et~al.} 2015, The
  Astrophysical Journal Supplement Series, 221, 8

\bibitem[{{Huertas-Company} {et~al.}(2016){Huertas-Company}, {Bernardi},
  {P{\'e}rez-Gonz{\'a}lez}, {Ashby}, {Barro}, {Conselice}, {Daddi}, {Dekel},
  {Dimauro}, {Faber}, {Grogin}, {Kartaltepe}, {Kocevski}, {Koekemoer}, {Koo},
  {Mei}, \& {Shankar}}]{huertas-company2016a}
{Huertas-Company}, M., {Bernardi}, M., {P{\'e}rez-Gonz{\'a}lez}, P.~G.,
  {et~al.} 2016, \mnras, 462, 4495

\bibitem[{{Huertas-Company} {et~al.}(2018){Huertas-Company}, {Primack},
  {Dekel}, {Koo}, {Lapiner}, {Ceverino}, {Simons}, {Snyder}, {Bernardi}, \&
  {Chen}}]{huertas-company2018a}
{Huertas-Company}, M., {Primack}, J.~R., {Dekel}, A., {et~al.} 2018, \apj, 858,
  114

\bibitem[{{Huertas-Company} {et~al.}(2019){Huertas-Company}, {Rodriguez-Gomez},
  {Nelson}, {Pillepich}, {Bottrell}, {Bernardi}, {Dom{\'\i}nguez-S{\'a}nchez},
  {Genel}, {Pakmor}, {Snyder}, \& {Vogelsberger}}]{huertas-company2019a}
{Huertas-Company}, M., {Rodriguez-Gomez}, V., {Nelson}, D., {et~al.} 2019,
  \mnras, 489, 1859

\bibitem[{{Hunter}(2007)}]{hunter2007}
{Hunter}, J.~D. 2007, Computing in Science Engineering, 9, 90

\bibitem[{{Illingworth} {et~al.}(2016){Illingworth}, {Magee}, {Bouwens},
  {Oesch}, {Labbe}, {van Dokkum}, {Whitaker}, {Holden}, {Franx}, \&
  {Gonzalez}}]{illingworth2016a}
{Illingworth}, G., {Magee}, D., {Bouwens}, R., {et~al.} 2016, arXiv e-prints,
  arXiv:1606.00841

\bibitem[{{Ioffe} \& {Szegedy}(2015)}]{ioffe2015a}
{Ioffe}, S., \& {Szegedy}, C. 2015, ArXiv e-prints, arXiv:1502.03167

\bibitem[{{Ivezi{\'c}} {et~al.}(2019){Ivezi{\'c}}, {Kahn}, {Tyson}, {Abel},
  {Acosta}, {Allsman}, {Alonso}, {AlSayyad}, {Anderson}, \&
  {Andrew}}]{ivezic2019a}
{Ivezi{\'c}}, {\v{Z}}., {Kahn}, S.~M., {Tyson}, J.~A., {et~al.} 2019, \apj,
  873, 111

\bibitem[{{Jiang} {et~al.}(2018){Jiang}, {Liu}, {Zheng}, {Yesuf}, {Koo},
  {Faber}, {Guo}, {Koekemoer}, {Wang}, {Fang}, {Barro}, {Jia}, {Tong}, {Liu},
  {Meng}, {Kocevski}, {McGrath}, \& {Hathi}}]{jiang2018a}
{Jiang}, D., {Liu}, F.~S., {Zheng}, X., {et~al.} 2018, \apj, 854, 70

\bibitem[{{Kartaltepe} {et~al.}(2015){Kartaltepe}, {Mozena}, {Kocevski},
  {McIntosh}, {Lotz}, {Bell}, {Faber}, {Ferguson}, {Koo}, {Bassett}, {Bernyk},
  {Blancato}, {Bournaud}, {Cassata}, {Castellano}, {Cheung}, {Conselice},
  {Croton}, {Dahlen}, {de Mello}, {DeGroot}, {Donley}, {Guedes}, {Grogin},
  {Hathi}, {Hilton}, {Hollon}, {Koekemoer}, {Liu}, {Lucas}, {Martig},
  {McGrath}, {McPartland}, {Mobasher}, {Morlock}, {O'Leary}, {Peth}, {Pforr},
  {Pillepich}, {Rosario}, {Soto}, {Straughn}, {Telford}, {Sunnquist}, {Trump},
  {Weiner}, {Wuyts}, {Inami}, {Kassin}, {Lani}, {Poole}, \&
  {Rizer}}]{kartaltepe2015a}
{Kartaltepe}, J.~S., {Mozena}, M., {Kocevski}, D., {et~al.} 2015, \apjs, 221,
  11

\bibitem[{{Kawinwanichakij} {et~al.}(2017){Kawinwanichakij}, {Papovich},
  {Quadri}, {Glazebrook}, {Kacprzak}, {Allen}, {Bell}, {Croton}, {Dekel},
  {Ferguson}, {Forrest}, {Grogin}, {Guo}, {Kocevski}, {Koekemoer}, {Labb{\'e}},
  {Lucas}, {Nanayakkara}, {Spitler}, {Straatman}, {Tran}, {Tomczak}, \& {van
  Dokkum}}]{kawinwanichakij2017a}
{Kawinwanichakij}, L., {Papovich}, C., {Quadri}, R.~F., {et~al.} 2017, \apj,
  847, 134

\bibitem[{{Kelly} \& {McKay}(2004)}]{kelly2004a}
{Kelly}, B.~C., \& {McKay}, T.~A. 2004, \aj, 127, 625

\bibitem[{Kelly \& McKay(2005)}]{kelly2005a}
Kelly, B.~C., \& McKay, T.~A. 2005, The Astronomical Journal, 129, 1287.
\newblock \url{http://stacks.iop.org/1538-3881/129/i=3/a=1287}

\bibitem[{{Kingma} \& {Ba}(2014)}]{Kingma2014a}
{Kingma}, D.~P., \& {Ba}, J. 2014, ArXiv e-prints, arXiv:1412.6980

\bibitem[{{Kocevski} {et~al.}(2012){Kocevski}, {Faber}, {Mozena}, {Koekemoer},
  {Nandra}, {Rangel}, {Laird}, {Brusa}, {Wuyts}, {Trump}, {Koo}, {Somerville},
  {Bell}, {Lotz}, {Alexander}, {Bournaud}, {Conselice}, {Dahlen}, {Dekel},
  {Donley}, {Dunlop}, {Finoguenov}, {Georgakakis}, {Giavalisco}, {Guo},
  {Grogin}, {Hathi}, {Juneau}, {Kartaltepe}, {Lucas}, {McGrath}, {McIntosh},
  {Mobasher}, {Robaina}, {Rosario}, {Straughn}, {van der Wel}, \&
  {Villforth}}]{kocevski2012a}
{Kocevski}, D.~D., {Faber}, S.~M., {Mozena}, M., {et~al.} 2012, \apj, 744, 148

\bibitem[{{Koekemoer} {et~al.}(2011){Koekemoer}, {Faber}, {Ferguson}, {Grogin},
  {Kocevski}, {Koo}, {Lai}, {Lotz}, {Lucas}, {McGrath}, {Ogaz}, {Rajan},
  {Riess}, {Rodney}, {Strolger}, {Casertano}, {Castellano}, {Dahlen},
  {Dickinson}, {Dolch}, {Fontana}, {Giavalisco}, {Grazian}, {Guo}, {Hathi},
  {Huang}, {van der Wel}, {Yan}, {Acquaviva}, {Alexander}, {Almaini}, {Ashby},
  {Barden}, {Bell}, {Bournaud}, {Brown}, {Caputi}, {Cassata}, {Challis},
  {Chary}, {Cheung}, {Cirasuolo}, {Conselice}, {Roshan Cooray}, {Croton},
  {Daddi}, {Dav{\'e}}, {de Mello}, {de Ravel}, {Dekel}, {Donley}, {Dunlop},
  {Dutton}, {Elbaz}, {Fazio}, {Filippenko}, {Finkelstein}, {Frazer}, {Gardner},
  {Garnavich}, {Gawiser}, {Gruetzbauch}, {Hartley}, {H{\"a}ussler},
  {Herrington}, {Hopkins}, {Huang}, {Jha}, {Johnson}, {Kartaltepe},
  {Khostovan}, {Kirshner}, {Lani}, {Lee}, {Li}, {Madau}, {McCarthy},
  {McIntosh}, {McLure}, {McPartland}, {Mobasher}, {Moreira}, {Mortlock},
  {Moustakas}, {Mozena}, {Nandra}, {Newman}, {Nielsen}, {Niemi}, {Noeske},
  {Papovich}, {Pentericci}, {Pope}, {Primack}, {Ravindranath}, {Reddy},
  {Renzini}, {Rix}, {Robaina}, {Rosario}, {Rosati}, {Salimbeni}, {Scarlata},
  {Siana}, {Simard}, {Smidt}, {Snyder}, {Somerville}, {Spinrad}, {Straughn},
  {Telford}, {Teplitz}, {Trump}, {Vargas}, {Villforth}, {Wagner}, {Wandro},
  {Wechsler}, {Weiner}, {Wiklind}, {Wild}, {Wilson}, {Wuyts}, \&
  {Yun}}]{koekemoer2011a}
{Koekemoer}, A.~M., {Faber}, S.~M., {Ferguson}, H.~C., {et~al.} 2011, \apjs,
  197, 36

\bibitem[{{Kormendy}(1977)}]{kormendy1977a}
{Kormendy}, J. 1977, \apj, 218, 333

\bibitem[{{Krist} {et~al.}(2011){Krist}, {Hook}, \& {Stoehr}}]{krist2011a}
{Krist}, J.~E., {Hook}, R.~N., \& {Stoehr}, F. 2011, in Society of
  Photo-Optical Instrumentation Engineers (SPIE) Conference Series, Vol. 8127,
  Optical Modeling and Performance Predictions V, 81270J

\bibitem[{{Laureijs} {et~al.}(2011){Laureijs}, {Amiaux}, {Arduini},
  {Augu{\`e}res}, {Brinchmann}, {Cole}, {Cropper}, {Dabin}, {Duvet}, \&
  {Ealet}}]{laureijs2011a}
{Laureijs}, R., {Amiaux}, J., {Arduini}, S., {et~al.} 2011, arXiv e-prints,
  arXiv:1110.3193

\bibitem[{{Lecun} {et~al.}(2015){Lecun}, {Bengio}, \& {Hinton}}]{lecun2015a}
{Lecun}, Y., {Bengio}, Y., \& {Hinton}, G. 2015, \nat, 521, 436

\bibitem[{{Lee} {et~al.}(2013){Lee}, {Giavalisco}, {Williams}, {Guo}, {Lotz},
  {Van der Wel}, {Ferguson}, {Faber}, {Koekemoer}, {Grogin}, {Kocevski},
  {Conselice}, {Wuyts}, {Dekel}, {Kartaltepe}, \& {Bell}}]{lee2013a}
{Lee}, B., {Giavalisco}, M., {Williams}, C.~C., {et~al.} 2013, \apj, 774, 47

\bibitem[{{Lintott} {et~al.}(2008){Lintott}, {Schawinski}, {Slosar}, {Land},
  {Bamford}, {Thomas}, {Raddick}, {Nichol}, {Szalay}, {Andreescu}, {Murray}, \&
  {Vandenberg}}]{lintott2008a}
{Lintott}, C.~J., {Schawinski}, K., {Slosar}, A., {et~al.} 2008, \mnras, 389,
  1179

\bibitem[{{Lofthouse} {et~al.}(2017){Lofthouse}, {Kaviraj}, {Conselice},
  {Mortlock}, \& {Hartley}}]{lofthouse2017a}
{Lofthouse}, E.~K., {Kaviraj}, S., {Conselice}, C.~J., {Mortlock}, A., \&
  {Hartley}, W. 2017, \mnras, 465, 2895

\bibitem[{{Lotz} {et~al.}(2004){Lotz}, {Primack}, \& {Madau}}]{lotz2004a}
{Lotz}, J.~M., {Primack}, J., \& {Madau}, P. 2004, \aj, 128, 163

\bibitem[{{Lotz} {et~al.}(2008){Lotz}, {Davis}, {Faber}, {Guhathakurta},
  {Gwyn}, {Huang}, {Koo}, {Le Floc'h}, {Lin}, {Newman}, {Noeske}, {Papovich},
  {Willmer}, {Coil}, {Conselice}, {Cooper}, {Hopkins}, {Metevier}, {Primack},
  {Rieke}, \& {Weiner}}]{lotz2008a}
{Lotz}, J.~M., {Davis}, M., {Faber}, S.~M., {et~al.} 2008, \apj, 672, 177

\bibitem[{{LSST Science Collaboration} {et~al.}(2009){LSST Science
  Collaboration}, {Abell}, {Allison}, {Anderson}, {Andrew}, {Angel}, {Armus},
  {Arnett}, {Asztalos}, {Axelrod}, \& et~al.}]{lsst2009a}
{LSST Science Collaboration}, {Abell}, P.~A., {Allison}, J., {et~al.} 2009,
  arXiv e-prints, arXiv:0912.0201

\bibitem[{{Margalef-Bentabol} {et~al.}(2016){Margalef-Bentabol}, {Conselice},
  {Mortlock}, {Hartley}, {Duncan}, {Ferguson}, {Dekel}, \&
  {Primack}}]{margalef-bentabol2016a}
{Margalef-Bentabol}, B., {Conselice}, C.~J., {Mortlock}, A., {et~al.} 2016,
  \mnras, 461, 2728

\bibitem[{{Masters} {et~al.}(2015){Masters}, {Capak}, {Stern}, {Ilbert},
  {Salvato}, {Schmidt}, {Longo}, {Rhodes}, {Paltani}, \&
  {Mobasher}}]{masters2015a}
{Masters}, D., {Capak}, P., {Stern}, D., {et~al.} 2015, \apj, 813, 53

\bibitem[{Merkel(2014)}]{merkel2014a}
Merkel, D. 2014, Linux J., 2014

\bibitem[{{Miller} {et~al.}(2019){Miller}, {van Dokkum}, {Mowla}, \& {van der
  Wel}}]{miller2019a}
{Miller}, T.~B., {van Dokkum}, P., {Mowla}, L., \& {van der Wel}, A. 2019,
  \apjl, 872, L14

\bibitem[{{Milletari} {et~al.}(2016){Milletari}, {Navab}, \&
  {Ahmadi}}]{milletari2016a}
{Milletari}, F., {Navab}, N., \& {Ahmadi}, S.-A. 2016, ArXiv e-prints,
  arXiv:1606.04797

\bibitem[{{Momcheva} {et~al.}(2016){Momcheva}, {Brammer}, {van Dokkum},
  {Skelton}, {Whitaker}, {Nelson}, {Fumagalli}, {Maseda}, {Leja}, {Franx},
  {Rix}, {Bezanson}, {Da Cunha}, {Dickey}, {F{\"o}rster Schreiber},
  {Illingworth}, {Kriek}, {Labb{\'e}}, {Ulf Lange}, {Lundgren}, {Magee},
  {Marchesini}, {Oesch}, {Pacifici}, {Patel}, {Price}, {Tal}, {Wake}, {van der
  Wel}, \& {Wuyts}}]{momcheva2016a}
{Momcheva}, I.~G., {Brammer}, G.~B., {van Dokkum}, P.~G., {et~al.} 2016, \apjs,
  225, 27

\bibitem[{{Morishita} {et~al.}(2014){Morishita}, {Ichikawa}, \&
  {Kajisawa}}]{morishta2014a}
{Morishita}, T., {Ichikawa}, T., \& {Kajisawa}, M. 2014, \apj, 785, 18

\bibitem[{{Morningstar} {et~al.}(2018){Morningstar}, {Hezaveh}, {Perreault
  Levasseur}, {Blandford}, {Marshall}, {Putzky}, \&
  {Wechsler}}]{morningstar2018a}
{Morningstar}, W.~R., {Hezaveh}, Y.~D., {Perreault Levasseur}, L., {et~al.}
  2018, arXiv e-prints, arXiv:1808.00011

\bibitem[{{Morningstar} {et~al.}(2019){Morningstar}, {Perreault Levasseur},
  {Hezaveh}, {Blandford}, {Marshall}, {Putzky}, {Rueter}, {Wechsler}, \&
  {Welling}}]{morningstar2019a}
{Morningstar}, W.~R., {Perreault Levasseur}, L., {Hezaveh}, Y.~D., {et~al.}
  2019, arXiv e-prints, arXiv:1901.01359

\bibitem[{{Novikov} {et~al.}(2017){Novikov}, {Lenis}, {Major}, {Hladυka},
  {Wimmer}, \& {Bühler}}]{novikov2015a}
{Novikov}, A.~A., {Lenis}, D., {Major}, D., {et~al.} 2017, ArXiv e-prints,
  arXiv:1701.08816

\bibitem[{{Oke} \& {Gunn}(1983)}]{oke1983a}
{Oke}, J.~B., \& {Gunn}, J.~E. 1983, \apj, 266, 713

\bibitem[{{Patel} {et~al.}(2013){Patel}, {Fumagalli}, {Franx}, {van Dokkum},
  {van der Wel}, {Leja}, {Labb{\'e}}, {Brammer}, {Skelton}, {Momcheva},
  {Whitaker}, {Lundgren}, {Muzzin}, {Quadri}, {Nelson}, {Wake}, \&
  {Rix}}]{patel2013a}
{Patel}, S.~G., {Fumagalli}, M., {Franx}, M., {et~al.} 2013, \apj, 778, 115

\bibitem[{Pedregosa {et~al.}(2011)Pedregosa, Varoquaux, Gramfort, Michel,
  Thirion, Grisel, Blondel, Prettenhofer, Weiss, Dubourg, Vanderplas, Passos,
  Cournapeau, Brucher, Perrot, \& Duchesnay}]{pedregosa2011a}
Pedregosa, F., Varoquaux, G., Gramfort, A., {et~al.} 2011, Journal of Machine
  Learning Research, 12, 2825

\bibitem[{{Peng} {et~al.}(2010){Peng}, {Ho}, {Impey}, \& {Rix}}]{peng2010a}
{Peng}, C.~Y., {Ho}, L.~C., {Impey}, C.~D., \& {Rix}, H.-W. 2010, \aj, 139,
  2097

\bibitem[{{Peth} {et~al.}(2016){Peth}, {Lotz}, {Freeman}, {McPartland},
  {Mortazavi}, {Snyder}, {Barro}, {Grogin}, {Guo}, {Hemmati}, {Kartaltepe},
  {Kocevski}, {Koekemoer}, {McIntosh}, {Nayyeri}, {Papovich}, {Primack}, \&
  {Simons}}]{peth2016a}
{Peth}, M.~A., {Lotz}, J.~M., {Freeman}, P.~E., {et~al.} 2016, \mnras, 458, 963

\bibitem[{{Powell} {et~al.}(2017){Powell}, {Urry}, {Cardamone}, {Simmons},
  {Schawinski}, {Young}, \& {Kawakatsu}}]{powell2017a}
{Powell}, M.~C., {Urry}, C.~M., {Cardamone}, C.~N., {et~al.} 2017, \apj, 835,
  22

\bibitem[{Pratt(1993)}]{pratt1993a}
Pratt, L.~Y. 1993, in Advances in Neural Information Processing Systems 5, ed.
  S.~J. Hanson, J.~D. Cowan, \& C.~L. Giles (Morgan-Kaufmann), 204--211.
\newblock
  \url{http://papers.nips.cc/paper/641-discriminability-based-transfer-between-neural-networks.pdf}

\bibitem[{{Price-Whelan} {et~al.}(2018){Price-Whelan}, {Sip{\H{o}}cz},
  {G{\"u}nther}, {Lim}, {Crawford}, {Conseil}, {Shupe}, {Craig}, {Dencheva},
  {Ginsburg}, {VanderPlas}, {Bradley}, {P{\'e}rez-Su{\'a}rez}, {de Val-Borro},
  {Paper Contributors}, {Aldcroft}, {Cruz}, {Robitaille}, {Tollerud},
  {Coordination Committee}, {Ardelean}, {Babej}, {Bach}, {Bachetti}, {Bakanov},
  {Bamford}, {Barentsen}, {Barmby}, {Baumbach}, {Berry}, {Biscani}, {Boquien},
  {Bostroem}, {Bouma}, {Brammer}, {Bray}, {Breytenbach}, {Buddelmeijer},
  {Burke}, {Calderone}, {Cano Rodr{\'\i}guez}, {Cara}, {Cardoso}, {Cheedella},
  {Copin}, {Corrales}, {Crichton}, {D{\textquoteright}Avella}, {Deil},
  {Depagne}, {Dietrich}, {Donath}, {Droettboom}, {Earl}, {Erben}, {Fabbro},
  {Ferreira}, {Finethy}, {Fox}, {Garrison}, {Gibbons}, {Goldstein}, {Gommers},
  {Greco}, {Greenfield}, {Groener}, {Grollier}, {Hagen}, {Hirst}, {Homeier},
  {Horton}, {Hosseinzadeh}, {Hu}, {Hunkeler}, {Ivezi{\'c}}, {Jain}, {Jenness},
  {Kanarek}, {Kendrew}, {Kern}, {Kerzendorf}, {Khvalko}, {King}, {Kirkby},
  {Kulkarni}, {Kumar}, {Lee}, {Lenz}, {Littlefair}, {Ma}, {Macleod},
  {Mastropietro}, {McCully}, {Montagnac}, {Morris}, {Mueller}, {Mumford},
  {Muna}, {Murphy}, {Nelson}, {Nguyen}, {Ninan}, {N{\"o}the}, {Ogaz}, {Oh},
  {Parejko}, {Parley}, {Pascual}, {Patil}, {Patil}, {Plunkett}, {Prochaska},
  {Rastogi}, {Reddy Janga}, {Sabater}, {Sakurikar}, {Seifert}, {Sherbert},
  {Sherwood-Taylor}, {Shih}, {Sick}, {Silbiger}, {Singanamalla}, {Singer},
  {Sladen}, {Sooley}, {Sornarajah}, {Streicher}, {Teuben}, {Thomas},
  {Tremblay}, {Turner}, {Terr{\'o}n}, {van Kerkwijk}, {de la Vega}, {Watkins},
  {Weaver}, {Whitmore}, {Woillez}, {Zabalza}, \& {Contributors}}]{price2018}
{Price-Whelan}, A.~M., {Sip{\H{o}}cz}, B.~M., {G{\"u}nther}, H.~M., {et~al.}
  2018, \aj, 156, 123

\bibitem[{{Rhodes} {et~al.}(2017){Rhodes}, {Nichol}, {Aubourg}, {Bean},
  {Boutigny}, {Bremer}, {Capak}, {Cardone}, {Carry}, \&
  {Conselice}}]{rhodes2017a}
{Rhodes}, J., {Nichol}, R.~C., {Aubourg}, {\'E}., {et~al.} 2017, \apjs, 233, 21

\bibitem[{{Roberts} \& {Haynes}(1994)}]{roberts1994a}
{Roberts}, M.~S., \& {Haynes}, M.~P. 1994, \araa, 32, 115

\bibitem[{{Robertson} {et~al.}(2017){Robertson}, {Banerji}, {Cooper}, {Davies},
  {Driver}, {Ferguson}, {Ferguson}, {Gawiser}, {Kaviraj}, {Knapen}, {Lintott},
  {Lotz}, {Newman}, {Norman}, {Padilla}, {Schmidt}, {Smith}, {Tyson}, {Verma},
  {Zehavi}, {Armus}, {Avestruz}, {Barrientos}, {Bowler}, {Bremer}, {Conselice},
  {Davies}, {Demarco}, {Dickinson}, {Galaz}, {Grazian}, {Holwerda}, {Jarvis},
  {Kasliwal}, {Lacerna}, {Loveday}, {Marshall}, {Merlin}, {Napolitano},
  {Puzia}, {Robotham}, {Salim}, {Sereno}, {Snyder}, {Stott}, {Tissera},
  {Werner}, {Yoachim}, {Borne}, \& {Members of the LSST Galaxies Science
  Collaboration}}]{robertson2017a}
{Robertson}, B.~E., {Banerji}, M., {Cooper}, M.~C., {et~al.} 2017, arXiv
  e-prints, arXiv:1708.01617

\bibitem[{Robertson {et~al.}(2019)Robertson, Banerji, Brough, Davies, Ferguson,
  Hausen, Kaviraj, Newman, Schmidt, Tyson, \& Wechsler}]{robertson2019a}
Robertson, B.~E., Banerji, M., Brough, S., {et~al.} 2019, Nature Reviews
  Physics, doi:10.1038/s42254-019-0067-x.
\newblock \url{https://doi.org/10.1038/s42254-019-0067-x}

\bibitem[{{Ronneberger} {et~al.}(2015){Ronneberger}, {Fischer}, \&
  {Brox}}]{ronneberger2015a}
{Ronneberger}, O., {Fischer}, P., \& {Brox}, T. 2015, ArXiv e-prints,
  arXiv:1505.04597

\bibitem[{Rossum(1995)}]{rossum1995}
Rossum, G. 1995, Python Reference Manual, Tech. rep., Amsterdam, The
  Netherlands, The Netherlands

\bibitem[{{Rumelhart} {et~al.}(1986){Rumelhart}, {Hinton}, \&
  {Williams}}]{rumelhart1986a}
{Rumelhart}, D.~E., {Hinton}, G.~E., \& {Williams}, R.~J. 1986, \nat, 323, 533

\bibitem[{Russakovsky {et~al.}(2015)Russakovsky, Deng, Su, Krause, Satheesh,
  Ma, Huang, Karpathy, Khosla, Bernstein, Berg, \& Fei-Fei}]{russakovsky2015a}
Russakovsky, O., Deng, J., Su, H., {et~al.} 2015, International Journal of
  Computer Vision (IJCV), 115, 211

\bibitem[{{Sersic}(1968)}]{sersic1968a}
{Sersic}, J.~L. 1968, {Atlas de Galaxias Australes} (Cordoba, Argentina:
  Observatorio Astronomico)

\bibitem[{{Shen} {et~al.}(2003){Shen}, {Mo}, {White}, {Blanton}, {Kauffmann},
  {Voges}, {Brinkmann}, \& {Csabai}}]{shen2003a}
{Shen}, S., {Mo}, H.~J., {White}, S.~D.~M., {et~al.} 2003, \mnras, 343, 978

\bibitem[{{Sheth} {et~al.}(2008){Sheth}, {Elmegreen}, {Elmegreen}, {Capak},
  {Abraham}, {Athanassoula}, {Ellis}, {Mobasher}, {Salvato}, {Schinnerer},
  {Scoville}, {Spalsbury}, {Strubbe}, {Carollo}, {Rich}, \&
  {West}}]{sheth2008a}
{Sheth}, K., {Elmegreen}, D.~M., {Elmegreen}, B.~G., {et~al.} 2008, \apj, 675,
  1141

\bibitem[{{Sheth} {et~al.}(2010){Sheth}, {Regan}, {Hinz}, {Gil de Paz},
  {Men{\'e}ndez-Delmestre}, {Mu{\~n}oz-Mateos}, {Seibert}, {Kim},
  {Laurikainen}, {Salo}, {Gadotti}, {Laine}, {Mizusawa}, {Armus},
  {Athanassoula}, {Bosma}, {Buta}, {Capak}, {Jarrett}, {Elmegreen},
  {Elmegreen}, {Knapen}, {Koda}, {Helou}, {Ho}, {Madore}, {Masters},
  {Mobasher}, {Ogle}, {Peng}, {Schinnerer}, {Surace}, {Zaritsky},
  {Comer{\'o}n}, {de Swardt}, {Meidt}, {Kasliwal}, \& {Aravena}}]{sheth2010a}
{Sheth}, K., {Regan}, M., {Hinz}, J.~L., {et~al.} 2010, \pasp, 122, 1397

\bibitem[{{Shibuya} {et~al.}(2015){Shibuya}, {Ouchi}, \&
  {Harikane}}]{shibuya2015a}
{Shibuya}, T., {Ouchi}, M., \& {Harikane}, Y. 2015, \apjs, 219, 15

\bibitem[{{Simmons} {et~al.}(2014){Simmons}, {Melvin}, {Lintott}, {Masters},
  {Willett}, {Keel}, {Smethurst}, {Cheung}, {Nichol}, {Schawinski},
  {Rutkowski}, {Kartaltepe}, {Bell}, {Casteels}, {Conselice}, {Almaini},
  {Ferguson}, {Fortson}, {Hartley}, {Kocevski}, {Koekemoer}, {McIntosh},
  {Mortlock}, {Newman}, {Ownsworth}, {Bamford}, {Dahlen}, {Faber},
  {Finkelstein}, {Fontana}, {Galametz}, {Grogin}, {Gr{\"u}tzbauch}, {Guo},
  {H{\"a}u{\ss}ler}, {Jek}, {Kaviraj}, {Lucas}, {Peth}, {Salvato}, {Wiklind},
  \& {Wuyts}}]{simmons2014a}
{Simmons}, B.~D., {Melvin}, T., {Lintott}, C., {et~al.} 2014, \mnras, 445, 3466

\bibitem[{{Simmons} {et~al.}(2017){Simmons}, {Lintott}, {Willett}, {Masters},
  {Kartaltepe}, {H{\"a}u{\ss}ler}, {Kaviraj}, {Krawczyk}, {Kruk}, {McIntosh},
  {Smethurst}, {Nichol}, {Scarlata}, {Schawinski}, {Conselice}, {Almaini},
  {Ferguson}, {Fortson}, {Hartley}, {Kocevski}, {Koekemoer}, {Mortlock},
  {Newman}, {Bamford}, {Grogin}, {Lucas}, {Hathi}, {McGrath}, {Peth}, {Pforr},
  {Rizer}, {Wuyts}, {Barro}, {Bell}, {Castellano}, {Dahlen}, {Dekel},
  {Ownsworth}, {Faber}, {Finkelstein}, {Fontana}, {Galametz}, {Gr{\"u}tzbauch},
  {Koo}, {Lotz}, {Mobasher}, {Mozena}, {Salvato}, \& {Wiklind}}]{simmons2017a}
{Simmons}, B.~D., {Lintott}, C., {Willett}, K.~W., {et~al.} 2017, \mnras, 464,
  4420

\bibitem[{{Skelton} {et~al.}(2014){Skelton}, {Whitaker}, {Momcheva}, {Brammer},
  {van Dokkum}, {Labb{\'e}}, {Franx}, {van der Wel}, {Bezanson}, {Da Cunha},
  {Fumagalli}, {F{\"o}rster Schreiber}, {Kriek}, {Leja}, {Lundgren}, {Magee},
  {Marchesini}, {Maseda}, {Nelson}, {Oesch}, {Pacifici}, {Patel}, {Price},
  {Rix}, {Tal}, {Wake}, \& {Wuyts}}]{skelton2014a}
{Skelton}, R.~E., {Whitaker}, K.~E., {Momcheva}, I.~G., {et~al.} 2014, \apjs,
  214, 24

\bibitem[{{Smail} {et~al.}(1997){Smail}, {Dressler}, {Couch}, {Ellis},
  {Oemler}, {Butcher}, \& {Sharples}}]{smail1997a}
{Smail}, I., {Dressler}, A., {Couch}, W.~J., {et~al.} 1997, \apjs, 110, 213

\bibitem[{{Smith} \& {Geach}(2019)}]{smith2019a}
{Smith}, M.~J., \& {Geach}, J.~E. 2019, arXiv e-prints, arXiv:1904.10286

\bibitem[{Sobel \& Feldman(1968)}]{sobel1968a}
Sobel, I., \& Feldman, G. 1968, presented at the Stanford Artificial
  Intelligence Project

\bibitem[{{Sudre} {et~al.}(2017){Sudre}, {Li}, {Vercauteren}, {Ourselin}, \&
  {Cardoso}}]{sudre2017a}
{Sudre}, C.~H., {Li}, W., {Vercauteren}, T., {Ourselin}, S., \& {Cardoso},
  M.~J. 2017, ArXiv e-prints, arXiv:1707.03237

\bibitem[{Szegedy {et~al.}(2016)Szegedy, Vanhoucke, Ioffe, Shlens, \&
  Wojna}]{szegedy2016a}
Szegedy, C., Vanhoucke, V., Ioffe, S., Shlens, J., \& Wojna, Z. 2016, in
  Proceedings of the IEEE conference on computer vision and pattern
  recognition, 2818--2826

\bibitem[{{Tarsitano} {et~al.}(2018){Tarsitano}, {Hartley}, {Amara}, {Bluck},
  {Bruderer}, {Carollo}, {Conselice}, {Melchior}, {Moraes}, {Refregier},
  {Sevilla-Noarbe}, {Woo}, {Abbott}, {Allam}, {Annis}, {Avila}, {Banerji},
  {Bertin}, {Brooks}, {Burke}, {Carnero Rosell}, {Carrasco Kind}, {Carretero},
  {Cunha}, {D'Andrea}, {da Costa}, {Davis}, {De Vicente}, {Desai}, {Doel},
  {Estrada}, {Frieman}, {Garc{\'{\i}}a-Bellido}, {Gruen}, {Gruendl},
  {Gutierrez}, {Hollowood}, {Honscheid}, {James}, {Jeltema}, {Krause}, {Kuehn},
  {Kuropatkin}, {Lahav}, {Maia}, {Menanteau}, {Miquel}, {Plazas}, {Romer},
  {Roodman}, {Sanchez}, {Santiago}, {Schindler}, {Smith}, {Smith},
  {Soares-Santos}, {Sobreira}, {Suchyta}, {Swanson}, {Tarle}, {Thomas},
  {Vikram}, \& {Walker}}]{tarsitano2018a}
{Tarsitano}, F., {Hartley}, W.~G., {Amara}, A., {et~al.} 2018, \mnras, 481,
  2018

\bibitem[{{Toft} {et~al.}(2007){Toft}, {van Dokkum}, {Franx}, {Labbe},
  {F{\"o}rster Schreiber}, {Wuyts}, {Webb}, {Rudnick}, {Zirm}, {Kriek}, {van
  der Werf}, {Blakeslee}, {Illingworth}, {Rix}, {Papovich}, \&
  {Moorwood}}]{toft2007a}
{Toft}, S., {van Dokkum}, P., {Franx}, M., {et~al.} 2007, \apj, 671, 285

\bibitem[{{Tremaine} {et~al.}(2002){Tremaine}, {Gebhardt}, {Bender}, {Bower},
  {Dressler}, {Faber}, {Filippenko}, {Green}, {Grillmair}, {Ho}, {Kormendy},
  {Lauer}, {Magorrian}, {Pinkney}, \& {Richstone}}]{tremaine2002a}
{Tremaine}, S., {Gebhardt}, K., {Bender}, R., {et~al.} 2002, \apj, 574, 740

\bibitem[{{Trujillo} {et~al.}(2004){Trujillo}, {Rudnick}, {Rix}, {Labb{\'e}},
  {Franx}, {Daddi}, {van Dokkum}, {F{\"o}rster Schreiber}, {Kuijken},
  {Moorwood}, {R{\"o}ttgering}, {van der Wel}, {van der Werf}, \& {van
  Starkenburg}}]{trujillo2004a}
{Trujillo}, I., {Rudnick}, G., {Rix}, H.-W., {et~al.} 2004, \apj, 604, 521

\bibitem[{{Trujillo} {et~al.}(2006){Trujillo}, {F{\"o}rster Schreiber},
  {Rudnick}, {Barden}, {Franx}, {Rix}, {Caldwell}, {McIntosh}, {Toft},
  {H{\"a}ussler}, {Zirm}, {van Dokkum}, {Labb{\'e}}, {Moorwood},
  {R{\"o}ttgering}, {van der Wel}, {van der Werf}, \& {van
  Starkenburg}}]{trujillo2006a}
{Trujillo}, I., {F{\"o}rster Schreiber}, N.~M., {Rudnick}, G., {et~al.} 2006,
  \apj, 650, 18

\bibitem[{{Tuccillo} {et~al.}(2018){Tuccillo}, {Huertas-Company},
  {Decenci{\`e}re}, {Velasco-Forero}, {Dom{\'{\i}}nguez S{\'a}nchez}, \&
  {Dimauro}}]{tuccillo2018a}
{Tuccillo}, D., {Huertas-Company}, M., {Decenci{\`e}re}, E., {et~al.} 2018,
  \mnras, 475, 894

\bibitem[{{van der Walt} {et~al.}(2011){van der Walt}, {Colbert}, \&
  {Varoquaux}}]{vanderwal2011a}
{van der Walt}, S., {Colbert}, S.~C., \& {Varoquaux}, G. 2011, Computing in
  Science Engineering, 13, 22

\bibitem[{van~der Walt {et~al.}(2014)van~der Walt, {S}ch\"onberger,
  {Nunez-Iglesias}, {B}oulogne, {W}arner, {Y}ager, {G}ouillart, {Y}u, \& the
  scikit-image contributors}]{scikit-image}
van~der Walt, S., {S}ch\"onberger, J.~L., {Nunez-Iglesias}, J., {et~al.} 2014,
  PeerJ, 2, e453.
\newblock \url{https://doi.org/10.7717/peerj.453}

\bibitem[{{van der Wel} {et~al.}(2012){van der Wel}, {Bell}, {H{\"a}ussler},
  {McGrath}, {Chang}, {Guo}, {McIntosh}, {Rix}, {Barden}, {Cheung}, {Faber},
  {Ferguson}, {Galametz}, {Grogin}, {Hartley}, {Kartaltepe}, {Kocevski},
  {Koekemoer}, {Lotz}, {Mozena}, {Peth}, \& {Peng}}]{vanderwel2012a}
{van der Wel}, A., {Bell}, E.~F., {H{\"a}ussler}, B., {et~al.} 2012, \apjs,
  203, 24

\bibitem[{{van der Wel} {et~al.}(2014){van der Wel}, {Franx}, {van Dokkum},
  {Skelton}, {Momcheva}, {Whitaker}, {Brammer}, {Bell}, {Rix}, {Wuyts},
  {Ferguson}, {Holden}, {Barro}, {Koekemoer}, {Chang}, {McGrath},
  {H{\"a}ussler}, {Dekel}, {Behroozi}, {Fumagalli}, {Leja}, {Lundgren},
  {Maseda}, {Nelson}, {Wake}, {Patel}, {Labb{\'e}}, {Faber}, {Grogin}, \&
  {Kocevski}}]{vanderwel2014a}
{van der Wel}, A., {Franx}, M., {van Dokkum}, P.~G., {et~al.} 2014, \apj, 788,
  28

\bibitem[{{van Dokkum} {et~al.}(2010){van Dokkum}, {Whitaker}, {Brammer},
  {Franx}, {Kriek}, {Labb{\'e}}, {Marchesini}, {Quadri}, {Bezanson},
  {Illingworth}, {Muzzin}, {Rudnick}, {Tal}, \& {Wake}}]{vandokkum2010a}
{van Dokkum}, P.~G., {Whitaker}, K.~E., {Brammer}, G., {et~al.} 2010, \apj,
  709, 1018

\bibitem[{{Vaucouleurs}(1959)}]{devaucouleurs1959a}
{Vaucouleurs}, G. 1959, Handbuch der Physik, 53, 275

\bibitem[{{Walmsley} {et~al.}(2019){Walmsley}, {Smith}, {Lintott}, {Gal},
  {Bamford}, {Dickinson}, {Fortson}, {Kruk}, {Masters}, \&
  {Scarlata}}]{walmsley2019a}
{Walmsley}, M., {Smith}, L., {Lintott}, C., {et~al.} 2019, arXiv e-prints,
  arXiv:1905.07424

\bibitem[{{Weigel} {et~al.}(2017){Weigel}, {Schawinski}, {Caplar}, {Carpineti},
  {Hart}, {Kaviraj}, {Keel}, {Kruk}, {Lintott}, {Nichol}, {Simmons}, \&
  {Smethurst}}]{weigel2017a}
{Weigel}, A.~K., {Schawinski}, K., {Caplar}, N., {et~al.} 2017, \apj, 845, 145

\bibitem[{{Wells} {et~al.}(1981){Wells}, {Greisen}, \& {Harten}}]{wells1981a}
{Wells}, D.~C., {Greisen}, E.~W., \& {Harten}, R.~H. 1981, \aaps, 44, 363

\bibitem[{{Whitaker} {et~al.}(2015){Whitaker}, {Franx}, {Bezanson}, {Brammer},
  {van Dokkum}, {Kriek}, {Labb{\'e}}, {Leja}, {Momcheva}, {Nelson}, {Rigby},
  {Rix}, {Skelton}, {van der Wel}, \& {Wuyts}}]{whitaker2015a}
{Whitaker}, K.~E., {Franx}, M., {Bezanson}, R., {et~al.} 2015, \apjl, 811, L12

\bibitem[{{Willett} {et~al.}(2013){Willett}, {Lintott}, {Bamford}, {Masters},
  {Simmons}, {Casteels}, {Edmondson}, {Fortson}, {Kaviraj}, {Keel}, {Melvin},
  {Nichol}, {Raddick}, {Schawinski}, {Simpson}, {Skibba}, {Smith}, \&
  {Thomas}}]{lintott2013a}
{Willett}, K.~W., {Lintott}, C.~J., {Bamford}, S.~P., {et~al.} 2013, \mnras,
  435, 2835

\bibitem[{{Willett} {et~al.}(2015){Willett}, {Schawinski}, {Simmons},
  {Masters}, {Skibba}, {Kaviraj}, {Melvin}, {Wong}, {Nichol}, {Cheung},
  {Lintott}, \& {Fortson}}]{willett2015a}
{Willett}, K.~W., {Schawinski}, K., {Simmons}, B.~D., {et~al.} 2015, \mnras,
  449, 820

\bibitem[{{Willett} {et~al.}(2017){Willett}, {Galloway}, {Bamford}, {Lintott},
  {Masters}, {Scarlata}, {Simmons}, {Beck}, {Cardamone}, {Cheung}, {Edmondson},
  {Fortson}, {Griffith}, {H{\"a}u{\ss}ler}, {Han}, {Hart}, {Melvin}, {Parrish},
  {Schawinski}, {Smethurst}, \& {Smith}}]{lintott2017a}
{Willett}, K.~W., {Galloway}, M.~A., {Bamford}, S.~P., {et~al.} 2017, \mnras,
  464, 4176

\bibitem[{{Williams} {et~al.}(2017){Williams}, {Giavalisco}, {Bezanson},
  {Cappelluti}, {Cassata}, {Liu}, {Lee}, {Tundo}, \&
  {Vanzella}}]{williams2017a}
{Williams}, C.~C., {Giavalisco}, M., {Bezanson}, R., {et~al.} 2017, \apj, 838,
  94

\bibitem[{{Williams} {et~al.}(2018){Williams}, {Curtis-Lake}, {Hainline},
  {Chevallard}, {Robertson}, {Charlot}, {Endsley}, {Stark}, {Willmer},
  {Alberts}, {Amorin}, {Arribas}, {Baum}, {Bunker}, {Carniani}, {Crandall},
  {Egami}, {Eisenstein}, {Ferruit}, {Husemann}, {Maseda}, {Maiolino}, {Rawle},
  {Rieke}, {Smit}, {Tacchella}, \& {Willott}}]{williams2018a}
{Williams}, C.~C., {Curtis-Lake}, E., {Hainline}, K.~N., {et~al.} 2018, \apjs,
  236, 33

\bibitem[{{Wuyts} {et~al.}(2011){Wuyts}, {F{\"o}rster Schreiber}, {van der
  Wel}, {Magnelli}, {Guo}, {Genzel}, {Lutz}, {Aussel}, {Barro}, {Berta},
  {Cava}, {Graci{\'a}-Carpio}, {Hathi}, {Huang}, {Kocevski}, {Koekemoer},
  {Lee}, {Le Floc'h}, {McGrath}, {Nordon}, {Popesso}, {Pozzi}, {Riguccini},
  {Rodighiero}, {Saintonge}, \& {Tacconi}}]{wuyts2011a}
{Wuyts}, S., {F{\"o}rster Schreiber}, N.~M., {van der Wel}, A., {et~al.} 2011,
  \apj, 742, 96

\bibitem[{{Yano} {et~al.}(2016){Yano}, {Kriek}, {van der Wel}, \&
  {Whitaker}}]{yano2016a}
{Yano}, M., {Kriek}, M., {van der Wel}, A., \& {Whitaker}, K.~E. 2016, \apjl,
  817, L21

\bibitem[{{Zhang} {et~al.}(2019){Zhang}, {Primack}, {Faber}, {Koo}, {Dekel},
  {Chen}, {Ceverino}, {Chang}, {Fang}, {Guo}, {Lin}, \& {Wel}}]{zhang2019a}
{Zhang}, H., {Primack}, J.~R., {Faber}, S.~M., {et~al.} 2019, \mnras, 484, 5170

\bibitem[{{Zirm} {et~al.}(2007){Zirm}, {van der Wel}, {Franx}, {Labb{\'e}},
  {Trujillo}, {van Dokkum}, {Toft}, {Daddi}, {Rudnick}, {Rix},
  {R{\"o}ttgering}, \& {van der Werf}}]{zirm2007a}
{Zirm}, A.~W., {van der Wel}, A., {Franx}, M., {et~al.} 2007, \apj, 656, 66

\end{thebibliography}

\end{document}